\documentclass[fleqn,usenatbib]{mnras}

\usepackage{newtxtext,newtxmath}

\usepackage[T1]{fontenc}
\newcommand{\code}{\texttt{Nullasar}}

\DeclareRobustCommand{\VAN}[3]{#2}
\let\VANthebibliography\thebibliography
\def\thebibliography{\DeclareRobustCommand{\VAN}[3]{##3}\VANthebibliography}

\usepackage{xcolor}
\definecolor{plotBlue}{RGB}{31,119,180}
\definecolor{plotOrange}{RGB}{255,127,14}
\definecolor{plotGreen}{RGB}{44,160,44}
\definecolor{plotRed}{RGB}{214,39,40}
\definecolor{plotPurple}{RGB}{148,103,189}
\definecolor{plotBrown}{RGB}{140,86,75}
\definecolor{plotPink}{RGB}{227,119,194}
\definecolor{plotGrey}{RGB}{128,128,128}
\definecolor{histBlue}{RGB}{100,149,237}
\definecolor{histRed}{RGB}{250,128,114}
\definecolor{referenceGreen}{RGB}{34,139,34}
\definecolor{referenceOrange}{RGB}{255,140,0}
\definecolor{ablationBlue}{RGB}{0,114,178}
\definecolor{ablationOrange}{RGB}{213,94,0}
\definecolor{ablationGreen}{RGB}{0,158,115}
\definecolor{ablationGrey}{RGB}{127,127,127}
\usepackage{graphicx}	
\usepackage{amsmath}	

\usepackage{float}
\usepackage{algorithm}
\usepackage{algpseudocode}
\usepackage{placeins}

\usepackage{lastpage}

\title[Moving Beyond the DRW Model]{Moving Beyond the DRW Model:\\Learning Quasar Variability with Generative Models}

\author[N. B. Sellam et al.]{
Nadav Bojan Sellam,$^{1}$\thanks{E-mail: Nadav.Sellam@ist.ac.at}
Kevin Park,$^{1}$
Alex Bronstein$^{1}$
and Zolt\'an Haiman$^{1,2,3}$
\\
$^{1}$Institute of Science and Technology Austria (ISTA), Am Campus 1, Klosterneuburg 3400, Austria\\
$^{2}$Department of Astronomy, Columbia University New York, NY 10027, USA\\
$^{3}$Department of Physics, Columbia University New York, NY 10027, USA
}

\pubyear{\the\year{}}

\begin{document}
\raggedbottom
\label{firstpage}
\pagerange{\pageref{firstpage}--\pageref{LastPage}}
\maketitle

\begin{abstract}
The optical continuum variability of quasars is often statistically described by a damped random walk (DRW) model, with characteristic damping timescales $\tau_{\rm DRW}$ of hundreds of days and amplitudes $\sigma_{\rm DRW}$ of O(0.1) magnitudes. This model has been extensively used as a null hypothesis to assess the significance of periodic signals in quasar light curves in the search for supermassive black hole binaries, the significance of extreme quasar variability like tidal disruption events (TDEs) and changing-look (CL) quasars, or to model time lags in reverberation mapping. However, the DRW model is known to be incomplete, and an incorrect null hypothesis can lead to biased significance estimates. 
  In this work, we investigate the impact of such null-hypothesis mis-specification on the detection of periodic signals in synthetic quasar light curves and introduce the generative model \code, capable of learning quasar variability directly from time-series data. We demonstrate that \code, trained using flow matching, can learn delicate statistical properties of stochastic light-curve variability without explicitly assuming any noise model. We further demonstrate that \code\ is capable of learning correlated distributions of noise parameters and can generate new light-curve realizations from a range of deviations from the Gaussian DRW noise model,
  effectively approximating parametric descriptions of stochastic processes. When used as a null hypothesis for detecting periodic signals in synthetic light curves, \code\ performs comparably to fitting the correct DRW model to each light curve, and outperforms parametric fitting when the variability follows a more flexible AR(2) model. Finally, we identify practical limitations, including sensitivity to contamination of the training set by non-null signals, and show that projecting the conditioning vector of the model onto its leading principal components largely mitigates it.
\code\ is publicly available on GitHub.
\end{abstract}

\section{Introduction}\label{sec:introduction}

Quasars are growing supermassive black holes (SMBHs) located at the centers of galaxies with inferred masses of 
$10^6-10^9\,{\rm M_\odot}$. 
The optical and ultraviolet emission from the accretion disks of these black holes typically outshine their host galaxy, and the full spectrum of their electromagnetic emission ranges from the radio to X-ray wavelengths. It has also been discovered that the optical emission of quasars is variable, although the exact physical mechanism within the accretion disk which generates this stochastic variability is not precisely understood. The advent of optical surveys like the Catalina Real-time Transient Survey (CRTS) or the Zwicky Transient Facility (ZTF), which have monitored large areas of the sky at short cadence of days to weeks has yielded quasar light curves with baselines of years. Statistical analyses of these time series have shown that the damped random walk (DRW) model, or an auto-regressive process of order 1 (hereafter AR(1)), well-describes quasar optical variability~\citep{Kelly+2009,MacLeod+2010,Kozlowski+2010}. Beyond its simplicity, its success has been driven by the fact that the DRW parameters $\sigma_{\rm DRW}, \tau_{\rm DRW}$ show significant correlations with physical properties like BH mass and Eddington ratio. On the other hand, it has been recognized that stochastic quasar variability is more complex and cannot be fully characterized by the DRW model, motivating the development of AR(2), non-stationary, and other alternative models \citep[see, e.g.][and references therein]{Kelly+2014,kepler_drw,mushotzky_kepler_drw,tachibana2020,Stone+2022,Kroupa+2026}.

Such models are valuable in a variety of astrophysical problems. In this paper, we focus on the search for supermassive black hole binaries (SMBHBs) in quasar light curves, for which characterizing stochastic variability is crucial and has been a major limitation.
SMBHBs have long been expected to be natural by-products of galaxy mergers~\citep{begelman1980}.  More recently,
hydrodynamical simulations have shown that these SMBHBs accrete efficiently from a circumbinary disk and exhibit bright electromagnetic emission long before their merger (see, e.g., \citealt{Westernacher-Schneider+2022} and the recent review \citealt{Bogdanovic+2022} and references therein). Due to their small angular size at cosmological distances, direct imaging of the orbital motion of compact, gravitationally bound binaries in galactic nuclei at sub-pc scales is currently not feasible, even for the very-long baseline interferometers (VLBIs) such as the Event Horizon Telescope 
 \citep{d_orazio_2023}. 
 An indirect way to find SMBHBs at the current stage is to look for periodicity in quasar/AGN light curves in time-domain surveys. Hundreds of SMBHB candidates have been found in time-domain surveys, including PTF  
  \citep{Charisi+2016}, CRTS \citep{Graham+2015b}, and ZTF \citep{Chen+2024} 
 and this number is expected to 
 vastly 
 increase with the upcoming data releases of the Vera Rubin Observatory's Legacy Survey of Space and Time (LSST; see predictions in \citealt{XinHaiman2021,XinHaiman2024}) and other forthcoming time-domain surveys by the Nancy Grace Roman Space Telescope~\citep{Roman} and by ULTRASAT~\citep{ULTRASAT}.
 
 In such searches, the significance of a periodicity is typically assessed relative to the DRW, i.e. the null hypothesis assuming a single-SMBH quasar. This red-noise variability is known to easily mimic sinusoidal modulations for a few periods 
\citep{Vaughan+2016}, 
 and reassessment of SMBHB candidates by either extending their light curves with more data or with a different null-hypothesis has disproved or cast doubt on the significance of many binary candidates 
\citep{Jakob_null, El-Badry_2025}. For example, 
\citet{Jakob_null} 
 use sinusoids with periods that are randomized each cycle as the null hypothesis. Under this modified null hypothesis, the periodicity of 33 binary candidates found in PTF by \citet{Charisi+2016} relative to the null hypothesis of DRW were shown to be statistically insignificant.

The above examples highlight the fact that if the null hypothesis is mis-specified, the resulting significance estimates can be systematically biased. In such cases, stochastic fluctuations generated by the true noise process may be incorrectly interpreted as periodic signals, leading to false positives, or genuine periodic components may be masked by an overly permissive null, leading to false negatives, especially when the signal shape is not sinusoidal~\citep{lin2025, park2025}.
Importantly, these effects can arise even from mild mismatches between the assumed and the true variability process, and assuming the wrong noise model can distort statistical conclusions in structured, frequency-dependent ways.
 
In this paper, our goal is to develop a tool that can characterize stochastic quasar variability and avoid the above limitations. Specifically, we build an emulator, \href{https://github.com/NadavSellamWork/Nullasar}{\code}, which can generate synthetic quasars light-curves with the same statistical properties as an input sample. \code\ learns the variability of an entire population of light curves and, given a single observed light curve, generates new realizations consistent with it. The population supplies information that a single short light curve cannot constrain on its own, without committing to a fixed parametric noise model. \code\ is therefore suitable for assessing false alarm probabilities in searching for periodicities, and more broadly other rare signals, among large catalogs of quasar light-curves.

The rest of this paper is organized as follows. In \S~\ref{sec:ar1_vs_ar2}, we provide
a simple example of null-hypothesis mis-specification using synthetic
AR(1) and AR(2) light curves. In \S~\ref{sec:generative_models}, we introduce our generative model. In
\S~\ref{sec:results}, we test the model on a series of synthetic experiments, and
\S~\ref{sec:discussion} and \S~\ref{sec:conclusions} discuss the results and their implications.

\section{AR(1) vs AR(2): the impact of null hypothesis mis-specification}\label{sec:ar1_vs_ar2}

\begin{table}
  \centering
  \caption{Terminology.  }
  \label{tab:terminology}
  \begin{tabular}{|p{0.1\textwidth}|p{0.3\textwidth}|}
    \hline
    \textbf{Term} & \textbf{Description} \\
    \hline
    AR(1) & Auto-regressive process of order 1; a normalized version of the "Damped Random Walk (DRW)" model for quasar variability. See Eq.~\ref{eq:ar1}\\
    \hline
   AR(2) & Auto-regressive process of order 2; a normalized version of the "Damped Harmonic Oscillator (DHO)" model for quasar variability. See Eq.~\ref{eq:ar2}\\
    \hline
  \end{tabular}
\end{table}

As an illustrative example, we demonstrate the significance of
null-hypothesis mis-specification in a controlled and fully
parametric setting: we construct 8,000 synthetic light curves of
length $T = 256$, uniformly sampled with unit time step ($\Delta t
= 1$), in which the true stochastic variability follows a discrete
autoregressive process of order two (AR(2)) given by
\begin{equation}\label{eq:ar2}
    x_t = \phi_1 x_{t-1} + \phi_2 x_{t-2} + \epsilon_t,
\end{equation}
where $\phi_1$ and $\phi_2$ are the autoregressive coefficients,
and $\epsilon_t$ is a discrete white-noise process with zero mean
and unit variance.

\begin{figure*}
    \centering
    \includegraphics[width=1.0\textwidth]{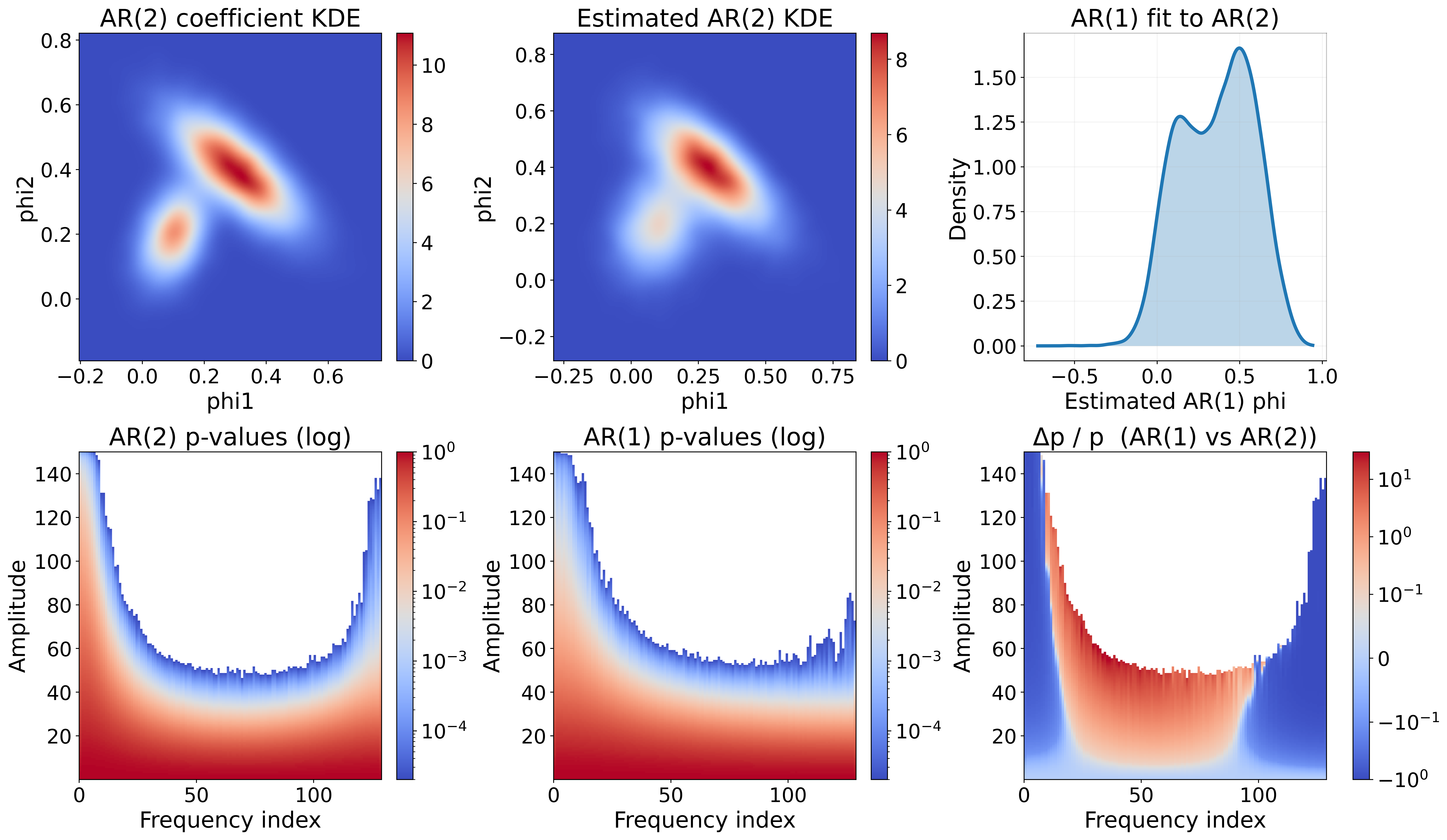}
    \caption{
    Illustration of null-hypothesis mis-specification in a controlled synthetic example, based on 8,000 AR(2) light curves.
    \emph{Top left:} true joint distribution of AR(2) coefficients $(\phi_1,\phi_2)$ used to generate the light curves.
    \emph{Top middle:} estimated coefficient distribution obtained by fitting an AR(2) model to the simulated data, accurately recovering the true structure.
    \emph{Top right:} distribution of AR(1) coefficients obtained by fitting an incorrect AR(1) noise model to the same light curves, collapsing the correlated AR(2) variability into a single effective parameter.
    \emph{Bottom left:} $p$-value, i.e. the probability that the noise alone produces a Fourier amplitude at least as large as the value on the vertical axis, as a function of discrete frequency, for light curves generated from the true AR(2) model, shown on a logarithmic color scale.
    \emph{Bottom middle:} corresponding $p$-values obtained from the fitted AR(1) null model, on the same logarithmic scale.
    In both panels, white pixels mark amplitudes that are not reached by any simulated noise realization at that frequency, so that the empirical $p$-value is zero (i.e. below the resolution set by the finite number of simulated light curves) and cannot be displayed on a logarithmic scale.
    \emph{Bottom right:} relative difference between the $p$-values computed under the correct AR(2) and the incorrect AR(1) null hypotheses, $\Delta p / p_{\mathrm{AR1}} = (p_{\mathrm{AR2}} - p_{\mathrm{AR1}})/p_{\mathrm{AR1}}$, shown on a symmetric logarithmic color scale (linear close to zero). Positive values indicate zones where the mis-specified AR(1) model gives a $p$-value that is too low, causing false positive detections, while negative values indicate zones where AR(1) gives a $p$-value that is too high, causing false negatives. Since $p_{\mathrm{AR2}} \geq 0$, negative values are bounded below by $-1$, which is approached where $p_{\mathrm{AR2}} \ll p_{\mathrm{AR1}}$, whereas positive values can greatly exceed unity where $p_{\mathrm{AR1}}$ is small. White pixels mark amplitudes at which $p_{\mathrm{AR1}} = 0$, where the ratio is undefined.}
    \label{fig:coeff_distributions}
\end{figure*}

To generate a heterogeneous population of stochastic processes, the coefficient pairs $(\phi_1,\phi_2)$ are drawn from a correlated mixture of two 2D Gaussian distributions characterized by their centers and covariance matrices
\begin{equation}\label{eq:cen1}
    \texttt{center1}=(0.1,0.2),\; \texttt{cov1}=\begin{bmatrix}
        0.004\; 0.002\\
        0.002\; 0.008
    \end{bmatrix},
\end{equation}
and
\begin{equation}\label{eq:cen2}
    \texttt{center2}=(0.3,0.4),\; \texttt{cov2}=\begin{bmatrix}
        0.015\; -0.01\\
        -0.01\; 0.012
    \end{bmatrix}.
\end{equation}
The first Gaussian component is assigned a weight of $0.3$, and the second a weight of $0.7$. The resulting joint distribution over $(\phi_1,\phi_2)$ is shown in the top-left panel of Fig.~\ref{fig:coeff_distributions}.

To efficiently estimate the coefficients $\phi_1$ and $\phi_2$, we use the Yule-Walker equations (\citealt{Yule1927,Walker1931}; see also Appendix \ref{appendix:YW_eq}), which accurately recover the underlying coefficient distribution, as shown in the top-middle panel of Fig.~\ref{fig:coeff_distributions}. Both the correlation structure and the relative weights of the mixture components are preserved, indicating that the true discrete noise model is identifiable when the correct parametric form is assumed to be known.

We also fit AR(1) noise models, defined as
\begin{equation}\label{eq:ar1}
    x_t=\phi x_{t-1}+\epsilon_t,
\end{equation}
to the same 8,000 AR(2) light curves, which produces an estimated distribution of $\phi$ shown in the top-right panel of Fig.~\ref{fig:coeff_distributions}.

The consequences of an incorrect noise model are evident in the frequency domain. We compute the discrete Fourier transforms of the light curves, and use them to compute the $p$-values shown in the bottom-left and middle panels of Fig.~\ref{fig:coeff_distributions}. The distributions of spectral amplitudes obtained from the true AR(2) model and from the fitted AR(1) model differ systematically as a function of frequency. When these distributions are used to compute frequency-dependent $p$-values, the resulting significance estimates diverge in a structured manner. Here, the $p$-value at a given frequency and amplitude quantifies the probability that a periodic signal with at least that Fourier amplitude could arise purely from the stochastic noise process. As shown in the bottom-right panel of Fig.~\ref{fig:coeff_distributions}, the incorrect AR(1) null hypothesis induces both spuriously significant detections (positive $\Delta p / p_{\mathrm{AR1}}$) and suppressed significance (negative $\Delta p / p_{\mathrm{AR1}}$) at specific frequencies, as seen in the per-frequency relative $p$-value difference (bottom right). In the intermediate-frequency band, the relative difference reaches values of order ten, i.e. the AR(1) null underestimates the $p$-value by about an order of magnitude, while near the lowest and highest frequencies it approaches $-1$, i.e. the AR(1) null assigns non-negligible $p$-values to amplitudes that are essentially never produced by the true AR(2) noise.

This example demonstrates that adopting an incorrect null hypothesis can lead to systematic and frequency-dependent errors in the statistical assessment of periodic signals, even when the assumed noise model provides locally plausible fits to individual light curves. In particular, after fitting AR(1) models to the AR(2) light curves, the mean absolute lag-1 residual autocorrelation is only $\sim 0.02$, well below the $1.96/\sqrt{T} \approx 0.12$ threshold
expected for white noise at $T=256$ \citep{BoxJenkinsReinsel1994}, indicating that the AR(1)
residuals show no statistically significant lag-1 structure, indicating that the residuals are close to white noise and that the AR(1) model captures much of the short-timescale temporal structure of the individual realizations. Nevertheless, despite this apparently reasonable local fit quality, the inferred frequency-dependent significance estimates remain systematically biased. In the context of large time-domain surveys and automated periodicity searches, such effects can substantially contaminate candidate samples and bias downstream inference.

Motivated by this observation, we seek an alternative approach to null-hypothesis modelling that does not rely on committing to a fixed, hand-specified noise model. In the remainder of this work, we explore a data-driven framework based on generative models that learn the effective stochastic noise structure directly from ensembles of light curves, while retaining the statistical interpretation of classical parametric null hypotheses.

\section{Generative models for stochastic light curves}\label{sec:generative_models}

\subsection{Notation}

We denote a light curve by $\mathbf{x} \in \mathbb{R}^{T}$, where $T$ is the number of data points.
Its discrete Fourier transform is denoted by $\tilde{\mathbf{x}} \in \mathbb{C}^{T}$. Samples from a base distribution are denoted $\mathbf{x}_0 \sim p_0(\mathbf{x})$, while samples from the target data distribution are denoted $\mathbf{x}_1 \sim p_{\mathrm{data}}(\mathbf{x})$. In this work, $p_0$, used as the source for the generative process is taken to be a standard multivariate normal distribution.

In a typical procedure, quasar variability is described by a parametric stochastic noise model, which is specified by the numerical values of its parameters (e.g.\ $\tau_{\rm DRW}, \sigma_{\rm DRW}$, or $\phi_1,\phi_2$ for an AR(2) process).  In our generative model we introduce the multi-dimensional latent vector
$\mathbf{z}$, which plays a role analogous to these noise-model parameters. However, $\mathbf{z}$ is not restricted to a fixed functional form and is learned directly from the data. 
Our generative model consists of two components:
\begin{itemize}
    \item A generator defined by a learned velocity field $\mathbf{v}_\theta(\mathbf{x}_t, t \mid \mathbf{z})$ where $\theta$ represents the model weights which are learned during training.

    \item An learned encoder (neural network) $E_\phi(\mathbf{x}) = \mathbf{z}$ that extracts a latent representation of the underlying stochastic process, with $\phi$ representing the model's learned weights.
\end{itemize}
We define a linear interpolation between $\mathbf{x}_0$ and $\mathbf{x}_1$:
\begin{equation}
    \mathbf{x}_t = (1 - t)\mathbf{x}_0 + t\mathbf{x}_1, \quad t \in [0,1].
\end{equation}

\subsection{Continuous normalizing flows and flow matching}

We model the distribution of light curves using a continuous normalizing flow (CNF; \citealt{Chen2018NeuralODEs}), defined by the ordinary differential equation
\begin{equation}
    \frac{\mathrm{d}\mathbf{x}_t}{\mathrm{d}t}
    =
    \mathbf{v}_\theta(\mathbf{x}_t, t \mid \mathbf{z}),
\end{equation}
where $\mathbf{v}_\theta$ transports samples from the base distribution to the data distribution. The dependence on $\mathbf{z}$ enables conditional generation, on the conditional properties that the encoder $E_\phi$ is trained to represent.

Training is performed using flow matching \citep{Lipman2023FlowMatching}. Given $\mathbf{x}_0 \sim p_0$ and $\mathbf{x}_1 \sim p_{\mathrm{data}}$, the target velocity along the interpolation path is
\begin{equation}
    \mathbf{v}^\star = \mathbf{x}_1 - \mathbf{x}_0.
\end{equation}
Note that while the interpolation path in Eq.~(5) has a trivial, constant instantaneous velocity $\mathrm{d}\mathbf{x}_t/\mathrm{d}t = \mathbf{x}_1 - \mathbf{x}_0$ for any single pair $(\mathbf{x}_0, \mathbf{x}_1)$, the velocity field $\mathbf{v}_\theta$ in Eq.~(6) is a learned function that, at generation time, must produce a useful direction of travel without knowledge of the (unobserved) target $\mathbf{x}_1$. Training $\mathbf{v}_\theta$ to match $\mathbf{x}_1 - \mathbf{x}_0$ in expectation over the data distribution is what allows integrating Eq.~(6) from pure noise $\mathbf{x}_0 \sim p_0$ to produce a realistic sample $\mathbf{x}_1$.

The model is trained by minimizing the regression objective
\begin{equation}
    \label{eq:fm_loss}
\mathcal{L}_{\mathrm{FM}} =
    \mathbb{E}_{\mathbf{x}_0, \mathbf{x}_1, t}
    \left[
        \left\|
        \mathbf{v}_\theta(\mathbf{x}_t, t \mid \mathbf{z})
        -
        (\mathbf{x}_1 - \mathbf{x}_0)
        \right\|^2
    \right].
\end{equation}
The full training objective, which combines conditional and unconditional generation, is given in Appendix~\ref{appendix:implementation}.

\subsection{Model architecture}

The velocity field $\mathbf{v}_\theta$ is parameterized by a one-dimensional U-Net \citep{Ronneberger2015UNet}. U-Nets are encoder-decoder neural network architectures designed to capture both local and global structure through a hierarchy of progressively coarser feature representations. The encoder compresses the input into increasingly abstract features with larger receptive fields, while the decoder reconstructs high-resolution representations from these compressed features. Skip connections between encoder and decoder layers preserve fine-scale information that may otherwise be lost during downsampling, making the architecture well suited for stochastic time series that contain structure across multiple temporal scales.

U-Net-based architectures are also widely used in modern generative models, including continuous normalizing flows, diffusion models, and score-based generative models, where they have proven effective at representing complex multi-scale structure in high-dimensional data. Their ability to combine local feature extraction with large receptive fields makes them a natural choice for modelling stochastic light curves, which contain both short-timescale variability and long-range temporal correlations.

The generative process is defined in the frequency domain, with the flow acting directly on the Fourier coefficients $\tilde{\mathbf{x}}$. This allows the model to explicitly represent global spectral structure, which is central to stochastic variability.
At the same time, the network processes both frequency-domain and time-domain representations. The input consists of $\tilde{\mathbf{x}}$ together with the corresponding time-domain signal $\mathbf{x}$. This joint representation enables the model to capture both global properties, such as power spectra and long-range correlations, and local temporal structure.

The network uses group normalization and SiLU activations for stable optimization, and includes a self-attention block at the bottleneck to capture long-range dependencies.
Conditioning is implemented via Feature-wise Linear Modulation (FiLM) \citep{Perez2018FiLM}, where the latent representation $\mathbf{z}$ modulates intermediate feature maps. The network size and training settings are given in Appendix~\ref{appendix:training}. The model is trained by minimizing the flow-matching loss described above (Eq.~\ref{eq:fm_loss}), using the Adam optimizer (in its AdamW variant).

\subsection{Encoder and representation learning}

The encoder $E_\phi$ maps a light curve to a latent representation $\mathbf{z} = E_\phi(\mathbf{x})$ that captures statistical features of the underlying stochastic process.
We train the encoder jointly with the generative model, so that gradients flow backward through both components during training. This joint optimization encourages the encoder to extract features reflecting the latent parameters governing the stochastic variability, while simultaneously allowing the generative model to learn how these features relate to observable time series realizations. During training, the light curve to be generated and the conditioning light curve are two different, non-overlapping segments of the same simulated light curve, while the light curves used for testing are simulated independently and are never used in training (see Appendix~\ref{appendix:training} for details). The flow-matching loss (Eq.~\ref{eq:fm_loss}) is the only training loss used for the models in this paper; additional losses that we explored are described in Appendix~\ref{appendix:ablation}.

\subsection{Generation}

Samples are generated by integrating the learned ODE:
\begin{equation}
    \frac{d\mathbf{x}_t}{dt} = \mathbf{v}_\theta(\mathbf{x}_t, t \mid \mathbf{z}),
\end{equation}
from $t=0$ to $t=1$, starting from $\mathbf{x}_0 \sim p_0$. We use 100 integration steps in all experiments, except for the conditional generation experiment (Section~\ref{subsec:conditional}), where we use 200 integration steps (see Appendix~\ref{appendix:implementation}).

We use Euler integration:
\begin{equation}
    \mathbf{x}_{t+\Delta t} = \mathbf{x}_t + \Delta t \cdot \mathbf{v}_\theta(\mathbf{x}_t, t \mid \mathbf{z}).
\end{equation}

\begin{algorithm}
\caption{Generation with Euler method}
\begin{algorithmic}[1]

\Procedure{Generate}{$\mathbf{x}$}

\State Sample $\tilde{\mathbf{x}}_0 \sim p_0$

\If{conditional generation}
    \State $\mathbf{z} \gets E(\mathbf{x}_{\mathrm{condition}})$
\EndIf

\For{$t = 0, \Delta t, 2\Delta t, \dots, 1$}
    \State $\mathbf{v} \gets \mathbf{v}_\theta(\tilde{\mathbf{x}}_t, t \mid \mathbf{z})$
    \State $\tilde{\mathbf{x}}_{t+\Delta t}
    \gets
    \tilde{\mathbf{x}}_t + \Delta t \cdot \mathbf{v}$
\EndFor

\State $\mathbf{x} \gets \mathcal{F}^{-1}(\tilde{\mathbf{x}}_1)$

\State \Return $\mathbf{x}$

\EndProcedure

\end{algorithmic}
\end{algorithm}

The model operates in the frequency domain. After integration, the final sample is transformed back:
\begin{equation}
    \mathbf{x} = \mathcal{F}^{-1}(\tilde{\mathbf{x}}_1).
\end{equation}

\section{Experiments}\label{sec:results}
All experiments presented in this section are based on synthetic datasets. This allows us to precisely control the underlying stochastic processes and evaluate the proposed framework under well-defined conditions. The experiments are designed to assess whether the model captures distributional structure beyond standard summary statistics, whether it can recover latent parametric structure, and whether this translates into improved statistical inference. Implementation and training details, including the use of several independently trained models to assess the sensitivity of the results to training randomness, are given in Appendix~\ref{appendix:training}.

\begin{table*}
  \centering
  \caption{High level summary of the five experiments performed in this work. }
  \label{tab:experiments}
  \begin{tabular}{|c|p{0.42\textwidth}|p{0.42\textwidth}|}
    \hline
    \textbf{Experiment} & \textbf{Description} & \textbf{Result} \\
    \hline
    1 & Can the generative model detect a small higher-order deviation from the DRW model (i.e. an AR(1) process)?  & Yes, even for the two groups of light curves in which the average DRW PSD is identical (Fig.~\ref{fig:psd_match}), the model detects and reproduces small deviations in the distribution of Fourier amplitudes (Fig.~\ref{fig:uniform_model}: uniform ablation vs Fig.~\ref{fig:rayleigh_model}: pure DRW)  \\
    \hline
    2 & Can the generative model learn and reproduce AR(2) light curves without being informed of the parametric form of the AR(2) model? & Yes, the model reproduces light curves whose fitted coefficients match the distribution of coefficients fitted to the data (Fig.~\ref{fig:ar2_recovery}). \\
    \hline
    3 & Can the generative model infer the parameters of an individual light curve, and generate new light curves with the same parameters? & Yes, the coefficients of the generated light curves follow those of the conditioning light curves, with an accuracy comparable to fitting the correct model and simulating new light curves, and a somewhat larger spread (Fig.~\ref{fig:conditional}). \\
    \hline
    4 & How do the generative model-generated light curves perform as a null hypothesis compared to other null hypotheses in a periodic signal detection exercise?  & In the case of DRW, the generative model performs comparably to a null hypothesis in which the DRW parameters are fitted to each light curve (Fig.~\ref{fig:roc_ar1}, ~\ref{fig:pvals_ar1}). In the case of a higher-order deviation to DRW, i.e. an AR(2) process, the generative model approaches the "oracle" scenario, in which the parameters are exactly known, outperforms fitting an AR(2) model to each light curve, and significantly outperforms a null hypothesis incorrectly assuming DRW (see Fig.~\ref{fig:roc_ar2}, ~\ref{fig:pvals_ar2}). \\
    \hline
    5 & If the training data is "contaminated" by periodic signals, does it affect detection performance of periodic signals, and can this be mitigated?  & When the training data contains $\geq 0.5\%$ periodic signals (the smallest fraction tested), the detection performance decreases substantially (Fig.~\ref{fig:contamination_roc}). Projecting the conditioning vector onto its leading principal components restores most of the detection performance, without retraining (Fig.~\ref{fig:pca_roc}). \\
    \hline
  \end{tabular}
\end{table*}

\subsection{Experiment 1: Learning ensemble-level noise structure beyond single realizations}

\paragraph*{Motivation.}
Many properties of stochastic processes are not identifiable from a single realization, and instead emerge only at the level of an ensemble. While an individual light curve provides a noisy estimate of its power spectral density (PSD), it does not determine the full distribution of Fourier amplitudes across realizations. These ensemble-level distributions encode higher-order statistical structure that cannot be recovered from any single light curve alone.
We therefore test whether the generative model can learn such ensemble-level structure beyond standard second-order statistics.
\begin{figure*}
    \centering
    \includegraphics[width=0.95\textwidth]{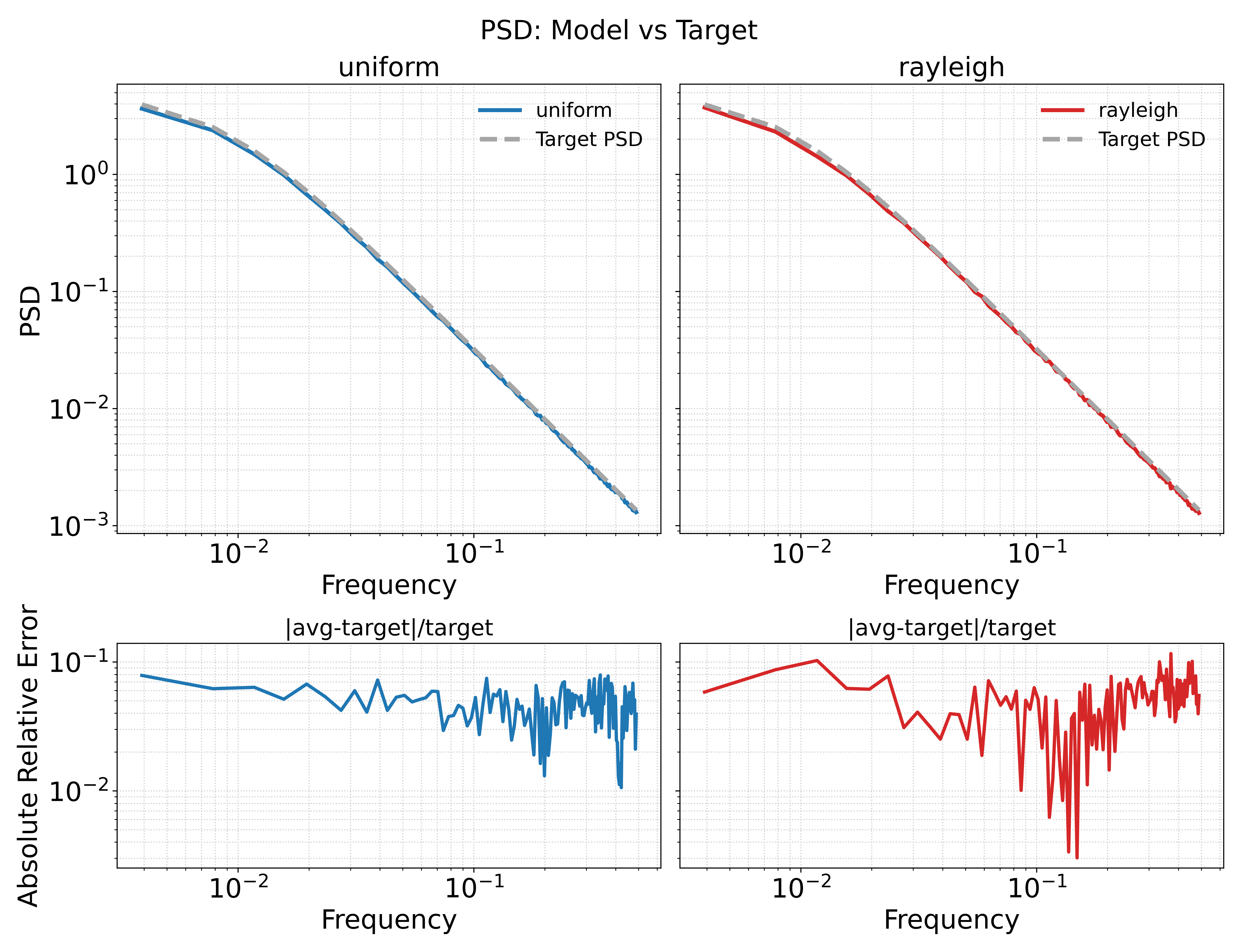}
    \caption{\textbf{Average PSDs of generated light curves compared to the target DRW PSD for both the DRW (Rayleigh) and Uniform-amplitude datasets.}
Top row: Solid curves show the empirical average PSD computed from 3,000 generated samples; dashed curves show the analytic target DRW PSD used during dataset construction. 
Bottom row: Absolute relative error between model and target PSD as a function of frequency.
Both datasets were constructed to have identical target PSDs but fundamentally different underlying Fourier-amplitude distributions (Rayleigh vs.\ Uniform). 
The model recovers the same PSD accurately for both datasets, as shown by the close agreement across all frequencies and the small errors in the bottom panels.
This result is crucial: it establishes that the two datasets are indistinguishable at the level of the power spectrum.}
    \label{fig:psd_match}
\end{figure*}

\begin{figure*}
    \centering
    \includegraphics[width=0.95\textwidth]{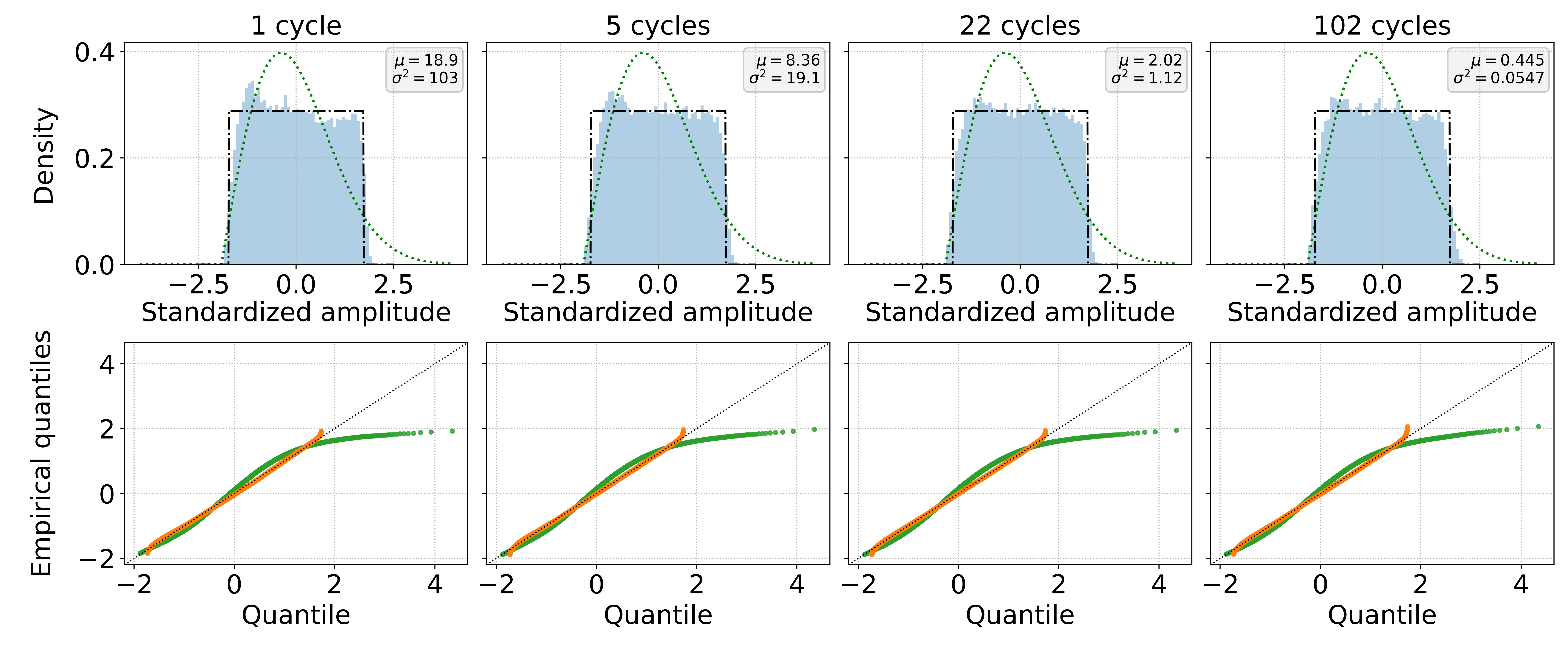}
    \caption{
    Fourier-amplitude statistics for the Uniform-amplitude (ablation) dataset, standardized per frequency (centered and normalized to zero mean, unit variance).
    \textit{Top row:} Empirical distributions of standardized Fourier amplitudes at representative Fourier modes corresponding to approximately 1, 5, 22, and 102 oscillation cycles across the full generated light curve. Lower cycle modes correspond to lower temporal frequencies and longer timescale variability, while higher cycle modes correspond to shorter timescale fluctuations. The boxed annotation in each panel gives the raw mean $\mu_k$ and variance $\sigma_k^2$ of the amplitude at that bin. The distributions exhibit the characteristic bounded shape of the Uniform amplitude construction.
\textit{Bottom row:} Quantile quantile (QQ) plots comparing generated amplitudes against reference distributions. A QQ plot tests whether two distributions match by comparing their quantiles: if they are identical, all points lie exactly on the identity line. Deviations from the line indicate differences in distributional shape or tails. Two comparisons are shown at each frequency: \textcolor[RGB]{255, 140, 0}{the orange line compares generated samples against a Uniform reference distribution}, while \textcolor[RGB]{34, 139, 34}{the green line compares against a Rayleigh reference distribution}. The close alignment of \textcolor[RGB]{255, 140, 0}{the orange line with the identity line} across all frequencies demonstrates that the model successfully reproduces the Uniform amplitude distribution. This shows the model captures higher order ensemble level statistics beyond the power spectrum—specifically, the bounded tail structure characteristic of Uniform amplitudes.}
    \label{fig:uniform_model}
\end{figure*}

\begin{figure*}
    \centering
    \includegraphics[width=0.95\textwidth]{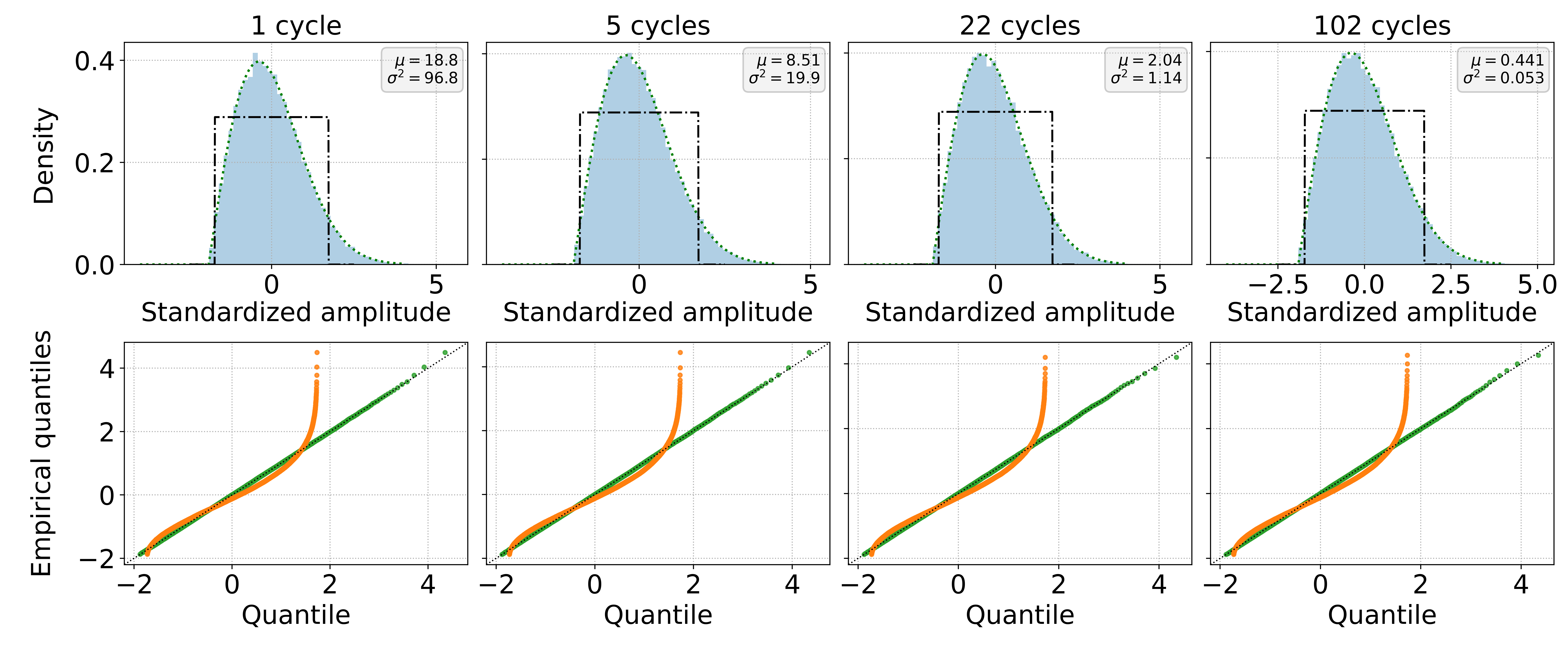}
    \caption{
    \textit{Top row:} Empirical distributions of standardized Fourier amplitudes at representative Fourier modes corresponding to approximately 1, 5, 22, and 102 oscillation cycles across the full generated light curve. Lower cycle modes correspond to lower temporal frequencies and longer timescale variability, while higher cycle modes correspond to shorter timescale fluctuations. The boxed annotation in each panel gives the raw mean $\mu_k$ and variance $\sigma_k^2$ of the amplitude at that bin. In contrast to Figure 3, the distributions exhibit the heavier, asymmetric tail characteristic of the Rayleigh distribution.
\textit{Bottom row:} Quantile quantile (QQ) plots comparing generated amplitudes against reference distributions. Two comparisons are shown at each frequency: \textcolor[RGB]{34, 139, 34}{the green line compares generated samples against a Rayleigh reference distribution}, while \textcolor[RGB]{255, 140, 0}{the orange line compares against a Uniform reference distribution}. The strong alignment of \textcolor[RGB]{34, 139, 34}{the green line with the identity line} across all frequencies and quantiles indicates that the model accurately reproduces the full Rayleigh ensemble distribution of the true DRW process. The clear deviation of \textcolor[RGB]{255, 140, 0}{the orange line from the identity line} further confirms that the generated data have Rayleigh, not Uniform, tails. This demonstrates the model captures the heavy tail behavior and higher order statistical structure characteristic of DRW variability across multiple temporal scales.
    }
    \label{fig:rayleigh_model}
\end{figure*}

\paragraph*{Experimental setup.}

We generate 100,000 DRW light curves directly in the Fourier domain. The power spectral density (PSD) describes how the variance of a stationary process is distributed across frequency: integrating $\mathrm{PSD}(f)$ over a frequency band gives the variance contributed by that band, so a target PSD fixes, at each frequency, the typical power a realization should carry. We use the DRW PSD from \cite{Kozlowski+2010},
\begin{equation}
    \mathrm{PSD}(f)
    =
    \frac{2 \hat{\sigma}^{2} \tau^{2}}
    {1 + (2\pi \tau f)^{2}},
\end{equation}
where $\tau$ is the characteristic damping timescale and $\hat{\sigma}$ controls the variability amplitude.
For a discrete signal of length $T = 256$, uniformly sampled with unit time step ($\Delta t = 1$), we build each realization one frequency bin at a time. At bin $k$, the target PSD sets the expected Fourier power $\mu_k \equiv \mathbb{E}[|\tilde{x}_k|^2] \propto \mathrm{PSD}(f_k)$; we draw an amplitude $R_k = |\tilde{x}_k|$ from a distribution with mean and variance set by $\mu_k$, and pair it with an independent random phase $\phi_k \sim \mathcal{U}(0,2\pi)$, giving
\begin{equation}
    \tilde{x}_k = R_k\, e^{i\phi_k}.
\end{equation}
The full light curve is then obtained by an inverse Fourier transform. The remaining choice is which distribution to draw $R_k$ from -- and here the DRW is not free: it must be Rayleigh.
The DRW is the stationary solution of a linear stochastic differential equation driven by Gaussian white noise, and is therefore itself a stationary Gaussian process: its Fourier coefficients, being linear functionals of $x(t)$, are Gaussian at every frequency. Concretely, at bin $k$ the real and imaginary parts of $\tilde{x}_k$ are independent and identically distributed,
\begin{equation}
    \mathrm{Re}(\tilde{x}_k),\ \mathrm{Im}(\tilde{x}_k) \;\sim\; \mathcal{N}\!\left(0,\ \tfrac{1}{2}\mu_k\right),
\end{equation}
so that $R_k = \sqrt{\mathrm{Re}(\tilde{x}_k)^2+\mathrm{Im}(\tilde{x}_k)^2}$ is the norm of an isotropic two-dimensional Gaussian vector, which by definition follows a Rayleigh distribution,
\begin{equation}
    R_k \sim \mathrm{Rayleigh}(\sigma_k), \qquad \sigma_k = \sqrt{\mu_k/2}.
    \label{eq:rayleigh}
\end{equation}
This result holds for any $\tau, \hat\sigma$: those parameters set the mean power $\mu_k$ at each bin, but the Rayleigh \emph{shape} of the amplitude distribution around that mean is fixed by the Gaussianity of the process alone, with a relative amplitude fluctuation of $\sqrt{4/\pi-1}\approx0.52$. We refer to a dataset built this way as the DRW (Rayleigh) dataset -- the statistically exact realization of the target process.
We then introduce a small, controlled deviation from this construction: at each bin we replace the Rayleigh amplitude distribution with a Uniform one, $R_k \sim \mathcal{U}(a_k,b_k)$, with $a_k, b_k$ chosen so that $R_k$ retains exactly the same mean and variance as $\mathrm{Rayleigh}(\sigma_k)$,
\begin{equation}
    a_k = \sigma_k\!\left(\sqrt{\tfrac{\pi}{2}} - \sqrt{3}\sqrt{2-\tfrac{\pi}{2}}\right), \qquad
    b_k = \sigma_k\!\left(\sqrt{\tfrac{\pi}{2}} + \sqrt{3}\sqrt{2-\tfrac{\pi}{2}}\right).
\end{equation}
Because $a_k>0$ for all $k$, the amplitudes remain non-negative by construction. This gives two datasets with an identical average PSD and identical per-frequency variance, differing only in the shape of the amplitude distribution -- Rayleigh for the true DRW, Uniform for the ablation -- which lets us test whether the generative model can detect such a subtle, higher-order deviation from the correct null hypothesis, rather than only matching the mean spectrum.

\paragraph*{Evaluation.}
We train the generative model separately on each dataset. After training, we generate new light curves (3,000 for the average PSDs and 10,000 for the amplitude distributions)
and compute both their average PSDs and the empirical distributions of Fourier amplitudes $|\tilde{x}_k|$ across realizations. Further details of the dataset construction and evaluation are given in Appendix~\ref{appendix:exp_details}.
To compare distributions, we use quantile--quantile (QQ) analysis. A QQ plot compares the quantiles of two distributions: if the distributions match exactly, the points lie on the identity line, while systematic deviations indicate differences in distributional shape, tails, or higher-order moments. Since $\sigma_k$ varies with frequency, we center and normalize the
distribution of the measured amplitudes in each frequency bin
to zero mean and unit variance before comparison, so that all frequencies can be overlaid against a single reference shape. Each histogram panel additionally reports, in a boxed annotation, the corresponding \emph{raw}, pre-standardization mean and variance of the amplitude at that frequency bin -- i.e.\ the empirical mean and variance of $R_k$ -- making explicit how strongly the absolute scale of the amplitude falls off with frequency even though the standardized shape shown in the histogram does not.
We therefore compare the Fourier-amplitude distributions of generated samples to those of the ground-truth datasets using both direct distribution visualizations and QQ plots.
\begin{figure*}
    \centering
    \includegraphics[width=1.00\textwidth]{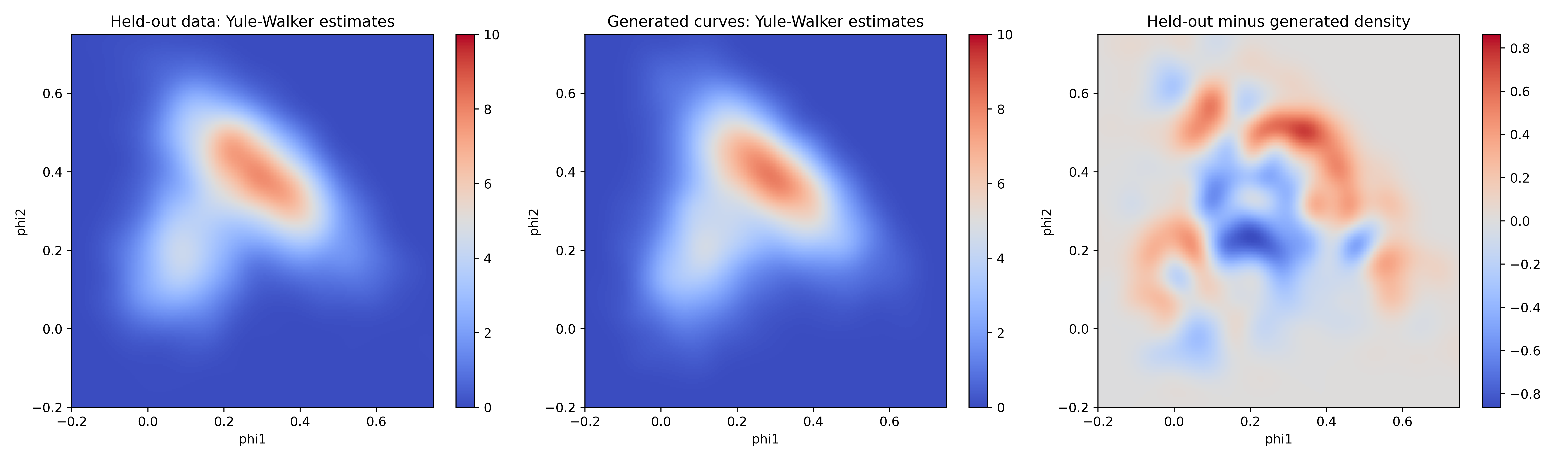}
    \caption{\textbf{Left:} Joint distribution of AR(2) coefficients $(\phi_1, \phi_2)$ estimated from 4,000 light curves of the dataset that were not used in training, using the Yule-Walker equations. 
\textbf{Center:} Model recovered coefficient density obtained by generating new light curves from the trained model, then estimating AR(2) coefficients from each generated sample using the Yule-Walker equations. 
The Yule-Walker method is a standard technique for fitting AR(2) models to time series data by matching the empirical autocovariances.
\textbf{Right:} Difference between the data and recovered coefficient densities, where red indicates regions where the data distribution has more mass and blue indicates regions where the recovered distribution has more mass.
The close agreement between left and center panels demonstrates that the generative model successfully captures the statistical structure of the AR(2) population, as measured by the joint distribution of fitted coefficients $(\phi_1, \phi_2)$, despite never being explicitly told about the AR(2) structure during training. 
This is quantified by a Wasserstein distance of $0.0168 \pm 0.0011$ between the generated and data coefficient distributions, comparable to the distance of $0.0164 \pm 0.0012$ between two independent samples of the data, indicating close agreement.}
    \label{fig:ar2_recovery}
\end{figure*}

\paragraph*{Results.}
Figure~\ref{fig:psd_match} shows that both datasets share the same target DRW PSD and that the generative model accurately reproduces this spectral structure in both cases.
However, Figures~\ref{fig:uniform_model} and~\ref{fig:rayleigh_model} demonstrate that the underlying distributions of Fourier amplitudes differ substantially between the two datasets, despite their identical mean and variance at every frequency bin. The Uniform-amplitude dataset exhibits bounded, non-Rayleigh distributions, whereas the DRW dataset exhibits the heavier, asymmetric tail intrinsic to the Rayleigh distribution.
The generative model successfully reproduces these distinct ensemble-level distributions in both settings. This is evident both from the direct distribution visualizations and from the QQ plots, which remain closely aligned with the identity line across frequencies and quantiles.
These results demonstrate that the model learns statistical structure beyond standard second-order statistics such as the PSD. In particular, the model captures the ensemble distribution of Fourier amplitudes at each frequency across realizations, including higher-order distributional properties that are not observable from individual light curves alone.

\paragraph*{Impact of the PSD error on detection.}
The generated PSD in Fig.~\ref{fig:psd_match} is slightly underestimated, by about 5\% on average. As a result, the $p$-values the model assigns to the Fourier amplitude at a given frequency, i.e. the probability that the noise alone produces an amplitude at least as large, are slightly too small. To estimate the size of this effect, we assume that the true null is exactly a DRW, or more generally a Gaussian process. The Fourier amplitude $A$ at each frequency is then Rayleigh distributed, and its $p$-value is $p = \exp(-A^2/\mu)$, where $\mu$ is the PSD at that frequency. A 5\% lower PSD gives $p_{\mathrm{model}} \approx p_{\mathrm{true}}^{1.053}$, so the error enters the exponent and grows only slowly, as $\ln(1/p)$. Even at $p = 10^{-9}$, roughly the threshold for a single reliable detection among $10^{7}$ quasars, the model $p$-value is too small by only a factor of $\approx 3$. At this threshold the expected number of false detections among $10^{7}$ quasars is therefore $\approx 0.03$ instead of $0.01$. This factor drops to about 1.5--2 when finer sampling reduces the PSD error to about 2\%. Moreover, the power in a single frequency bin fluctuates from one realization to the next by about 100\% of its mean, far more than the few per cent error of the model. This intrinsic scatter is a much bigger challenge for detection. Signals with true $p$-values well below the threshold therefore remain detected, and only signals lying right at the detection threshold can be misclassified. Such signals are difficult to detect reliably in any case. Finally, this comparison is the most favorable case for the parametric approach. If the true null is exactly a DRW, a DRW null carries no such error. For real quasars, however, the true null might not be exactly a DRW at every frequency, and a mis-specified parametric null can produce much larger, frequency-dependent errors (Section~\ref{sec:ar1_vs_ar2}).
\subsection{Experiment 2: Recovering population-level parametric structure}
\label{subsec:exp2_latent_structure}

\paragraph*{Motivation.}
In many stochastic processes, each realization is governed by a set of latent parameters that vary across the population. These parameters are not directly observed, and must typically be inferred from individual time series under an assumed parametric model. We test whether the generative model can instead reproduce the population-level statistics of these parameters directly from data, without being given the parametric form or parameter values.

\paragraph*{Experimental setup.}
We generate 100,000 synthetic light curves, each 2048 data-points in length, using an autoregressive process of order two (AR(2)):
\begin{equation}
    x_t = \phi_1 x_{t-1} + \phi_2 x_{t-2} + \epsilon_t,
\end{equation}
where $\epsilon_t$ is Gaussian white noise with standard deviation 0.2. As mentioned above, for training we used random 256-sample-long slices drawn from these light curves (see Appendix~\ref{appendix:exp2_details} for further details).
For each light curve, the coefficients $(\phi_1, \phi_2)$ are sampled independently from a fixed, correlated mixture of two Gaussian distributions. This defines a population of stochastic processes with heterogeneous dynamics.
The generative model is trained on this dataset of light curves without access to the underlying AR(2) structure or the coefficients $(\phi_1, \phi_2)$.

\paragraph*{Evaluation.}
After training, we generate 4,000 new light curves from the model. For each generated sample, we estimate AR(2) coefficients using the Yule--Walker equations.
This produces an empirical distribution of coefficients inferred from generated data, which we compare to the distribution obtained by applying the same estimator to 4,000 light curves of the same length (256 data points) from the dataset, which were not used in training. Since the coefficients of both samples are estimated in the same way from light curves of the same length, the estimation noise is the same on both sides of the comparison.

\paragraph*{Results.}
Figure~\ref{fig:ar2_recovery} shows that the distribution of coefficients inferred from the generated samples closely matches the distribution inferred from the data. 
To quantify this agreement, we compute the empirical Wasserstein distance between the generated and data coefficient distributions in $(\phi_1,\phi_2)$ space. The Wasserstein distance measures the minimal transport cost required to transform one probability distribution into another~\citep{Villani2009}. 

The generated coefficient distribution achieves a small Wasserstein distance of
\[
0.0168 \pm 0.0011,
\]
indicating close agreement with the distribution inferred from the data. For reference, even two independent samples of 4,000 light curves drawn from the dataset itself differ by a Wasserstein distance of $0.0164\pm0.0012$, due to the finite number of light curves; the generated light curves are therefore as close to the data as a second, independent sample of the data would be. Two additional, independently trained models give $0.0144\pm0.0010$ and $0.0235\pm0.0013$ (see Appendix~\ref{appendix:exp2_details}).

This indicates that the model has learned the statistical structure of the underlying population, as reflected in the joint distribution of inferred AR(2) coefficients. Importantly, this structure is never explicitly provided during training. The model therefore captures the relationship between stochastic variability and its population-level parameter distribution.

\subsection{Experiment 3: Conditional generation and consistency}\label{subsec:conditional}

\begin{figure*}
    \centering
    \includegraphics[width=0.83\textwidth]{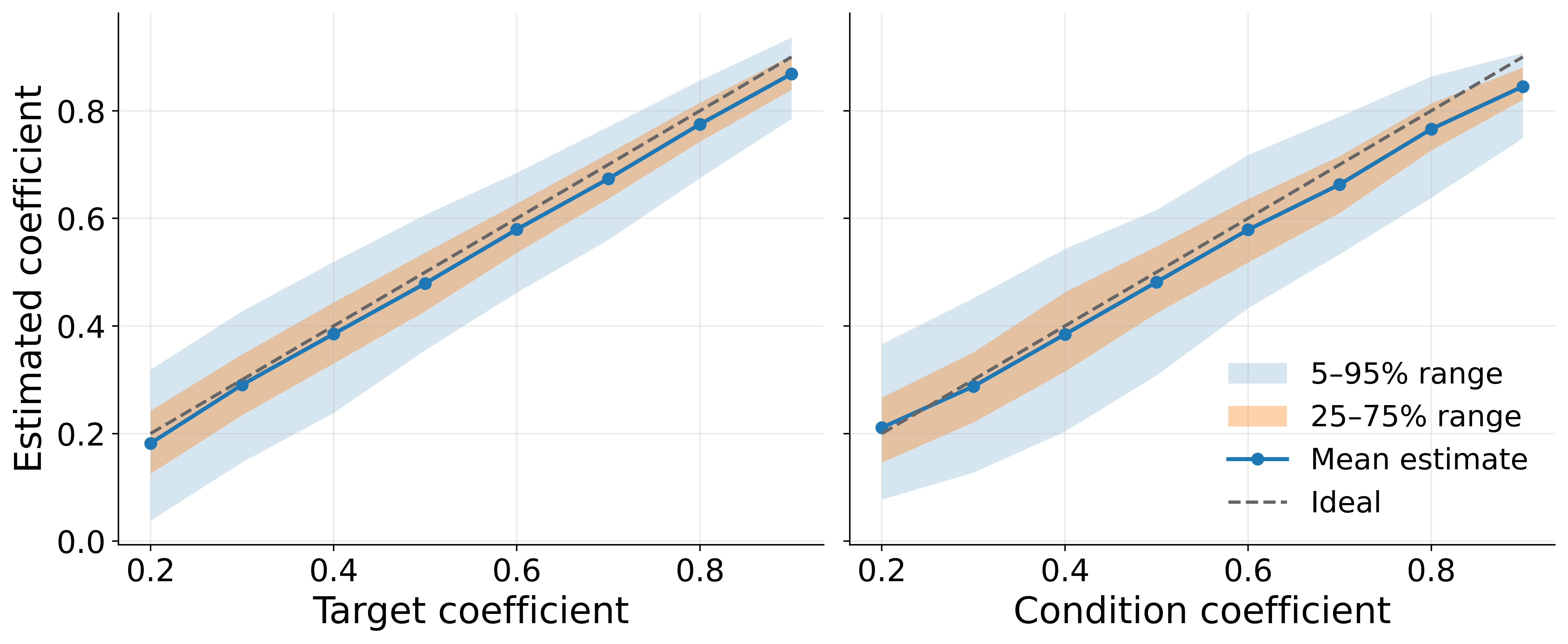}
\caption{\textbf{Recovery of AR(1) damping coefficients under conditional generation.}
Each panel shows the estimated coefficient $\hat{\phi}$ as a function of the true coefficient $\phi$ of the conditioning light curve. The dashed line indicates perfect recovery. The shaded regions show the 25–75\% and 5–95\% quantile ranges of estimates, and the solid line shows the mean estimate, over 500 conditioning light curves per coefficient.
\textbf{Left panel (Reference):} For each conditioning light curve, the AR(1) coefficient is estimated, a new light curve is simulated with the estimated coefficient, and the coefficient is estimated again from the new light curve. This shows the intrinsic uncertainty when fitting AR(1) coefficients to finite length time series and simulating new realizations from the fitted model.
\textbf{Right panel (Generative model):} For each conditioning light curve, the generative model generates a new light curve, from which the coefficient is estimated. The mean estimates closely follow those of the reference across the full range of true coefficients, with a somewhat larger spread. The displayed model is the best of three independently trained models (see Appendix~\ref{appendix:conditional_details}).
All light curves have 256 data points.}
\label{fig:conditional}
\end{figure*}

\paragraph*{Motivation.}
In many astrophysical applications, stochastic noise models are used as part of a statistical inference pipeline. A common procedure is to first estimate the parameters of a parametric model from an observed light curve, and then generate synthetic realizations under this model to assess the significance of observed features. This requires not only accurate parameter estimation, but also the ability to generate new samples that faithfully preserve the underlying stochastic process.
We therefore evaluate whether the generative model can (i) infer the latent stochastic parameters from a conditioning signal, and (ii) generate new light curves consistent with these parameters.

\paragraph*{Experimental setup.}
We generate 100,000 synthetic light curves, each 2048 data-points in length, using an AR(1) process
\begin{equation}
    x_t = \phi x_{t-1} + \epsilon_t,
\end{equation}
where $\phi$ is the damping coefficient, which varies between light curves.
The model is trained on a dataset of such light curves without access to the true values of $\phi$.
For evaluation, we generate 500 new synthetic light curves (256 data points each) at each of the coefficients $\phi = 0.2, 0.3, \dots, 0.9$, and use these as conditioning signals. For each conditioning light curve, we generate a new sample from the model, and estimate its damping coefficient $\hat{\phi}$ using the Yule--Walker equations.
This allows us to measure how well the model preserves the underlying stochastic parameter under conditional generation.

\paragraph*{Evaluation.}
For each target coefficient $\phi$, we compute the distribution of estimated coefficients $\hat{\phi}$ across generated samples. We compare these estimates to the ground-truth coefficient and analyze both the mean estimate and the spread of the distribution.
As a reference, we apply the standard parametric procedure to the same conditioning light curves: we estimate $\phi$ from each conditioning light curve, simulate a new AR(1) light curve with the estimated coefficient, and estimate $\hat{\phi}$ again from this new light curve. This mimics the common practice of fitting a noise model to an observed light curve and simulating new realizations from it. Since both procedures start from the same conditioning light curves and end with the same estimation step, the comparison does not favor either of them. Further details, including the per-coefficient results, are given in Appendix~\ref{appendix:conditional_details}.

\paragraph*{Results.}
The left panel in Fig.~\ref{fig:conditional} shows the reference case, where coefficients are estimated from light curves simulated with the coefficient fitted to each conditioning light curve. This reflects the intrinsic uncertainty of the estimation procedure: for finite-length stochastic signals, multiple nearby parameter values can produce similar realizations, leading to a non-zero spread around the true coefficient.

The right panel shows the performance of the generative model. The model captures the relationship between the conditioning signal and the underlying parameter, generating samples whose inferred coefficients follow the correct trend. The mean estimates of the model and of the reference differ by less than 0.02 at every coefficient, with mean absolute deviations from the true coefficient of 0.022 and 0.020, respectively. The spread of the model estimates is somewhat larger than that of the reference (by $\approx 17\%$ in standard deviation on average). Both procedures underestimate the coefficient at high $\phi$ (by $\approx 0.03$ for the reference and $\approx 0.05$ for the model at $\phi=0.9$), mostly due to the known bias of the Yule--Walker estimator for short light curves.

These results indicate that the model captures information about the latent stochastic parameters from a conditioning signal and generates new realizations consistent with them, with a performance comparable to the standard parametric procedure, but without being told the parametric form of the noise model.

\subsection{Experiment 4: Learning data-driven null hypotheses for periodic signal detection}\label{subsec:detection}

\begin{figure*}
\centering
\includegraphics[width=0.75\textwidth]{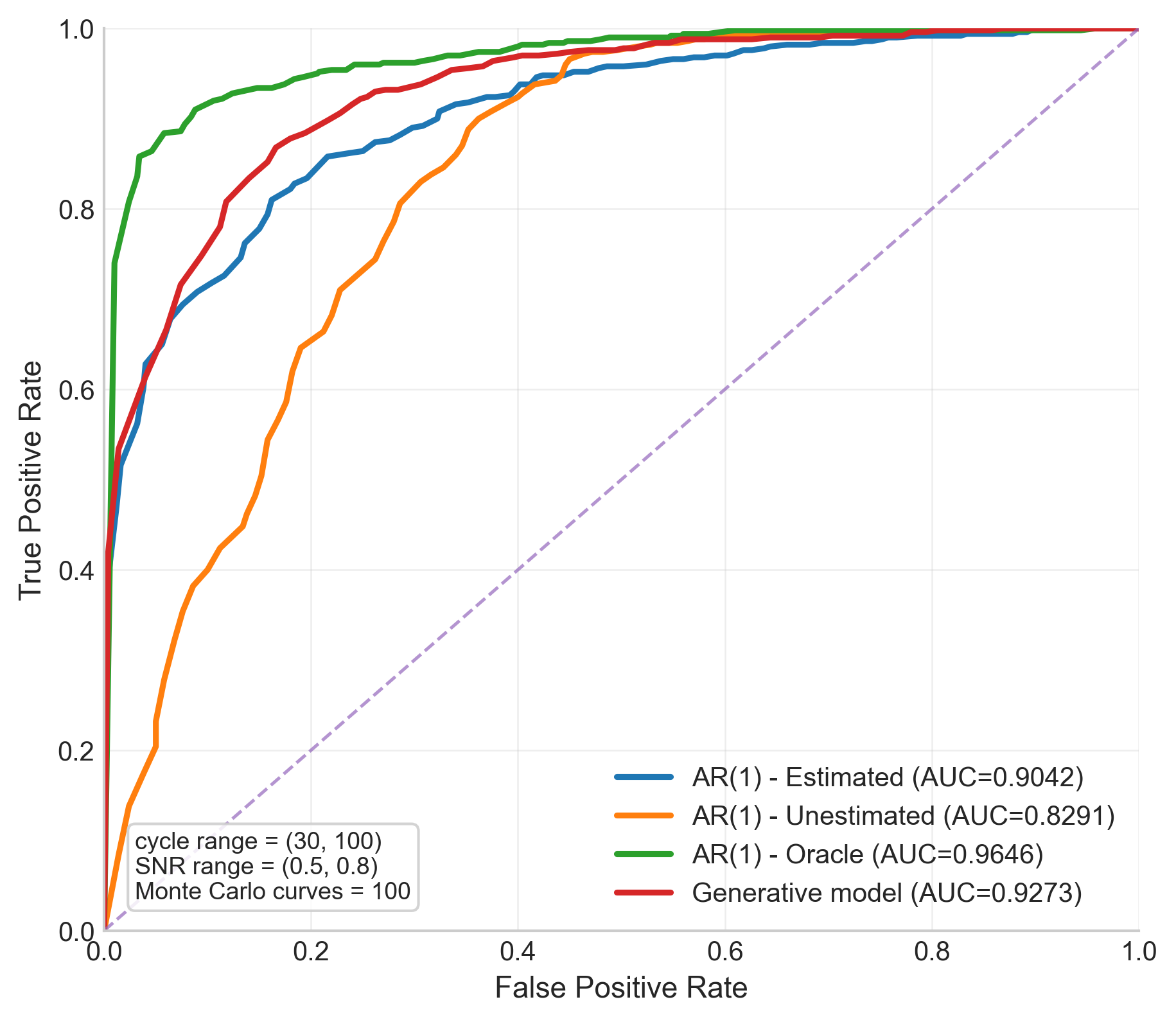}
\caption{\textbf{Detection performance for periodic signal detection when the true stochastic variability follows an AR(1) process.}
\textit{Experimental setting:} 1,000 synthetic AR(1) light curves were generated with damping coefficients $\phi$ uniformly distributed in $[0.1, 0.9]$. Sinusoidal signals with randomly varying phase and amplitude were injected into half of them, with cycle ranges of 30 to 100 (one complete oscillation per 30–100 data points), signal to noise ratios between 0.5 and 0.8, and 100 Monte Carlo realizations simulated per light curve.
\textit{Detection procedure:} The detection statistic is the maximum periodogram power $T(x) = \max_k |{\tilde{x}}_k|^2$. For each light curve, a p-value is computed by generating $N=100$ synthetic realizations from a noise model and computing $p=(b+1)/(N+1)$, where $b$ is the number of realizations with $T(x_{\mathrm{null}}) \geq T(x_{\mathrm{obs}})$. An ROC curve is constructed by varying the detection threshold and computing the true positive rate versus false positive rate across many light curves. The area under the ROC curve (AUC) measures performance: 1.0 indicates perfect separation between periodic and non-periodic signals, 0.5 indicates random guessing.
\textcolor{plotRed}{The generative model} achieves an AUC of 0.9273, matching \textcolor{plotBlue}{the fitted AR(1) reference} (AUC 0.9042) and approaching \textcolor{plotGreen}{the oracle bound} (AUC 0.9646). \textcolor{plotOrange}{The random-coefficient null}, which does not use the observed light curve, achieves an AUC of 0.8291. The corresponding false-positive rates are given in Table~\ref{tab:ar1_calibration} (Appendix~\ref{appendix:calibration}). This demonstrates the model successfully infers stochastic parameters from the observed signal and constructs an appropriate null distribution.}
\label{fig:roc_ar1}
\end{figure*}

\begin{figure*}[t]
\centering
\includegraphics[width=0.77\textwidth]{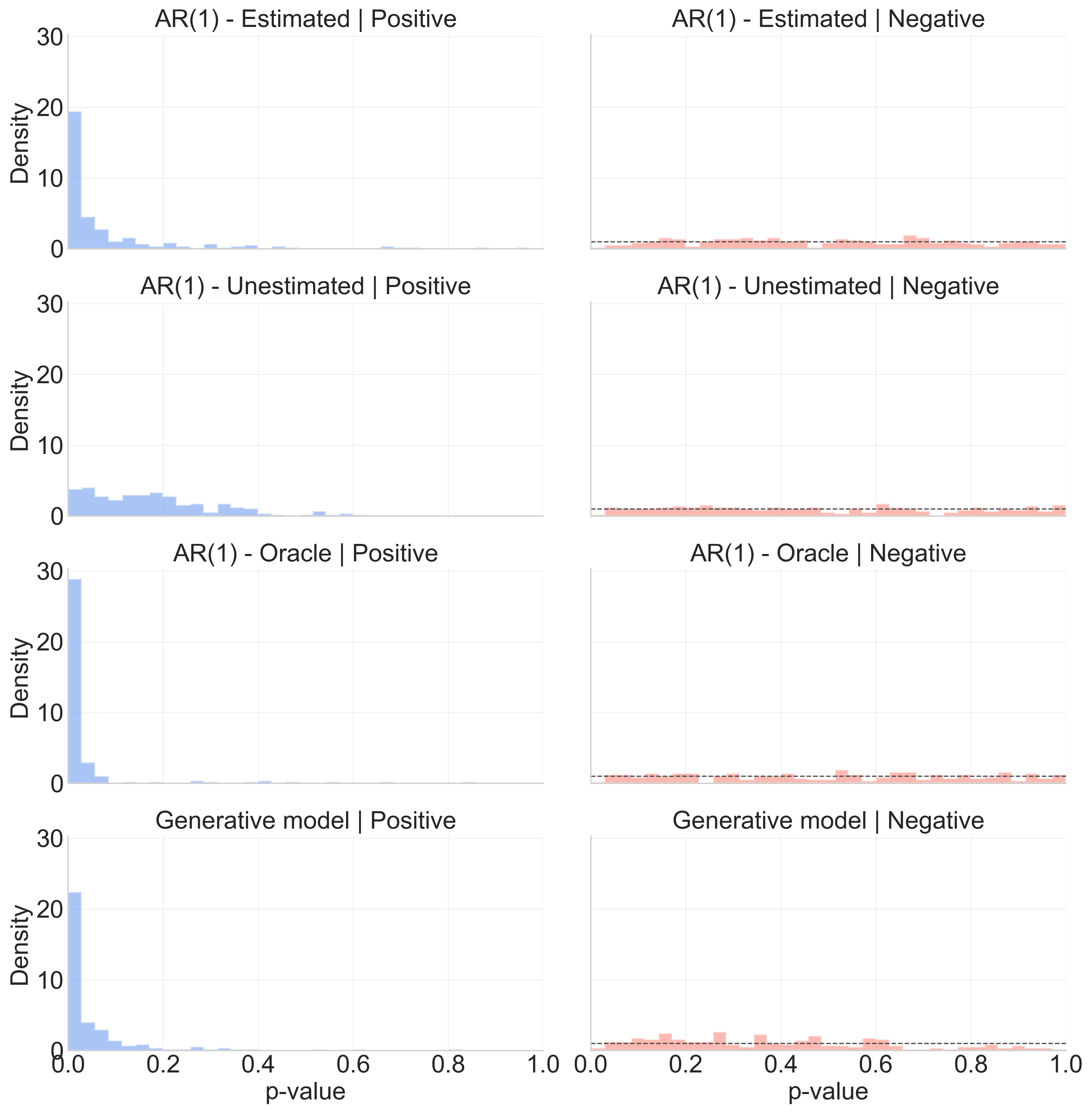}
\caption{\textbf{Distribution of p-values for the AR(1) experiment.}
For each light curve, p-values were computed under different null hypotheses using the detection procedure from Fig.~\ref{fig:roc_ar1}, here for a separate set of 400 light curves (200 with and 200 without an injected signal). Under the null hypothesis (no periodic signal), p-values should be uniformly distributed between 0 and 1, indicating the null model is properly calibrated. When a periodic signal is present, p-values should be concentrated near 0, indicating strong rejection of the null.
The AR(1) Estimated and Generative model methods both produce approximately uniform distributions under the null (negative) and p-values concentrated near 0 when signals are present (positive). The AR(1) Unestimated method shows degraded performance, with a broad distribution of p-values when signals are present. The Oracle method provides the best performance with strongest separation between null and signal-present cases. The generative model's p-value distributions closely match the fitted AR(1) reference. The corresponding false-positive rates are given in Appendix~\ref{appendix:calibration}.}
\label{fig:pvals_ar1}
\end{figure*}

\paragraph*{Motivation.}
The purpose of a null hypothesis in periodicity searches is to determine whether an observed periodic signal is unlikely to arise from stochastic variability alone. In practice, this is typically done by fitting a stochastic variability model to an observed light curve, generating synthetic realizations from the fitted model, and comparing a periodicity statistic to the resulting null distribution.

Even in the absence of any measurement errors or biases, 
this procedure can fail for several reasons. First, the assumed stochastic model may not accurately describe the true variability process, leading to biased significance estimates. Second, even when the correct model family is known, estimates its parameters from a single light curve can be difficult and is prone to large uncertainties and biases~\citep[e.g.][]{Charisi+2016}. Third, in sufficiently flexible models, the fitting procedure may partially absorb periodic structure into the noise model itself, reducing sensitivity to genuine periodic signals.
We therefore investigate whether synthetic light-curves from our generative model can serve as an alternative null hypothesis. Rather than assuming a specific parametric form, the model learns the statistical properties of stochastic variability directly during training on a large sample of input data.

\begin{figure*}
\centering
\includegraphics[width=0.8\textwidth]{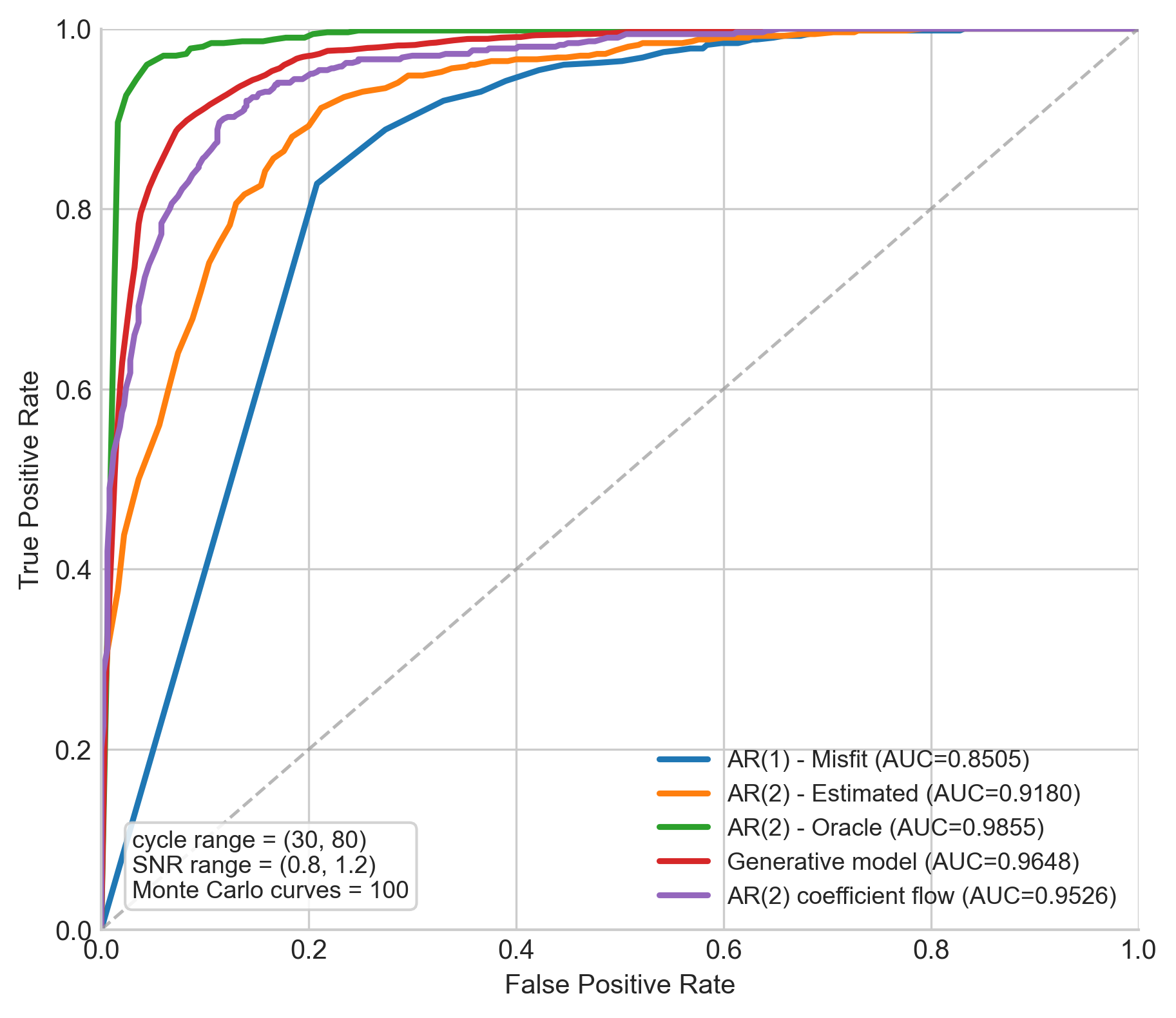}
\caption{\textbf{Detection performance for periodic signal detection when the true stochastic variability follows an AR(2) process.}
\textit{Experimental setting:} Synthetic AR(2) light curves were generated with coefficient pairs $(\phi_1, \phi_2)$ drawn from a correlated mixture of two Gaussian distributions (as in Section~\ref{subsec:exp2_latent_structure}). Sinusoidal signals with randomly varying phase and amplitude were injected into 500 of 1,000 light curves, with cycle ranges of 30 to 80, signal to noise ratios between 0.8 and 1.2, and 100 Monte Carlo realizations simulated per light curve (300 for the AR(2) coefficient flow).
\textit{Detection procedure:} Identical to Fig.~\ref{fig:roc_ar1}: the detection statistic is the maximum periodogram power, p-values are computed from 100 synthetic null realizations, and ROC curves measure detection performance.
The AR(2) setting reveals a critical challenge with parametric null hypotheses: AR(2) processes are flexible enough to generate oscillatory fluctuations, so fitting AR(2) coefficients independently from each light curve causes the fitting procedure to absorb part of the injected sinusoidal signal as stochastic variability. This makes \textcolor{plotOrange}{the fitted AR(2) null} overly permissive (AUC 0.9180), degrading detection. \textcolor{plotRed}{The generative model} achieves superior performance (AUC 0.9648, averaged over three independently trained models), approaching \textcolor{plotGreen}{the oracle} (AUC 0.9855), by learning the distribution of AR(2) coefficients across the training set and using this as a regularizing constraint: unlikely coefficient values are not favored merely because they can absorb a periodic component. \textcolor{plotPurple}{The AR(2) coefficient flow}, which explicitly combines the population distribution of coefficients with the observed light curve, performs similarly (AUC 0.9526). \textcolor{plotBlue}{The mis-specified AR(1) null} performs worst (AUC 0.8505). The corresponding false-positive rates are given in Table~\ref{tab:ar2_calibration} (Appendix~\ref{appendix:calibration}). This demonstrates that data-driven null hypotheses can outperform parametric fitting when models are sufficiently flexible to absorb genuine signals.}
\label{fig:roc_ar2}
\end{figure*}

\paragraph*{Experimental setup.}
We evaluate the use of our generative model in this context with synthetic datasets, for which the true stochastic process is known.
Two complementary scenarios are considered. \begin{enumerate}
    \item Light curves are generated from an AR(1) process with coefficients that vary across the dataset.
    \item Light curves are generated from an AR(2) process (with coefficients drawn from the distribution described in Eq.~\ref{eq:cen1}, ~\ref{eq:cen2})  
\end{enumerate} This experiment primarily tests the consequences of model mis-specification. Because AR(2) models are more prone to producing oscillatory behavior \citep{Moreno2019}, 
this setting additionally tests the challenges associated with parameter estimation in flexible stochastic models.
In both cases, sinusoidal signals with random phase and varying amplitude are injected into the light curves. The amplitude of each injected sinusoid was set so that the signal-to-noise ratio (SNR), defined as the ratio between the root-mean-square (RMS) amplitude of the sinusoid and the RMS variability (i.e. the standard deviation) of the stochastic light curve it is added to, is $\rm SNR\approx 0.5-1.2$ (see the specific ranges in Figs.~\ref{fig:roc_ar1} and~\ref{fig:roc_ar2}). Defining the SNR relative to the variability of each light curve ensures that the signal strength is comparable across light curves with different coefficients (see Appendix~\ref{appendix:calibration} for the exact definition).  This provides a controlled benchmark for evaluating different approaches to null-hypothesis construction that is also consistent with previous variability studies \citep{witt2022, davis2024, davis2026}.

\paragraph*{Detection procedure.}
For each light curve, we subtract its mean, multiply it by a Hann window (to reduce spectral leakage from the edges of the light curve), compute its discrete Fourier transform $\tilde{x}_k$, and define the detection statistic
\begin{equation}
T(x)=\max_k |\tilde{x}_k|^2,
\end{equation}
corresponding to the largest periodogram power present in the signal. This statistic does not assume prior knowledge of the injected frequency and therefore mimics a blind periodicity search.
Given an observed light curve $x_{\rm obs}$, a p-value is then computed by generating $N$ realizations from a chosen noise model and counting the number $b$ of noise realizations that have maximum Fourier power $T(x_{\rm null})$ larger than the observed value $T(x_{\rm obs})$
\begin{equation}\label{eq:mc_p}
p =
\frac{b+1}{N+1},
\qquad
b=\sum_{i=1}^{N}
\mathbf{1}
\left[
T(x^{(i)}_{\rm null})
\ge
T(x_{\rm obs})
\right].
\end{equation}
The resulting p-value represents the probability that a detection statistic of equal or greater apparent periodicity could arise under the assumed null hypothesis. Adding one to the numerator and the denominator is the standard choice for such Monte Carlo tests: it counts the observed light curve itself as one of the realizations, and avoids assigning $p=0$ when none of the $N$ noise realizations exceeds the observed value. With $N=100$, the smallest possible p-value is therefore $1/101$.

\paragraph*{Comparing null hypotheses.}
For the experiments, we compare the following approaches:

\begin{itemize}

\item \textbf{Oracle parametric null.} The correct autoregressive model family is used with the true coefficients that generated each light curve. This provides a reference as the best possible performance achievable within the assumed parametric model.

\item \textbf{Fitted parametric null.} The correct autoregressive model family is used, but its coefficients are estimated from each observed light curve. This mimics a simpler and more common approach in existing quasar periodicity searches.

\item \textbf{Random-coefficient parametric null.} The correct autoregressive model family is used, but instead of estimating coefficients from the observed light curve, coefficients are drawn directly from a prior distribution ($\phi \in U[0.1,0.9]$ for AR(1), $\phi_1,\phi_2 $ from the distribution of Eq.~\ref{eq:cen1}, ~\ref{eq:cen2}) used to generate the dataset.

\item \textbf{Mis-specified fitted null.} An incorrect autoregressive model family is fit to each observed light curve and used to generate null realizations.
This addresses biases that arise from model mis-specification.

\item \textbf{Learned generative null.} The flow model is trained on a collection of light curves and used to generate null realizations conditioned on the observed signal. No information about the underlying autoregressive model or coefficient distribution is provided.

\item \textbf{AR(2) coefficient flow} (AR(2) case only). An intermediate approach between the parametric and the learned nulls, described below.

\end{itemize}

\begin{figure*}
\centering
\includegraphics[width=0.77\textwidth]{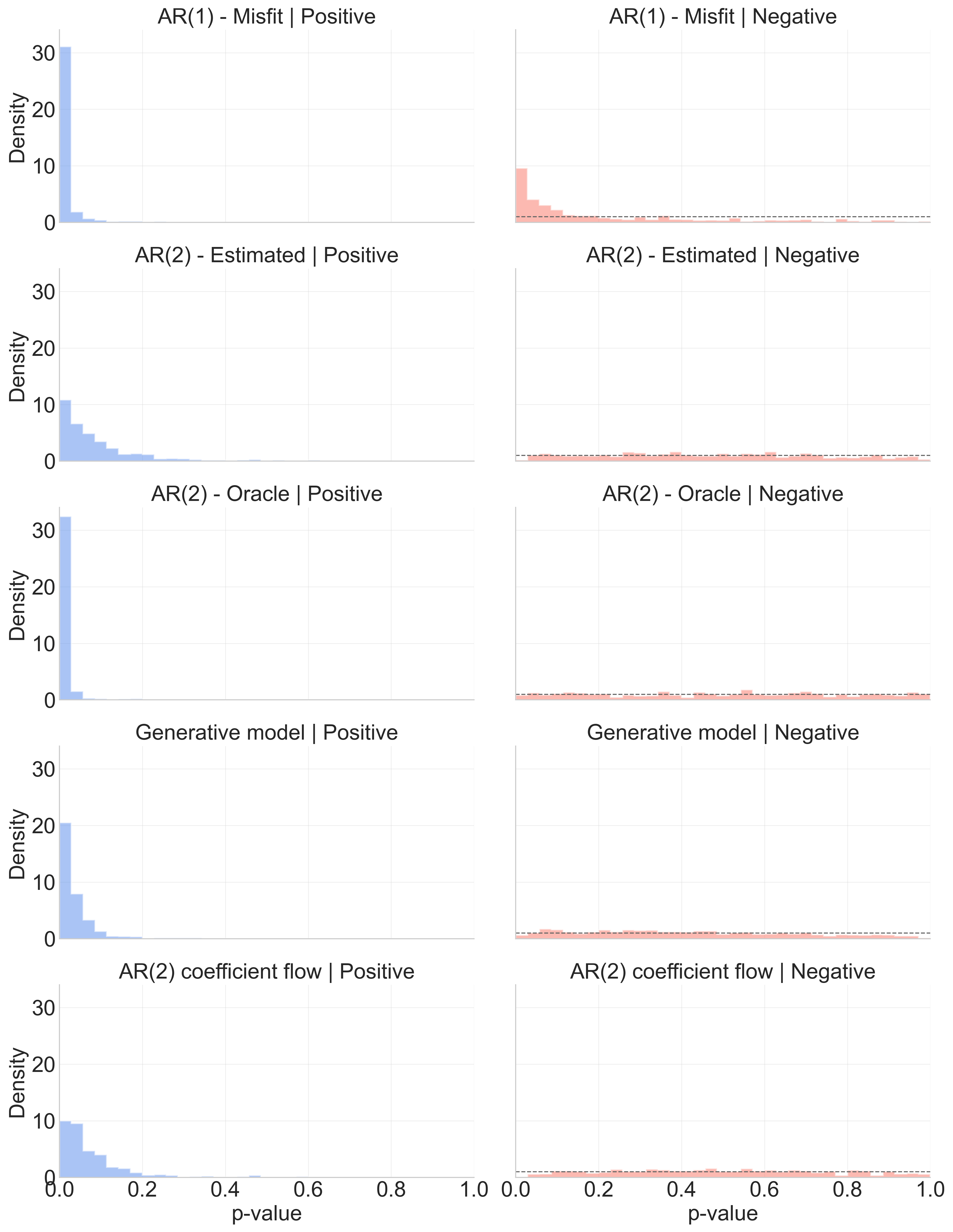}
\caption{\textbf{Distribution of p-values for the AR(2) experiment.}
For each light curve, p-values were computed under different null hypotheses using the detection procedure from Fig.~\ref{fig:roc_ar2}. Ideally, p-values for signal-present cases should be concentrated near 0 (strong rejection of null), while p-values for null cases should be uniformly distributed between 0 and 1 (a properly calibrated null).
The AR(1) Misfit method fails catastrophically: p-values are concentrated near 0 for both null and signal-present cases, causing it to reject the null far too often. This produces widespread false positives (42\% of the null cases at the 5\% significance level).
The AR(2) Estimated method shows a uniform distribution for null cases (correct calibration) but the peak for signal-present cases is broad and not sharply concentrated near 0. This overlap between null and signal distributions causes both false positives and false negatives.
The AR(2) Oracle method approaches the ideal: p-values for signal-present cases form a sharp peak near 0, while null cases show a relatively uniform distribution. This provides good separation.
The Generative model (averaged over three independently trained models) achieves near-optimal performance: p-values for signal-present cases concentrate sharply near 0, while p-values for null cases are approximately uniform, as for the oracle. This provides good separation with a false-positive rate close to the nominal level (Appendix~\ref{appendix:calibration}). The AR(2) coefficient flow behaves similarly, with a somewhat broader peak for signal-present cases.}
\label{fig:pvals_ar2}
\end{figure*}

\paragraph*{The AR(2) coefficient flow.}
In the AR(2) case, we additionally test an intermediate null hypothesis, designed to probe how the learned generative null works. As we discuss below, we suspect that the learned generative null benefits from having learned the population distribution of AR(2) coefficients during training: when conditioned on a new light curve, it effectively estimates the coefficients from the light curve, while taking into account which coefficient values are common in the population, and then generates new light curves accordingly. To test this interpretation, we construct an explicitly parametric version of this procedure. We train a small, separate flow model on simulated AR(2) light curves whose coefficients are drawn from the population distribution (Eqs.~\ref{eq:cen1}--\ref{eq:cen2}) and are known. Given a light curve, this model outputs the range of coefficient pairs $(\phi_1,\phi_2)$ compatible with it, weighted by how common they are in the population, i.e. the posterior distribution $p(\phi_1,\phi_2\mid \mathbf{x})$. For each null realization, we draw a coefficient pair from this distribution and simulate an AR(2) light curve with it. Unlike the fitted AR(2) null, which uses a single best-fit coefficient pair, this null therefore accounts both for the uncertainty in the coefficients and for their population distribution. If the learned generative null indeed works in this way, the two should perform similarly. Note that the AR(2) coefficient flow requires the parametric form of the noise model and the true coefficients of the training light curves, neither of which is available for real quasars; the learned generative null requires neither. Details of this model are given in Appendix~\ref{appendix:calibration}.


\begin{figure*}
\centering
\includegraphics[width=0.8\textwidth]{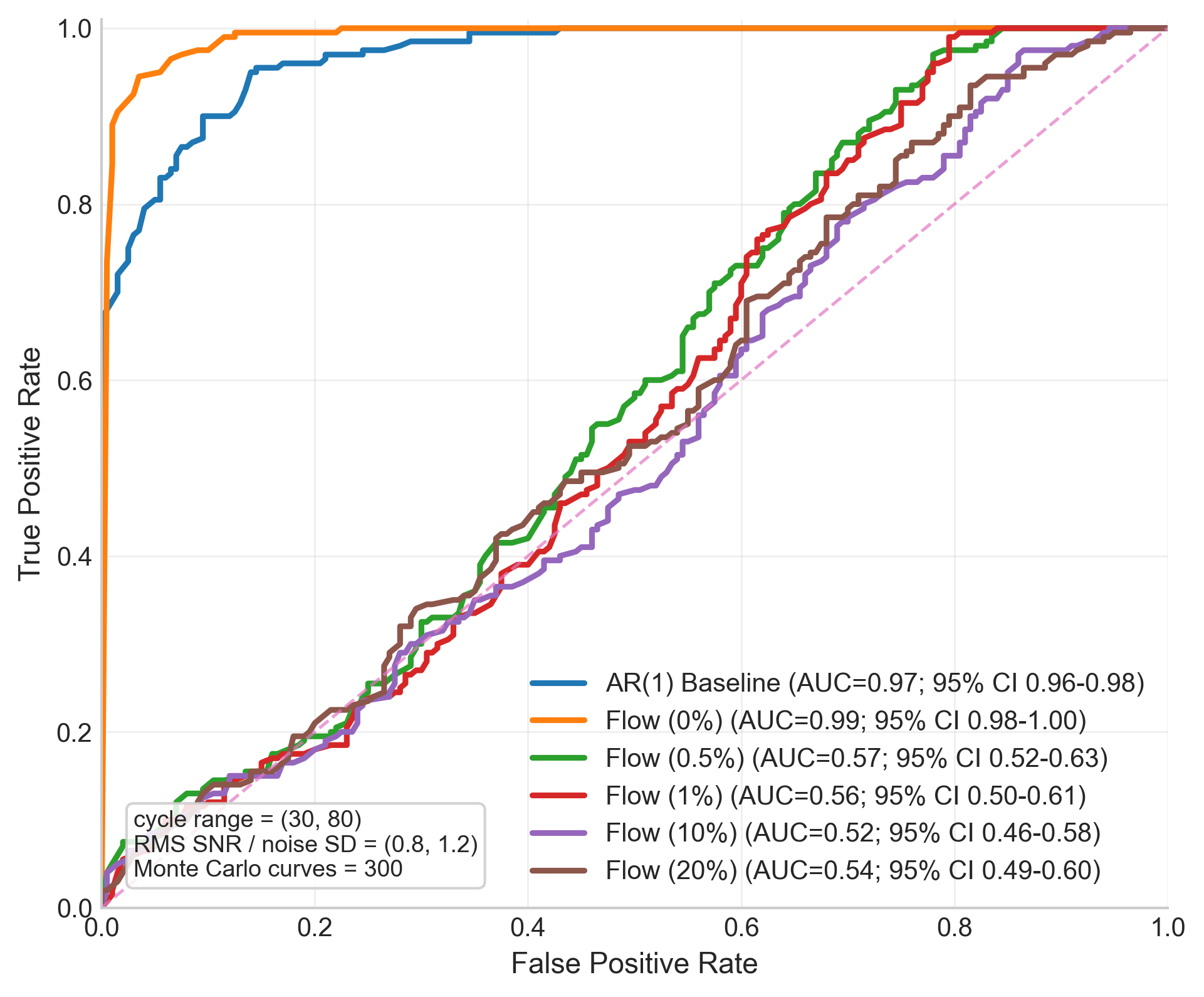}
\caption{\textbf{Detection performance as a function of contamination level in the training data.}
\textit{Experimental design:} The generative model's performance depends critically on the purity of its training data. To test this, synthetic AR(1) light curves were generated with damping coefficients $\phi$ uniformly distributed in $[0.2, 0.9]$. Sinusoidal signals with randomly varying phase and amplitude were injected into an increasing fraction of the training set (0\%, 0.5\%, 1\%, 5\%, 10\%, and 20\% contamination). The injected signals have cycle ranges of 30 to 80, signal to noise ratios between 0.8 and 1.2, and 300 Monte Carlo realizations were simulated per light curve (100 from each of three independently trained models). New models were trained for each contamination level, and detection performance was then evaluated on clean test data using the AR(1) detection procedure from Fig.~\ref{fig:roc_ar1}. For clarity, the 5\% level is shown only in Appendix~\ref{appendix:contamination}, together with the false-positive rates and the results of each individual training run. \textcolor{plotBlue}{Blue: fitted AR(1) baseline}; \textcolor{plotOrange}{orange: clean model (0\%)}; \textcolor{plotGreen}{green: 0.5\% contamination}; \textcolor{plotRed}{red: 1\%}; \textcolor{plotPurple}{purple: 10\%}; \textcolor{plotBrown}{brown: 20\%}. The \textcolor{plotPink}{pink dashed diagonal} corresponds to random guessing.
\textit{Results:} When the training set is clean (0\% contamination), the model achieves excellent performance (AUC 0.9904, compared with 0.9699 for the fitted AR(1) null). However, already at 0.5\% contamination, the smallest level tested, performance drops sharply (AUC 0.5735), and at higher contamination levels detection performance remains close to random guessing levels (AUC 0.52--0.56), without a monotonic trend.
This degradation occurs because the generative model learns statistical features of the training data, including periodic structure when present. When periodic structure is incorporated into the training set, the model treats it as part of the null distribution. During inference, when the model encounters similar periodic structure, it assigns it high probability, making the null hypothesis overly permissive. This highlights a fundamental limitation of data-driven null hypotheses: their validity depends critically on careful dataset curation and freedom from contamination by the signals we seek to detect.}
\label{fig:contamination_roc}
\end{figure*}

\paragraph*{Results.}
Detection performance is evaluated using receiver operating characteristic (ROC) curves and their corresponding area under the curve (AUC), together with the calibration of the resulting p-values. The ROC curve measures the trade-off between true-positive and false-positive rates as the detection threshold is varied. The AUC summarizes this curve into a single number between 0 and 1, where 0.5 corresponds to random guessing and 1 corresponds to perfect separation between periodic and non-periodic light curves. Larger AUC values therefore indicate better detection performance. The calibration of the p-values, i.e. whether the fraction of light curves without a signal that are falsely flagged as periodic at a given significance level matches that level, is quantified in Appendix~\ref{appendix:calibration}.

In the AR(1) setting (Figs.~\ref{fig:roc_ar1} and~\ref{fig:pvals_ar1}), performance is mainly controlled by the quality of coefficient estimation. 
When the AR(1) coefficient is estimated inaccurately, the resulting null realizations do not match the variability of the observed light curve, leading to degraded detection performance and biased p-values. The fitted AR(1) null provides the relevant parametric reference in this experiment, since it uses the correct model class and estimates the coefficient directly from the observed curve. The learned generative null (AUC 0.9273) performs on par with this fitted AR(1) reference (AUC 0.9042), showing that the generative model has learned to infer the effective stochastic parameters from the conditioning light curve with sufficient accuracy to construct an appropriate null distribution.

The AR(2) setting (Figs.~\ref{fig:roc_ar2} and~\ref{fig:pvals_ar2}) is more challenging. 
Using the wrong noise model leads to worse detection performance, confirming that model mis-specification can strongly bias periodicity searches. However, even with the correct AR(2) model family, fitting the coefficients independently from each observed light curve is problematic. AR(2) processes are expressive enough to produce oscillatory fluctuations, so the fitting procedure can explain part of the injected sinusoidal signal as stochastic variability. This makes the fitted null too permissive and reduces the ability to detect periodic signals.

The learned generative null performs better than the fitted AR(2) null (AUC 0.9648, averaged over three independently trained models, compared with 0.9180; Appendix~\ref{appendix:calibration}), which we attribute to the fact that it does not rely only on coefficient estimation from a single light curve. Instead, it has learned the distribution of AR(2) coefficients across the training set and uses this information when conditioning on a new observed curve. This acts as a regularizing constraint on the inferred stochastic process: coefficient values that are unlikely under the learned distribution are not favored merely because they can absorb a periodic component. As a result, the learned null better separates stochastic variability from injected periodic signals, producing stronger detection performance, while its false-positive rates remain close to the nominal significance levels (Appendix~\ref{appendix:calibration}).

This interpretation is supported by the AR(2) coefficient flow, which explicitly follows this procedure: it combines the population distribution of coefficients with the information in the observed light curve, and simulates null realizations from the resulting range of coefficients. It performs similarly to the learned generative null (AUC 0.9526 compared with 0.9648), and both outperform the fitted AR(2) null. This suggests that, as we suspected, the learned generative null has learned to estimate the stochastic parameters from the conditioning light curve, taking into account the distribution of AR(2) coefficients it has seen during training, and then uses them to generate new realizations, without ever being told the AR(2) form or the coefficient values.

Taken together, these results show that the learned generative null is useful in two complementary regimes. In the simpler AR(1) case, it matches the performance of explicit coefficient estimation, indicating that it can recover the relevant stochastic parameter from the observed light curve. In the more flexible AR(2) case, it improves over direct parametric fitting by using information learned across many realizations, which reduces the tendency of the fitted noise model to absorb periodic signal structure.

\subsection{Experiment 5: Contamination of the training set and its mitigation}\label{subsec:contamination}

\paragraph*{Motivation.}
The results of the previous section suggest that the learned generative null can improve periodicity detection by learning the distribution of variability across a large collection of light curves. However, this advantage relies on a critical assumption: the training data must be representative of the null hypothesis that we wish to model.
In practice, this assumption may not hold. Real astronomical datasets may contain a mixture of stochastic variability and genuine signals. Because the generative model learns directly from the training data, contamination by non-null signals can alter the learned distribution and bias the resulting null hypothesis. We therefore investigate how contamination of the training set affects the statistical validity of the learned null model, and then propose a simple correction that largely removes this sensitivity.
\begin{figure*}
\centering
\includegraphics[width=0.8\textwidth]{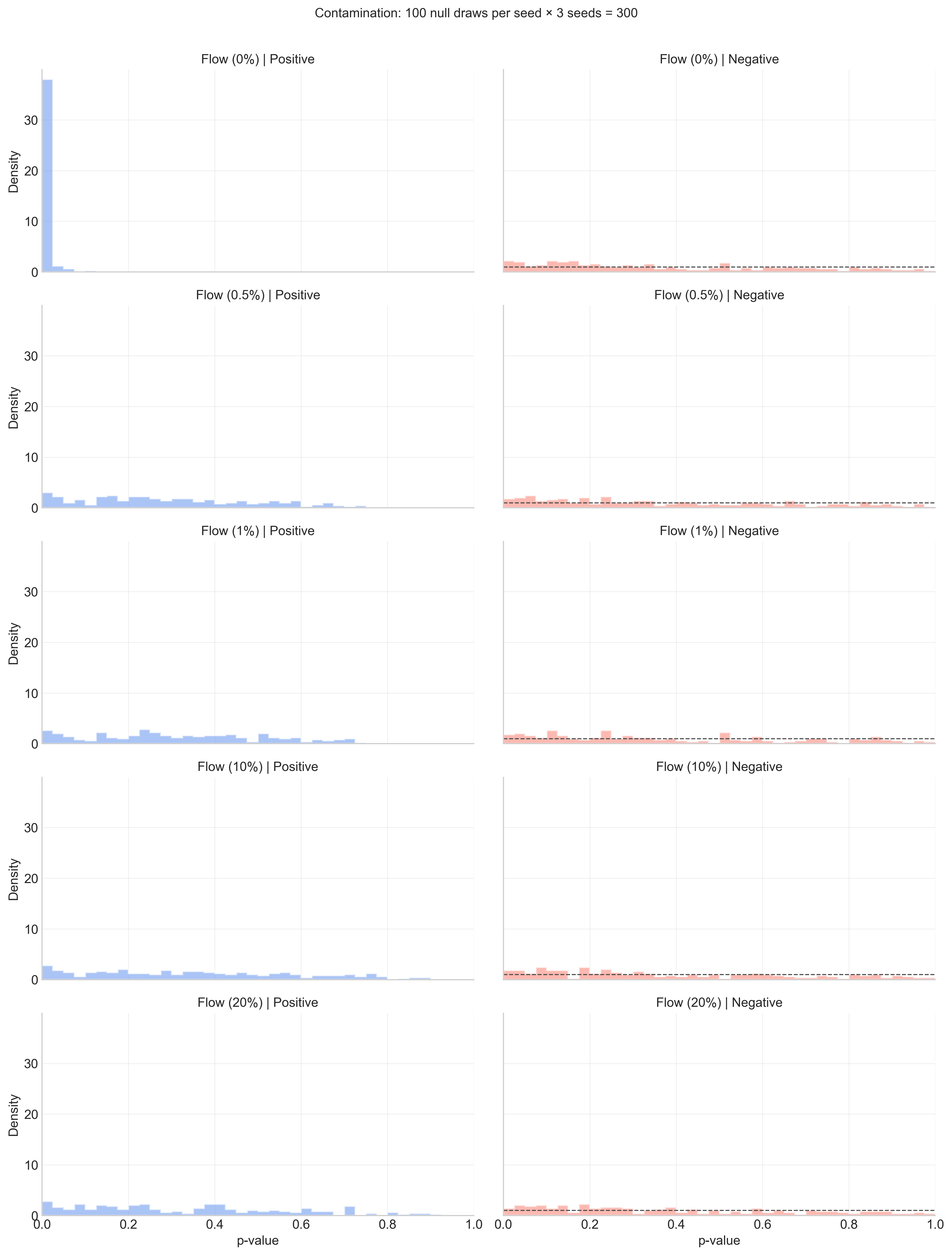}
\caption{\textbf{Distribution of p-values under different contamination levels.}
For each contamination level shown in Fig.~\ref{fig:contamination_roc}, p-value distributions are shown for null cases (negative) and signal-present cases (positive).
\textit{Clean training data (Flow 0\%):} The distributions are sharply separated. Signal-present cases produce p-values concentrated near 0, while null cases spread across all values, enabling reliable detection.
\textit{Mechanism of contamination:} As the training set becomes contaminated with periodic signals, the generative model learns that periodic structure is part of the normal data distribution. During conditional generation, the model is therefore inclined to generate signal-like realizations when conditioned on a light curve containing a periodic signal. Specifically, the model interprets the observed periodic component as a normal feature of the underlying stochastic process and reproduces similar signal-like patterns in its null realizations. This causes the p-values of signal-present cases to increase: the model now treats periodic signals as plausible stochastic variability rather than rare deviations, degrading the detection of true signals.
\textit{Effect on null distribution:} The null case p-values are similar for all contamination levels, with a mild excess at low values (false-positive rates of 7.5--10.5\% at the nominal 5\% level, including for the clean model; Appendix~\ref{appendix:contamination}). Already at 0.5\% contamination, null and signal distributions largely overlap, rendering the null hypothesis useless for detection.
This demonstrates a fundamental vulnerability of data-driven methods: they adapt directly to whatever is present in the training data, including unwanted signals. Careful dataset curation is essential for maintaining the validity of learned null hypotheses.}
\label{fig:contamination_pvals}
\end{figure*}

\paragraph*{Experimental setup.}
We consider the AR(1) setting and introduce controlled contamination into the training dataset by injecting periodic signals into a fraction of the light curves. The contamination level is defined as the fraction of training samples that contain an injected sinusoidal component (0, 0.5, 1, 5, 10, and 20\%).
For each contamination level, new generative models are trained and evaluated using the periodic signal detection procedure described in the previous section. The models are evaluated on the same 400 test light curves (200 with and 200 without an injected signal), which are simulated independently after training and are never used in training. To reduce the dependence on a particular training run, for each contamination level we train three models independently and combine 100 null realizations from each of them (Appendix~\ref{appendix:contamination}). Detection performance is quantified using the AUC of the ROC curve and the calibration of the resulting p-values.
\begin{figure*}
\centering
\includegraphics[width=0.8\textwidth]{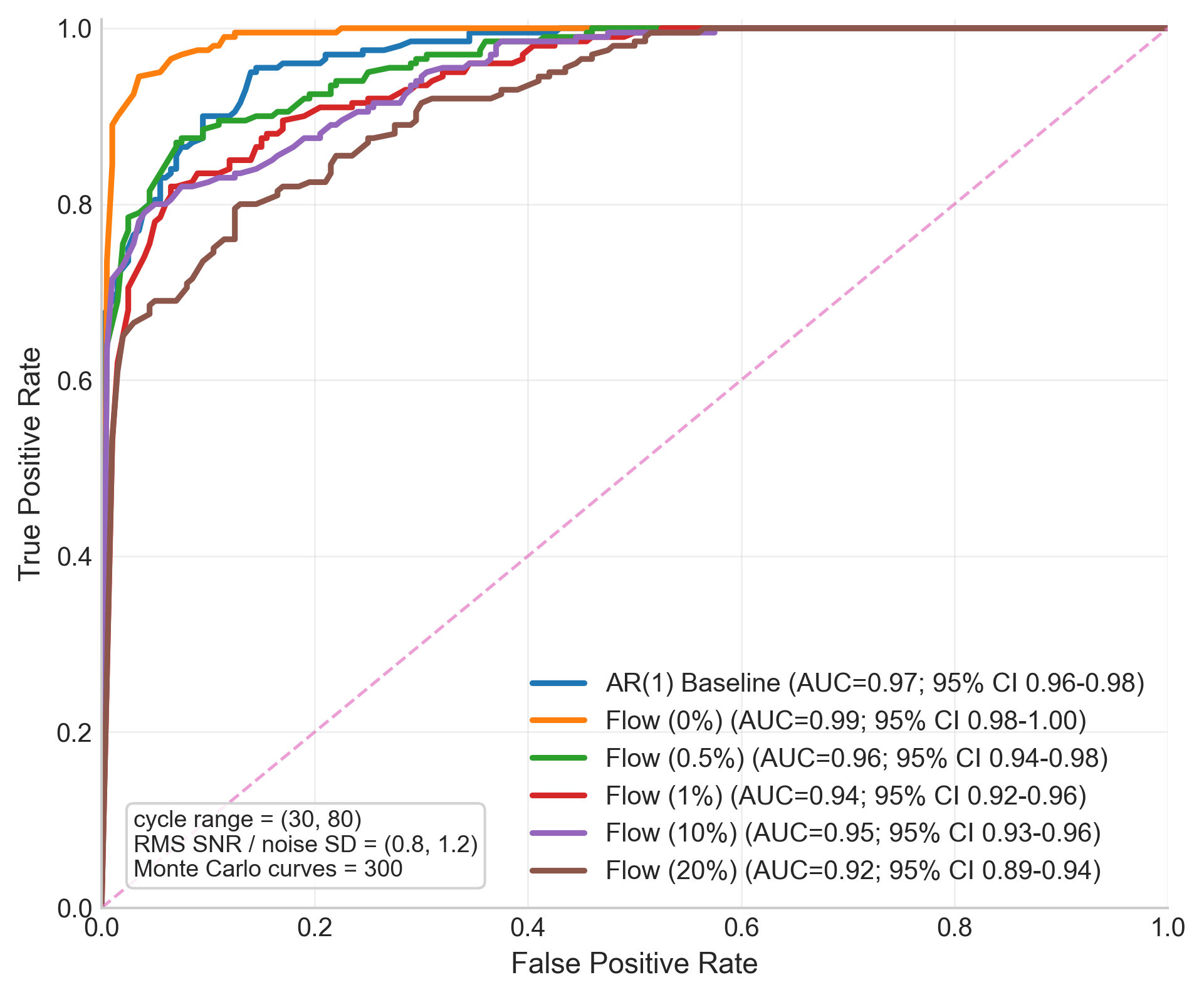}
\caption{\textbf{Detection performance of contaminated models after projecting the conditioning vector onto its six leading principal directions.}
\textit{Experimental design:} The trained models are the same as in Fig.~\ref{fig:contamination_roc}, with no retraining, and the test light curves and detection procedure are identical: training contamination of 0, 0.5, 1, 10 and 20\%, injected signals with cycle ranges of 30 to 80 and signal to noise ratios between 0.8 and 1.2, and 300 Monte Carlo realizations per light curve, pooled from 100 realizations from each of the three independently trained models into a single null. The only change is that, for each trained model, the conditioning vector of every observed light curve is replaced by its projection onto the six leading principal directions of the encodings of that model's training set (Eq.~\ref{eq:pca_projection}). \textcolor{plotBlue}{Blue: fitted AR(1) baseline}; \textcolor{plotOrange}{orange: clean model (0\%)}; \textcolor{plotGreen}{green: 0.5\% contamination}; \textcolor{plotRed}{red: 1\%}; \textcolor{plotPurple}{purple: 10\%}; \textcolor{plotBrown}{brown: 20\%}. The \textcolor{plotPink}{pink dashed diagonal} corresponds to random guessing.
\textit{Results:} The projection does not harm the clean model (AUC 0.99, as without projection). For the contaminated models, the AUC rises from 0.52--0.57 without projection (Fig.~\ref{fig:contamination_roc}) to 0.92--0.96, close to the fitted AR(1) null (0.97). A mild decrease of performance with increasing contamination remains.}
\label{fig:pca_roc}
\end{figure*}

\begin{figure*}
\centering
\includegraphics[width=0.8\textwidth]{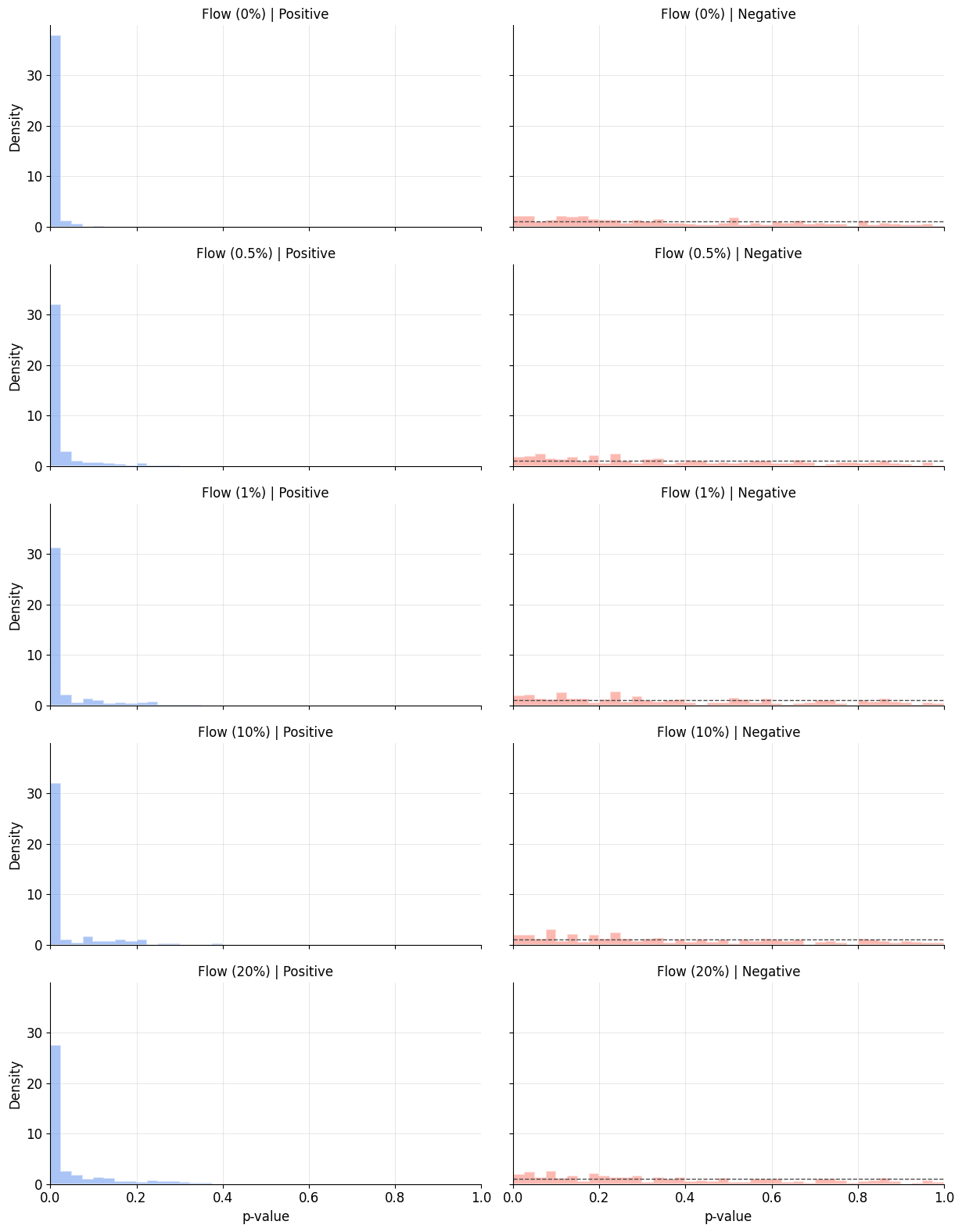}
\caption{\textbf{Distribution of p-values after low-rank projection of the conditioning vector.}
For each contamination level shown in Fig.~\ref{fig:pca_roc}, using the same trained models as in Fig.~\ref{fig:contamination_pvals} and a null of 300 realizations pooled from the three independently trained models, p-value distributions are shown for \textcolor{histBlue}{signal-present cases (positive, blue)} and \textcolor{histRed}{null cases (negative, red)}. The \textcolor{plotGrey}{dashed line} marks the uniform distribution expected for a calibrated null.
\textit{Signal-present cases:} Unlike without projection (Fig.~\ref{fig:contamination_pvals}), the p-values remain sharply concentrated near 0 at all contamination levels, with a slightly broader tail at higher contamination.
\textit{Null cases:} The p-values remain spread across all values, approximately uniformly, as for the unprojected models. The projection therefore restores detection power without making the null overly strict.}
\label{fig:pca_pvals}
\end{figure*}

\paragraph*{Results.}
Detection performance is evaluated using receiver operating characteristic (ROC) curves and their corresponding area under the curve (AUC), together with the calibration of the resulting p-values. Figure~\ref{fig:contamination_roc} shows that detection performance deteriorates sharply once the training set is contaminated. When the training data consists primarily of stochastic variability, the learned generative null provides strong separation between periodic and non-periodic light curves. However, already when 0.5\% of the training set contains periodic signals, the AUC decreases substantially, indicating reduced detection power; at higher contamination levels, the AUC remains close to 0.5 and does not decrease further in a monotonic way.
A similar trend is visible in the p-value distributions shown in Fig.~\ref{fig:contamination_pvals}: for contaminated training sets, the p-values of light curves containing a periodic signal are no longer concentrated near 0, indicating that the learned null no longer accurately represents the underlying null distribution.

The higher AUC of the clean model compared with the fitted AR(1) null, even though AR(1) is the true process, is likely due to the information about the population of light curves that the generative model has learned, which the fitted AR(1) null does not use. Since the test light curves are simulated independently after training, it cannot be due to an overlap between the training and the test data.

This behavior follows directly from the role of the generative model in our framework. The model is not trained to learn a stochastic noise process in isolation, but rather the full distribution of light curves present in the training dataset. When the training set consists entirely of representative null samples, the model can learn an effective approximation to the desired null distribution. However, injected periodic signals become part of the learned distribution, so the model assigns them relatively high probability at inference time and increasingly treats signal-like structure as part of the null. Once the training set is contaminated, the resulting null becomes more permissive, reducing the significance assigned to genuine periodic signals. The generative model is therefore only as reliable as the dataset used to train it. The model is therefore sensitive to even slight contamination: periodic signals in the training set are learned as part of the dataset and incorporated into the null. Below, we show that this sensitivity can be largely removed by a low-rank projection of the conditioning vector.

\paragraph*{Discussion.}
This highlights a fundamental limitation of data-driven null hypotheses: their validity depends on the quality and purity of the training data. Unlike parametric models, which impose a fixed functional form, generative models adapt to the data distribution and therefore inherit its biases.

At the same time, this limitation reflects a general challenge in statistical inference: any method that estimates a null model from contaminated data will suffer from similar effects. The key difference is that generative models make this dependence explicit, as they directly learn the data distribution.

These results emphasize the importance of careful dataset curation and suggest that robustness to contamination is a key direction for improving data-driven null hypothesis models. We present one such approach below.

\paragraph*{Mitigation by low-rank projection of the conditioning vector.}\label{subsec:pca}
The results above show that the generative model is sensitive to even slight contamination of its training set: periodic signals present in the training data are learned as part of the dataset and are therefore incorporated into the null. Removing the contaminated light curves before training would require identifying them, which is precisely the detection problem we wish to solve. We therefore propose a simple correction that is applied at inference time to the already trained, contaminated models described above, and requires no retraining.

\paragraph*{Principal component analysis (PCA).}
Generation is conditioned on the latent vector $\mathbf{z} = E_\phi(\mathbf{x})$ (Section~\ref{sec:generative_models}), a 128-dimensional summary of the observed light curve. Given a large collection of such vectors, principal component analysis (PCA) finds the orthogonal directions along which the vectors vary the most. The first principal direction is the direction of maximal variance across the collection, the second is the direction of maximal variance orthogonal to the first, and so on. Keeping only the $k$ leading directions gives the best $k$-dimensional (low-rank) approximation of the vectors in the least-squares sense.

These directions describe variations of the \emph{encoding}, not of the light curves themselves. Because the encoder summarizes the stochastic process that generated a light curve, the leading directions capture the dominant ways in which the noise model varies across the training population.

\paragraph*{Assumption.}
We assume that the contamination is of low rank, and that it contributes much less structural variation than the noise model, since it is much less common in the training data. Under this assumption, the encoding directions associated with periodic signals and other contaminations reside in the less dominant principal directions, and discarding those directions removes them from the condition.

\paragraph*{Method.}
We use the same trained models as in the first part of this experiment (three independently trained models per contamination level), without any retraining. For each of these models, we encode every light curve $\mathbf{x}_i$ of its (contaminated) training set, $\mathbf{z}_i = E_\phi(\mathbf{x}_i)$, and compute the mean $\bar{\mathbf{z}}$ and the $k$ leading principal directions of these encodings, stored as the orthonormal columns of a matrix $\mathbf{U}_k$. At inference, the conditioning vector of an observed light curve is replaced by its projection
\begin{equation}
\mathbf{z}_{\rm proj} = \bar{\mathbf{z}} + \mathbf{U}_k \mathbf{U}_k^{\top}\left(\mathbf{z} - \bar{\mathbf{z}}\right),
\label{eq:pca_projection}
\end{equation}
and the generator is conditioned on $\mathbf{z}_{\rm proj}$ instead of $\mathbf{z}$. All other steps of the detection procedure are unchanged. As before, for each observed light curve we pool 100 realizations from each of the three trained models, each with its own projection, into a single null of 300 realizations, and compute one p-value from it. Here we use $k=6$.

Note that the projection does not make the model forget the contamination. The model has learned the periodic signals in its training set and is still able to generate signal-like realizations. However, the generator is faithful to its conditioning vector: it produces realizations consistent with what the condition encodes. When the observed light curve contains a periodic signal, the imprint of the signal on $\mathbf{z}$ lies mostly in the discarded directions. Projecting it away leaves a condition that describes only the stochastic variability, so the model generates realizations of the null, without the signal, against which the observed signal stands out.

\paragraph*{Results of the projection.}
Figures~\ref{fig:pca_roc} and~\ref{fig:pca_pvals} show the effect of the projection. For the clean model, detection performance is unchanged, indicating that the six leading directions retain the information needed to describe the AR(1) noise. For the contaminated models, the projection improves detection drastically: the AUC increases from close to random guessing to 0.92--0.96, and the p-values of light curves containing a periodic signal are again concentrated near 0.

\paragraph*{Choosing the number of directions.}
The number of retained principal directions $k$ can be tailored to the situation. Retaining too many directions keeps the contamination in the condition, while retaining too few may discard genuine variations of the noise model. When the appropriate $k$ is unknown, as for real survey data, we suggest the following calibration. A separate model is trained on the same data with additional injected signals of known form and location. On this model, the detection of the injected signals is tested for different values of $k$, and the value with the best detection performance is selected. This value of $k$ is then used with the model trained without the injected signals.

\section{Discussion}\label{sec:discussion}

\subsection{Uniform sampling assumption.}

All experiments presented in this paper assume light curves that are uniformly sampled in time, which simplifies the model architecture. Real astronomical time-domain surveys, however, typically produce non-uniformly sampled light curves, due to observational constraints such as weather, seasonal visibility, and telescope scheduling, in addition to having measurement errors. We believe the same framework and conclusions should extend to non-uniformly sampled data, for instance by incorporating time-stamp information directly into the model conditioning or architecture, but we leave a systematic evaluation of this extension for future work.

\subsection{Fixed sequence length.}
Our model is trained to generate light curves of a fixed length, $T=256$ samples, and therefore cannot learn or reproduce stochastic structure that only becomes apparent on timescales longer than this window. This limits the model's ability to capture very long-timescale variability, such as the low-frequency tail of the DRW power spectrum or trends that would require baselines longer than the training sequence length to constrain. Extending the model to longer or variable-length sequences is an important direction for future work.

\subsection{Application to quasar light curves from time-domain surveys}

Given the two caveats discussed above, we also discuss next steps required to train the model on quasar light curves from large time-domain surveys like CRTS~\cite{crts_2009, tachibana2020}, ZTF~\cite{ztf2019}, and in the future, LSST~\cite{lsst2019}. Periodic quasars powered by SMBHBs are expected to be rare, but our experiments show a substantial loss of detection performance already at $0.5\%$ contamination, the smallest rate we tested (see Fig.~\ref{fig:contamination_roc}). For example, the inspiral time of a $q=1, M=10^9 M_\odot$ with an orbital period of 2 years is roughly $10^4$ years~\cite{haiman2009}, and if the average quasar lifetime~\cite{qso_lifetimes} is about $10^7$ years, then the fraction of bright SMBHBs present in quasar light curves is expected to be about $10^4/10^7=0.1\%$. Whether such a low level of contamination affects the performance of our model remains to be tested. However, given the large impact that contamination by periodic signals can have in SMBHB detection, if such signals can be identified independently, excluding or down-weighting them and retraining the model could be a useful strategy. Alternatively, the low-rank projection of the conditioning vector (Section~\ref{subsec:pca}) reduces this sensitivity without requiring the contaminating signals to be identified.

A limitation of our model is that variability with timescales longer than the length of the light curve cannot be captured. The length of the training data can be extended by combining data from two surveys like CRTS and ZTF to create 20-year light curves~\cite{crts_ztf_binaries, CAZ_catalog}. As addressed in \cite{crts_ztf_binaries}, the challenge lies in the calibration of the light curves because the passbands of CRTS and ZTF differ. If this can be achieved, current surveys, in which periodic signals have already been identified and can be excluded from the training set, can be used to train \code\ and can be applied to future LSST data releases to find periodic signals. 

Rich information is also contained in the multi-band (g,r,i) data of ZTF and is expected of LSST (u,g,r,i,z,y), and should be exploited given the wavelength-dependent signatures of SMBHBs \cite{pg1302_dan, xin2020, qhx}. Our current model assumes only a single filter from a single survey, and we plan to extend the scope of \code\ in future works.

\section{Conclusions and prospects}\label{sec:conclusions}

Our experiments demonstrate that \code\ learns the statistical structure of the underlying variability as assessed by the properties examined here. These tests probe detailed distributional features, including the joint distribution of AR(2) coefficients inferred from generated light curves and the shapes of Fourier-amplitude distributions after centering and normalization to unit variance. In particular, the latter distinguish processes with identical mean power spectra and matched amplitude means and variances, demonstrating sensitivity to higher-order statistical structure. The model may also learn additional statistical characteristics that we have not explicitly tested. Thus, our results establish agreement in the statistical behavior examined here, while leaving the extent of agreement in other properties open to future investigation.

When used as a null hypothesis for detecting periodic signals in synthetic light curves, \code\ performs comparably to fitting the correct DRW model to each light curve, and outperforms fitting an AR(2) model when the variability follows the more flexible AR(2) process, by taking into account the population distribution of the noise parameters. At the same time, its performance is sensitive to contamination of the training set by periodic signals, highlighting the importance of careful dataset curation. Projecting the conditioning vector onto its leading principal components largely restores the detection performance of contaminated models.

\section{Data Availability}\label{sec:data_avail}
\code\ is publicly available on GitHub at \url{https://github.com/NadavSellamWork/Nullasar} and can also be used to reproduce the data used in this article.

\section*{Acknowledgements}
This research was supported by ERC grant (Bright-BHBs, grant agreement no.101200084). Views and opinions expressed are however those of the author(s) only and do not necessarily reflect those of the European Union or the European Research Council Executive Agency. Neither the European Union nor the granting authority can be held responsible for them.

\bibliographystyle{mnras}
\bibliography{bibliography}

@ARTICLE{xin2020,
       author = {{Xin}, Chengcheng and {Charisi}, Maria and {Haiman}, Zolt{\'a}n and {Schiminovich}, David and {Graham}, Matthew J. and {Stern}, Daniel and {D'Orazio}, Daniel J.},
        title = "{Testing the relativistic Doppler boost hypothesis for the binary candidate quasar PG1302-102 with multiband Swift data}",
      journal = {\mnras},
         year = 2020,
        month = aug,
       volume = {496},
       number = {2},
        pages = {1683-1696},
          doi = {10.1093/mnras/staa1643},
archivePrefix = {arXiv},
       eprint = {1907.11246},
 primaryClass = {astro-ph.GA},
       adsurl = {https://ui.adsabs.harvard.edu/abs/2020MNRAS.496.1683X}
}

@ARTICLE{pg1302_dan,
       author = {{D'Orazio}, Daniel J. and {Haiman}, Zolt{\'a}n and {Schiminovich}, David},
        title = "{Relativistic boost as the cause of periodicity in a massive black-hole binary candidate}",
      journal = {\nat},
         year = 2015,
        month = sep,
       volume = {525},
       number = {7569},
        pages = {351-353},
          doi = {10.1038/nature15262},
archivePrefix = {arXiv},
       eprint = {1509.04301},
 primaryClass = {astro-ph.HE},
       adsurl = {https://ui.adsabs.harvard.edu/abs/2015Natur.525..351D}
}

@ARTICLE{CAZ_catalog,
       author = {{Kouch}, Pouya M. and {Lindfors}, Elina and {Hovatta}, Talvikki and {Liodakis}, Ioannis and {Koljonen}, Karri I.~I. and {Paggi}, Alessandro and {Nilsson}, Kari and {Jormanainen}, Jenni and {Fallah Ramazani}, Vandad and {Kankkunen}, Sofia and {Wierda}, Folkert and {Wagner}, Sarah M. and {Graham}, Matthew J.},
        title = "{CAZ catalog and optical light curves of 7918 blazar-selected active galactic nuclei}",
      journal = {\aap},
         year = 2026,
        month = apr,
       volume = {708},
          eid = {A382},
        pages = {A382},
          doi = {10.1051/0004-6361/202557582},
archivePrefix = {arXiv},
       eprint = {2510.16584},
 primaryClass = {astro-ph.HE},
       adsurl = {https://ui.adsabs.harvard.edu/abs/2026A&A...708A.382K}
}

@ARTICLE{crts_ztf_binaries,
       author = {{Foustoul}, V. and {Webb}, N.~A. and {Mignon-Risse}, R. and {Kammoun}, E. and {Volonteri}, M. and {Dong-P{\'a}ez}, C.~A.},
        title = "{A catalogue of candidate milliparsec-separation massive black hole binaries from long-term optical photometric monitoring}",
      journal = {\aap},
         year = 2025,
        month = jul,
       volume = {699},
          eid = {A55},
        pages = {A55},
          doi = {10.1051/0004-6361/202452818},
archivePrefix = {arXiv},
       eprint = {2505.06656},
 primaryClass = {astro-ph.GA},
       adsurl = {https://ui.adsabs.harvard.edu/abs/2025A&A...699A..55F}
}

@ARTICLE{qhx,
       author = {{Kova{\v{c}}evi{\'c}}, Andjelka B. and {Ili{\'c}}, Dragana and {Me{\v{s}}tri{\'c}}, Uro{\v{s}} and {Stankovi{\'c} Cerovi{\'c}}, Milena and {De Bonis}, Natale and {Popovi{\'c}}, Luka {\v{C}}. and {Simi{\'c}}, Sa{\v{s}}a and {Fatovi{\'c}}, Marta and {Hernandez-Garc{\'\i}a}, Lorena and {Schneider}, Donald P. and et al.},
        title = "{QhX Catalog of Coherent and Quasi-coherent Active Galactic Nucleus Variability from Vera C. Rubin LSST DP1}",
      journal = {\apjs},
         year = 2026,
        month = aug,
       volume = {285},
       number = {2},
          eid = {61},
        pages = {61},
          doi = {10.3847/1538-4365/ae8444},
       adsurl = {https://ui.adsabs.harvard.edu/abs/2026ApJS..285...61K}
}

@ARTICLE{haiman2009,
       author = {{Haiman}, Zolt{\'a}n and {Kocsis}, Bence and {Menou}, Kristen},
        title = "{The Population of Viscosity- and Gravitational Wave-driven Supermassive Black Hole Binaries Among Luminous Active Galactic Nuclei}",
      journal = {\apj},
         year = 2009,
        month = aug,
       volume = {700},
       number = {2},
        pages = {1952-1969},
          doi = {10.1088/0004-637X/700/2/1952},
archivePrefix = {arXiv},
       eprint = {0904.1383},
 primaryClass = {astro-ph.CO},
       adsurl = {https://ui.adsabs.harvard.edu/abs/2009ApJ...700.1952H}
}

@INPROCEEDINGS{qso_lifetimes,
       author = {{Martini}, Paul},
        title = "{QSO Lifetimes}",
    booktitle = {Coevolution of Black Holes and Galaxies},
         year = 2004,
       editor = {{Ho}, Luis C.},
        month = jan,
        pages = {169},
          doi = {10.48550/arXiv.astro-ph/0304009},
archivePrefix = {arXiv},
       eprint = {astro-ph/0304009},
 primaryClass = {astro-ph},
       adsurl = {https://ui.adsabs.harvard.edu/abs/2004cbhg.symp..169M}
}

@ARTICLE{lsst2019,
       author = {{Ivezi{\'c}}, {\v{Z}}eljko and {Kahn}, Steven M. and {Tyson}, J. Anthony and {Abel}, Bob and {Acosta}, Emily and {Allsman}, Robyn and {Alonso}, David and {AlSayyad}, Yusra and {Anderson}, Scott F. and {Andrew}, John and {Angel}, James Roger P. and {Angeli}, George Z. and {Ansari}, Reza and {Antilogus}, Pierre and {Araujo}, Constanza and {Armstrong}, Robert and {Arndt}, Kirk T. and {Astier}, Pierre and {Aubourg}, {\'E}ric and {Auza}, Nicole and {Axelrod}, Tim S. and {Bard}, Deborah J. and {Barr}, Jeff D. and {Barrau}, Aurelian and {Bartlett}, James G. and {Bauer}, Amanda E. and {Bauman}, Brian J. and {Baumont}, Sylvain and {Bechtol}, Ellen and {Bechtol}, Keith and {Becker}, Andrew C. and {Becla}, Jacek and {Beldica}, Cristina and {Bellavia}, Steve and {Bianco}, Federica B. and {Biswas}, Rahul and {Blanc}, Guillaume and {Blazek}, Jonathan and {Blandford}, Roger D. and {Bloom}, Josh S. and {Bogart}, Joanne and {Bond}, Tim W. and {Booth}, Michael T. and {Borgland}, Anders W. and {Borne}, Kirk and {Bosch}, James F. and {Boutigny}, Dominique and {Brackett}, Craig A. and {Bradshaw}, Andrew and {Brandt}, William Nielsen and {Brown}, Michael E. and {Bullock}, James S. and {Burchat}, Patricia and {Burke}, David L. and {Cagnoli}, Gianpietro and {Calabrese}, Daniel and {Callahan}, Shawn and {Callen}, Alice L. and {Carlin}, Jeffrey L. and {Carlson}, Erin L. and {Chandrasekharan}, Srinivasan and {Charles-Emerson}, Glenaver and {Chesley}, Steve and {Cheu}, Elliott C. and {Chiang}, Hsin-Fang and {Chiang}, James and {Chirino}, Carol and {Chow}, Derek and {Ciardi}, David R. and {Claver}, Charles F. and {Cohen-Tanugi}, Johann and {Cockrum}, Joseph J. and {Coles}, Rebecca and {Connolly}, Andrew J. and {Cook}, Kem H. and {Cooray}, Asantha and {Covey}, Kevin R. and {Cribbs}, Chris and {Cui}, Wei and {Cutri}, Roc and {Daly}, Philip N. and {Daniel}, Scott F. and {Daruich}, Felipe and {Daubard}, Guillaume and {Daues}, Greg and {Dawson}, William and {Delgado}, Francisco and {Dellapenna}, Alfred and {de Peyster}, Robert and {de Val-Borro}, Miguel and {Digel}, Seth W. and {Doherty}, Peter and {Dubois}, Richard and {Dubois-Felsmann}, Gregory P. and {Durech}, Josef and {Economou}, Frossie and {Eifler}, Tim and {Eracleous}, Michael and {Emmons}, Benjamin L. and {Fausti Neto}, Angelo and {Ferguson}, Henry and {Figueroa}, Enrique and {Fisher-Levine}, Merlin and {Focke}, Warren and {Foss}, Michael D. and {Frank}, James and {Freemon}, Michael D. and {Gangler}, Emmanuel and {Gawiser}, Eric and {Geary}, John C. and {Gee}, Perry and {Geha}, Marla and {Gessner}, Charles J.~B. and {Gibson}, Robert R. and {Gilmore}, D. Kirk and {Glanzman}, Thomas and {Glick}, William and {Goldina}, Tatiana and {Goldstein}, Daniel A. and {Goodenow}, Iain and {Graham}, Melissa L. and {Gressler}, William J. and {Gris}, Philippe and {Guy}, Leanne P. and {Guyonnet}, Augustin and {Haller}, Gunther and {Harris}, Ron and {Hascall}, Patrick A. and {Haupt}, Justine and {Hernandez}, Fabio and {Herrmann}, Sven and {Hileman}, Edward and {Hoblitt}, Joshua and {Hodgson}, John A. and {Hogan}, Craig and {Howard}, James D. and {Huang}, Dajun and {Huffer}, Michael E. and {Ingraham}, Patrick and {Innes}, Walter R. and {Jacoby}, Suzanne H. and {Jain}, Bhuvnesh and {Jammes}, Fabrice and {Jee}, M. James and {Jenness}, Tim and {Jernigan}, Garrett and {Jevremovi{\'c}}, Darko and {Johns}, Kenneth and {Johnson}, Anthony S. and {Johnson}, Margaret W.~G. and {Jones}, R. Lynne and {Juramy-Gilles}, Claire and {Juri{\'c}}, Mario and {Kalirai}, Jason S. and {Kallivayalil}, Nitya J. and {Kalmbach}, Bryce and {Kantor}, Jeffrey P. and {Karst}, Pierre and {Kasliwal}, Mansi M. and {Kelly}, Heather and {Kessler}, Richard and {Kinnison}, Veronica and {Kirkby}, David and {Knox}, Lloyd and {Kotov}, Ivan V. and {Krabbendam}, Victor L. and {Krughoff}, K. Simon and {Kub{\'a}nek}, Petr and {Kuczewski}, John and {Kulkarni}, Shri and {Ku}, John and {Kurita}, Nadine R. and {Lage}, Craig S. and {Lambert}, Ron and {Lange}, Travis and {Langton}, J. Brian and {Le Guillou}, Laurent and {Levine}, Deborah and {Liang}, Ming and {Lim}, Kian-Tat and {Lintott}, Chris J. and {Long}, Kevin E. and {Lopez}, Margaux and {Lotz}, Paul J. and {Lupton}, Robert H. and {Lust}, Nate B. and {MacArthur}, Lauren A. and {Mahabal}, Ashish and {Mandelbaum}, Rachel and {Markiewicz}, Thomas W. and {Marsh}, Darren S. and {Marshall}, Philip J. and {Marshall}, Stuart and {May}, Morgan and {McKercher}, Robert and {McQueen}, Michelle and {Meyers}, Joshua and {Migliore}, Myriam and {Miller}, Michelle and {Mills}, David J.},
        title = "{LSST: From Science Drivers to Reference Design and Anticipated Data Products}",
      journal = {\apj},
         year = 2019,
        month = mar,
       volume = {873},
       number = {2},
          eid = {111},
        pages = {111},
          doi = {10.3847/1538-4357/ab042c},
archivePrefix = {arXiv},
       eprint = {0805.2366},
 primaryClass = {astro-ph},
       adsurl = {https://ui.adsabs.harvard.edu/abs/2019ApJ...873..111I}
}

@ARTICLE{ztf2019,
       author = {{Masci}, Frank J. and {Laher}, Russ R. and {Rusholme}, Ben and {Shupe}, David L. and {Groom}, Steven and {Surace}, Jason and {Jackson}, Edward and {Monkewitz}, Serge and {Beck}, Ron and {Flynn}, David and {Terek}, Scott and {Landry}, Walter and {Hacopians}, Eugean and {Desai}, Vandana and {Howell}, Justin and {Brooke}, Tim and {Imel}, David and {Wachter}, Stefanie and {Ye}, Quan-Zhi and {Lin}, Hsing-Wen and {Cenko}, S. Bradley and {Cunningham}, Virginia and {Rebbapragada}, Umaa and {Bue}, Brian and {Miller}, Adam A. and {Mahabal}, Ashish and {Bellm}, Eric C. and {Patterson}, Maria T. and {Juri{\'c}}, Mario and {Golkhou}, V. Zach and {Ofek}, Eran O. and {Walters}, Richard and {Graham}, Matthew and {Kasliwal}, Mansi M. and {Dekany}, Richard G. and {Kupfer}, Thomas and {Burdge}, Kevin and {Cannella}, Christopher B. and {Barlow}, Tom and {Van Sistine}, Angela and {Giomi}, Matteo and {Fremling}, Christoffer and {Blagorodnova}, Nadejda and {Levitan}, David and {Riddle}, Reed and {Smith}, Roger M. and {Helou}, George and {Prince}, Thomas A. and {Kulkarni}, Shrinivas R.},
        title = "{The Zwicky Transient Facility: Data Processing, Products, and Archive}",
      journal = {\pasp},
         year = 2019,
        month = jan,
       volume = {131},
       number = {995},
        pages = {018003},
          doi = {10.1088/1538-3873/aae8ac},
archivePrefix = {arXiv},
       eprint = {1902.01872},
 primaryClass = {astro-ph.IM},
       adsurl = {https://ui.adsabs.harvard.edu/abs/2019PASP..131a8003M}
}

@ARTICLE{crts_2009,
       author = {{Drake}, A.~J. and {Djorgovski}, S.~G. and {Mahabal}, A. and {Beshore}, E. and {Larson}, S. and {Graham}, M.~J. and {Williams}, R. and {Christensen}, E. and {Catelan}, M. and {Boattini}, A. and {Gibbs}, A. and {Hill}, R. and {Kowalski}, R.},
        title = "{First Results from the Catalina Real-Time Transient Survey}",
      journal = {\apj},
         year = 2009,
        month = may,
       volume = {696},
       number = {1},
        pages = {870-884},
          doi = {10.1088/0004-637X/696/1/870},
archivePrefix = {arXiv},
       eprint = {0809.1394},
 primaryClass = {astro-ph},
       adsurl = {https://ui.adsabs.harvard.edu/abs/2009ApJ...696..870D}
}

@ARTICLE{tachibana2020,
       author = {{Tachibana}, Yutaro and {Graham}, Matthew J. and {Kawai}, Nobuyuki and {Djorgovski}, S.~G. and {Drake}, Andrew J. and {Mahabal}, Ashish A. and {Stern}, Daniel},
        title = "{Deep Modeling of Quasar Variability}",
      journal = {\apj},
         year = 2020,
        month = nov,
       volume = {903},
       number = {1},
          eid = {54},
        pages = {54},
          doi = {10.3847/1538-4357/abb9a9},
archivePrefix = {arXiv},
       eprint = {2003.01241},
 primaryClass = {astro-ph.GA},
       adsurl = {https://ui.adsabs.harvard.edu/abs/2020ApJ...903...54T}
}

@ARTICLE{mushotzky_kepler_drw,
       author = {{Mushotzky}, R.~F. and {Edelson}, R. and {Baumgartner}, W. and {Gandhi}, P.},
        title = "{Kepler Observations of Rapid Optical Variability in Active Galactic Nuclei}",
      journal = {\apjl},
         year = 2011,
        month = dec,
       volume = {743},
       number = {1},
          eid = {L12},
        pages = {L12},
          doi = {10.1088/2041-8205/743/1/L12},
archivePrefix = {arXiv},
       eprint = {1111.0672},
 primaryClass = {astro-ph.GA},
       adsurl = {https://ui.adsabs.harvard.edu/abs/2011ApJ...743L..12M}
}

@ARTICLE{kepler_drw,
       author = {{Kasliwal}, Vishal P. and {Vogeley}, Michael S. and {Richards}, Gordon T.},
        title = "{Are the variability properties of the Kepler AGN light curves consistent with a damped random walk?}",
      journal = {\mnras},
         year = 2015,
        month = aug,
       volume = {451},
       number = {4},
        pages = {4328-4345},
          doi = {10.1093/mnras/stv1230},
archivePrefix = {arXiv},
       eprint = {1505.00360},
 primaryClass = {astro-ph.GA},
       adsurl = {https://ui.adsabs.harvard.edu/abs/2015MNRAS.451.4328K}
}

@ARTICLE{davis2026,
       author = {{Davis}, Megan C. and {Trump}, Jonathan R. and {Charisi}, Maria and {Runnoe}, Jessie C. and {Brandt}, W.~N. and {Grace}, Kaylee E. and {Willson}, London E.},
        title = "{The Consequences of Rubin Observatory Time-domain Survey Design and Host-galaxy Contamination on the Identification of Binary Supermassive Black Holes}",
      journal = {\apjs},
         year = 2026,
        month = feb,
       volume = {282},
       number = {2},
          eid = {30},
        pages = {30},
          doi = {10.3847/1538-4365/ae2c53},
archivePrefix = {arXiv},
       eprint = {2508.05742},
 primaryClass = {astro-ph.HE},
       adsurl = {https://ui.adsabs.harvard.edu/abs/2026ApJS..282...30D}
}

@ARTICLE{davis2024,
       author = {{Davis}, Megan C. and {Grace}, Kaylee E. and {Trump}, Jonathan R. and {Runnoe}, Jessie C. and {Henkel}, Amelia and {Blecha}, Laura and {Brandt}, W.~N. and {Casey-Clyde}, J. Andrew and {Charisi}, Maria and {Witt}, Caitlin A.},
        title = "{Reliable Identification of Binary Supermassive Black Holes from Rubin Observatory Time-domain Monitoring}",
      journal = {\apj},
         year = 2024,
        month = apr,
       volume = {965},
       number = {1},
          eid = {34},
        pages = {34},
          doi = {10.3847/1538-4357/ad276e},
archivePrefix = {arXiv},
       eprint = {2311.10851},
 primaryClass = {astro-ph.HE},
       adsurl = {https://ui.adsabs.harvard.edu/abs/2024ApJ...965...34D}
}

@ARTICLE{witt2022,
       author = {{Witt}, Caitlin A. and {Charisi}, Maria and {Taylor}, Stephen R. and {Burke-Spolaor}, Sarah},
        title = "{Quasars with Periodic Variability: Capabilities and Limitations of Bayesian Searches for Supermassive Black Hole Binaries in Time-domain Surveys}",
      journal = {\apj},
         year = 2022,
        month = sep,
       volume = {936},
       number = {1},
          eid = {89},
        pages = {89},
          doi = {10.3847/1538-4357/ac8356},
archivePrefix = {arXiv},
       eprint = {2110.07465},
 primaryClass = {astro-ph.GA},
       adsurl = {https://ui.adsabs.harvard.edu/abs/2022ApJ...936...89W}
}

@ARTICLE{Roman,
       author = {{Sanderson}, Robyn E. and {Hickox}, Ryan and {Hirata}, Christopher M. and {Holman}, Matthew J. and {Lu}, Jessica R. and {Villar}, Ashley},
        title = "{Recommendations for Early Definition Science with the Nancy Grace Roman Space Telescope}",
      journal = {arXiv e-prints},
         year = 2024,
        month = apr,
          eid = {arXiv:2404.14342},
        pages = {arXiv:2404.14342},
          doi = {10.48550/arXiv.2404.14342},
archivePrefix = {arXiv},
       eprint = {2404.14342},
 primaryClass = {astro-ph.GA},
       adsurl = {https://ui.adsabs.harvard.edu/abs/2024arXiv240414342S}
}

@ARTICLE{ULTRASAT,
       author = {{Shvartzvald}, Y. and {Waxman}, E. and {Gal-Yam}, A. and {Ofek}, E.~O. and {Ben-Ami}, S. and {Berge}, D. and {Kowalski}, M. and {B{\"u}hler}, R. and {Worm}, S. and {Rhoads}, J.~E. and {Arcavi}, I. and {Maoz}, D. and {Polishook}, D. and {Stone}, N. and {Trakhtenbrot}, B. and {Ackermann}, M. and {Aharonson}, O. and {Birnholtz}, O. and {Chelouche}, D. and {Guetta}, D. and {Hallakoun}, N. and {Horesh}, A. and {Kushnir}, D. and {Mazeh}, T. and {Nordin}, J. and {Ofir}, A. and {Ohm}, S. and {Parsons}, D. and {Pe'er}, A. and {Perets}, H.~B. and {Perdelwitz}, V. and {Poznanski}, D. and {Sadeh}, I. and {Sagiv}, I. and {Shahaf}, S. and {Soumagnac}, M. and {Tal-Or}, L. and {Santen}, J. Van and {Zackay}, B. and {Guttman}, O. and {Rekhi}, P. and {Townsend}, A. and {Weinstein}, A. and {Wold}, I.},
        title = "{ULTRASAT: A Wide-field Time-domain UV Space Telescope}",
      journal = {\apj},
         year = 2024,
        month = mar,
       volume = {964},
       number = {1},
          eid = {74},
        pages = {74},
          doi = {10.3847/1538-4357/ad2704},
archivePrefix = {arXiv},
       eprint = {2304.14482},
 primaryClass = {astro-ph.IM},
       adsurl = {https://ui.adsabs.harvard.edu/abs/2024ApJ...964...74S}
}

@ARTICLE{XinHaiman2024,
       author = {{Xin}, Chengcheng and {Haiman}, Zolt{\'a}n},
        title = "{Identifying the electromagnetic counterparts of LISA massive black hole binaries in archival LSST data}",
      journal = {\mnras},
         year = 2024,
        month = sep,
       volume = {533},
       number = {3},
        pages = {3164-3173},
          doi = {10.1093/mnras/stae2009},
archivePrefix = {arXiv},
       eprint = {2403.18751},
 primaryClass = {astro-ph.HE},
       adsurl = {https://ui.adsabs.harvard.edu/abs/2024MNRAS.533.3164X}
}

@ARTICLE{XinHaiman2021,
       author = {{Xin}, Chengcheng and {Haiman}, Zolt{\'a}n},
        title = "{Ultra-short-period massive black hole binary candidates in LSST as LISA 'verification binaries'}",
      journal = {\mnras},
         year = 2021,
        month = sep,
       volume = {506},
       number = {2},
        pages = {2408-2417},
          doi = {10.1093/mnras/stab1856},
archivePrefix = {arXiv},
       eprint = {2105.00005},
 primaryClass = {astro-ph.HE},
       adsurl = {https://ui.adsabs.harvard.edu/abs/2021MNRAS.506.2408X}
}

@ARTICLE{Bogdanovic+2022,
       author = {{Bogdanovi{\'c}}, Tamara and {Miller}, M. Coleman and {Blecha}, Laura},
        title = "{Electromagnetic counterparts to massive black-hole mergers}",
      journal = {Living Reviews in Relativity},
         year = 2022,
        month = dec,
       volume = {25},
       number = {1},
          eid = {3},
        pages = {3},
          doi = {10.1007/s41114-022-00037-8},
archivePrefix = {arXiv},
       eprint = {2109.03262},
 primaryClass = {astro-ph.HE},
       adsurl = {https://ui.adsabs.harvard.edu/abs/2022LRR....25....3B}
}

@ARTICLE{Westernacher-Schneider+2022,
       author = {{Westernacher-Schneider}, John Ryan and {Zrake}, Jonathan and {MacFadyen}, Andrew and {Haiman}, Zolt{\'a}n},
        title = "{Multiband light curves from eccentric accreting supermassive black hole binaries}",
      journal = {\prd},
         year = 2022,
        month = nov,
       volume = {106},
       number = {10},
          eid = {103010},
        pages = {103010},
          doi = {10.1103/PhysRevD.106.103010},
archivePrefix = {arXiv},
       eprint = {2111.06882},
 primaryClass = {astro-ph.HE},
       adsurl = {https://ui.adsabs.harvard.edu/abs/2022PhRvD.106j3010W}
}

@ARTICLE{Kroupa+2026,
       author = {{Kroupa}, Namu and {Yallup}, David and {Handley}, Will},
        title = "{Time delays and stationarity in quasar light curves}",
      journal = {\prd},
         year = 2026,
        month = mar,
       volume = {113},
       number = {6},
          eid = {063549},
        pages = {063549},
          doi = {10.1103/87nn-klh9},
archivePrefix = {arXiv},
       eprint = {2602.11264},
 primaryClass = {astro-ph.CO},
       adsurl = {https://ui.adsabs.harvard.edu/abs/2026PhRvD.113f3549K}
}

@ARTICLE{Stone+2022,
       author = {{Stone}, Zachary and {Shen}, Yue and {Burke}, Colin J. and {Chen}, Yu-Ching and {Yang}, Qian and {Liu}, Xin and {Gruendl}, R.~A. and {Adam{\'o}w}, M. and {Andrade-Oliveira}, F. and {Annis}, J. and {Bacon}, D. and {Bertin}, E. and {Bocquet}, S. and {Brooks}, D. and {Burke}, D.~L. and {Carnero Rosell}, A. and {Carrasco Kind}, M. and {Carretero}, J. and {da Costa}, L.~N. and {Pereira}, M.~E.~S. and {De Vicente}, J. and {Desai}, S. and {Diehl}, H.~T. and {Doel}, P. and {Ferrero}, I. and {Friedel}, D.~N. and {Frieman}, J. and {Garc{\'\i}a-Bellido}, J. and {Gaztanaga}, E. and {Gruen}, D. and {Gutierrez}, G. and {Hinton}, S.~R. and {Hollowood}, D.~L. and {Honscheid}, K. and {James}, D.~J. and {Kuehn}, K. and {Kuropatkin}, N. and {Lidman}, C. and {Maia}, M.~A.~G. and {Menanteau}, F. and {Miquel}, R. and {Morgan}, R. and {Paz-Chinch{\'o}n}, F. and {Pieres}, A. and {Plazas Malag{\'o}n}, A.~A. and {Rodriguez-Monroy}, M. and {Sanchez}, E. and {Scarpine}, V. and {Serrano}, S. and {Sevilla-Noarbe}, I. and {Smith}, M. and {Suchyta}, E. and {Swanson}, M.~E.~C. and {Tarl{\'e}}, G. and {To}, C. and {DES Collaboration}},
        title = "{Optical variability of quasars with 20-yr photometric light curves}",
      journal = {\mnras},
         year = 2022,
        month = jul,
       volume = {514},
       number = {1},
        pages = {164-184},
          doi = {10.1093/mnras/stac1259},
archivePrefix = {arXiv},
       eprint = {2201.02762},
 primaryClass = {astro-ph.GA},
       adsurl = {https://ui.adsabs.harvard.edu/abs/2022MNRAS.514..164S}
}

@ARTICLE{Kelly+2014,
       author = {{Kelly}, Brandon C. and {Becker}, Andrew C. and {Sobolewska}, Malgosia and {Siemiginowska}, Aneta and {Uttley}, Phil},
        title = "{Flexible and Scalable Methods for Quantifying Stochastic Variability in the Era of Massive Time-domain Astronomical Data Sets}",
      journal = {\apj},
         year = 2014,
        month = jun,
       volume = {788},
       number = {1},
          eid = {33},
        pages = {33},
          doi = {10.1088/0004-637X/788/1/33},
archivePrefix = {arXiv},
       eprint = {1402.5978},
 primaryClass = {astro-ph.IM},
       adsurl = {https://ui.adsabs.harvard.edu/abs/2014ApJ...788...33K}
}

@ARTICLE{Kelly+2009,
       author = {{Kelly}, Brandon C. and {Bechtold}, Jill and {Siemiginowska}, Aneta},
        title = "{Are the Variations in Quasar Optical Flux Driven by Thermal Fluctuations?}",
      journal = {\apj},
         year = 2009,
        month = jun,
       volume = {698},
       number = {1},
        pages = {895-910},
          doi = {10.1088/0004-637X/698/1/895},
archivePrefix = {arXiv},
       eprint = {0903.5315},
 primaryClass = {astro-ph.CO},
       adsurl = {https://ui.adsabs.harvard.edu/abs/2009ApJ...698..895K}
}

@ARTICLE{MacLeod+2010,
       author = {{MacLeod}, C.~L. and {Ivezi{\'c}}, {\v{Z}}. and {Kochanek}, C.~S. and {Koz{\l}owski}, S. and {Kelly}, B. and {Bullock}, E. and {Kimball}, A. and {Sesar}, B. and {Westman}, D. and {Brooks}, K. and {Gibson}, R. and {Becker}, A.~C. and {de Vries}, W.~H.},
        title = "{Modeling the Time Variability of SDSS Stripe 82 Quasars as a Damped Random Walk}",
      journal = {\apj},
         year = 2010,
        month = oct,
       volume = {721},
       number = {2},
        pages = {1014-1033},
          doi = {10.1088/0004-637X/721/2/1014},
archivePrefix = {arXiv},
       eprint = {1004.0276},
 primaryClass = {astro-ph.CO},
       adsurl = {https://ui.adsabs.harvard.edu/abs/2010ApJ...721.1014M}
}

@ARTICLE{Kozlowski+2010,
       author = {{Koz{\l}owski}, Szymon and {Kochanek}, Christopher S. and {Udalski}, A. and {Wyrzykowski}, {\L}. and {Soszy{\'n}ski}, I. and {Szyma{\'n}ski}, M.~K. and {Kubiak}, M. and {Pietrzy{\'n}ski}, G. and {Szewczyk}, O. and {Ulaczyk}, K. and {Poleski}, R. and {OGLE Collaboration}},
        title = "{Quantifying Quasar Variability as Part of a General Approach to Classifying Continuously Varying Sources}",
      journal = {\apj},
         year = 2010,
        month = jan,
       volume = {708},
       number = {2},
        pages = {927-945},
          doi = {10.1088/0004-637X/708/2/927},
archivePrefix = {arXiv},
       eprint = {0909.1326},
 primaryClass = {astro-ph.CO},
       adsurl = {https://ui.adsabs.harvard.edu/abs/2010ApJ...708..927K}
}

@article{Chen2018NeuralODEs,
  author       = {Chen, Ricky T. Q. and Rubanova, Yulia and Bettencourt, Jesse and Duvenaud, David},
  title        = {Neural Ordinary Differential Equations},
  journal      = {Advances in Neural Information Processing Systems},
  volume       = {31},
  year         = {2018},
}

@article{Lipman2023FlowMatching,
  author       = {Lipman, Yaron and Chen, Ricky T. Q. and Ben-Hamu, Heli and Nickel, Maximilian and Le, Matt},
  title        = {Flow Matching for Generative Modeling},
  journal      = {International Conference on Learning Representations},
  year         = {2023},
}

@article{Ronneberger2015UNet,
  author       = {Ronneberger, Olaf and Fischer, Philipp and Brox, Thomas},
  title        = {U-Net: Convolutional Networks for Biomedical Image Segmentation},
  journal      = {International Conference on Medical Image Computing and Computer-Assisted Intervention},
  pages        = {234--241},
  year         = {2015},
}

@article{Perez2018FiLM,
  author       = {Perez, Ethan and Strub, Florian and de Vries, Harm and Dumoulin, Vincent and Courville, Aaron},
  title        = {FiLM: Visual Reasoning with a General Conditioning Layer},
  journal      = {Proceedings of the AAAI Conference on Artificial Intelligence},
  volume       = {32},
  number       = {1},
  year         = {2018},
}

@article{begelman1980,
	title = {Massive Black Hole Binaries in Active Galactic Nuclei},
	author = {Begelman, M. C. and Blandford, R. D. and Rees, M. J.},
	year = {1980},
	month = {sep},
	journal = {Nature},
	volume = {287},
	number = {5780},
	pages = {307--309},
	publisher = {{Nature Publishing Group}},
	issn = {1476-4687},
	doi = {10.1038/287307a0},
	urldate = {2023-03-11},
	copyright = {1980 Nature Publishing Group},
	langid = {english}
}

@misc{d_orazio_2023,
      title={Observational Signatures of Supermassive Black Hole Binaries}, 
      author={Daniel J. D'Orazio and Maria Charisi},
      year={2023},
      eprint={2310.16896},
      archivePrefix={arXiv},
      primaryClass={astro-ph.HE},
      url={https://arxiv.org/abs/2310.16896}, 
}

@ARTICLE{Graham+2015b,
   author = {{Graham}, M.~J. and {Djorgovski}, S.~G. and {Stern}, D. and 
	{Drake}, A.~J. and {Mahabal}, A.~A. and {Donalek}, C. and {Glikman}, E. and 
	{Larson}, S. and {Christensen}, E.},
    title = "{A systematic search for close supermassive black hole binaries in the Catalina Real-time Transient Survey}",
  journal = {\mnras},
archivePrefix = "arXiv",
   eprint = {1507.07603},
     year = 2015,
    month = oct,
   volume = 453,
    pages = {1562-1576},
      doi = {10.1093/mnras/stv1726},
   adsurl = {http://adsabs.harvard.edu/abs/2015MNRAS.453.1562G}
}

@ARTICLE{Chen+2024,
       author = {{Chen}, Yong-Jie and {Zhai}, Shuo and {Liu}, Jun-Rong and {Guo}, Wei-Jian and {Peng}, Yue-Chang and {Li}, Yan-Rong and {Songsheng}, Yu-Yang and {Du}, Pu and {Hu}, Chen and {Wang}, Jian-Min},
        title = "{Searching for quasar candidates with periodic variations from the Zwicky Transient Facility: results and implications}",
      journal = {\mnras},
         year = 2024,
        month = feb,
       volume = {527},
       number = {4},
        pages = {12154-12177},
          doi = {10.1093/mnras/stad3981},
archivePrefix = {arXiv},
       eprint = {2206.11497},
 primaryClass = {astro-ph.GA},
       adsurl = {https://ui.adsabs.harvard.edu/abs/2024MNRAS.52712154C}
}

@ARTICLE{Charisi+2016,
   author = {{Charisi}, M. and {Bartos}, I. and {Haiman}, Z. and {Price-Whelan}, A.~M. and 
	{Graham}, M.~J. and {Bellm}, E.~C. and {Laher}, R.~R. and {M{\'a}rka}, S.
	},
    title = "{A population of short-period variable quasars from PTF as supermassive black hole binary candidates}",
  journal = {\mnras},
archivePrefix = "arXiv",
   eprint = {1604.01020},
     year = 2016,
    month = dec,
   volume = 463,
    pages = {2145-2171},
      doi = {10.1093/mnras/stw1838},
   adsurl = {http://adsabs.harvard.edu/abs/2016MNRAS.463.2145C}
}

@article{Jakob_null,
    author = {Robnik, Jakob and Bayer, Adrian E and Charisi, Maria and Haiman, Zoltán and Lin, Allison and Seljak, Uroš},
    title = {Periodicity significance testing with null-signal templates: reassessment of PTF’s SMBH binary candidates},
    journal = {Monthly Notices of the Royal Astronomical Society},
    volume = {534},
    number = {2},
    pages = {1609-1620},
    year = {2024},
    month = {09},
    issn = {0035-8711},
    doi = {10.1093/mnras/stae2220},
    url = {https://doi.org/10.1093/mnras/stae2220},
    eprint = {https://academic.oup.com/mnras/article-pdf/534/2/1609/59577103/stae2220.pdf},
}

@article{El-Badry_2025,
       author = {{El-Badry}, Kareem and {Hogg}, David W. and {Rix}, Hans-Walter},
        title = "{Active Galactic Nuclei Do Not Exhibit Strictly Sinusoidal Brightness Variations}",
      journal = {\pasp},
         year = 2026,
        month = feb,
       volume = {138},
       number = {2},
          eid = {024102},
        pages = {024102},
          doi = {10.1088/1538-3873/ae3f56},
archivePrefix = {arXiv},
       eprint = {2509.10601},
 primaryClass = {astro-ph.GA},
       adsurl = {https://ui.adsabs.harvard.edu/abs/2026PASP..138b4102E}
}

@ARTICLE{Vaughan+2016,
       author = {{Vaughan}, S. and {Uttley}, P. and {Markowitz}, A.~G. and {Huppenkothen}, D. and {Middleton}, M.~J. and {Alston}, W.~N. and {Scargle}, J.~D. and {Farr}, W.~M.},
        title = "{False periodicities in quasar time-domain surveys}",
      journal = {\mnras},
         year = 2016,
        month = sep,
       volume = {461},
       number = {3},
        pages = {3145-3152},
          doi = {10.1093/mnras/stw1412},
archivePrefix = {arXiv},
       eprint = {1606.02620},
 primaryClass = {astro-ph.IM},
       adsurl = {https://ui.adsabs.harvard.edu/abs/2016MNRAS.461.3145V}
}

@misc{lin2025,
      title={Lomb-Scargle periodograms struggle with non-sinusoidal supermassive BH binary signatures in quasar lightcurves}, 
      author={Allison Lin and Maria Charisi and Zoltan Haiman},
      year={2025},
      eprint={2505.14778},
      archivePrefix={arXiv},
      primaryClass={astro-ph.HE},
      url={https://arxiv.org/abs/2505.14778}, 
}

@misc{park2025,
      title={Self-lensing flares from black hole binaries V: systematic searches in LSST}, 
      author={Kevin Park and Zoltan Haiman and Chengcheng Xin and Tzuken Shen and Ashley Villar and Jordy Davelaar},
      year={2025},
      eprint={2512.08427},
      archivePrefix={arXiv},
      primaryClass={astro-ph.HE},
      url={https://arxiv.org/abs/2512.08427}, 
}

@book{Villani2009,
  author = {Villani, C{\'e}dric},
  title = {Optimal Transport: Old and New},
  publisher = {Springer},
  year = {2009},
  series = {Grundlehren der mathematischen Wissenschaften},
  volume = {338},
  doi = {10.1007/978-3-540-71050-9}
}

@article{Yule1927,
  author  = {Yule, G. Udny},
  title   = {On a Method of Investigating Periodicities in Disturbed Series, with Special Reference to Wolfer's Sunspot Numbers},
  journal = {Philosophical Transactions of the Royal Society A},
  volume  = {226},
  pages   = {267--298},
  year    = {1927},
  doi     = {10.1098/rsta.1927.0007}
}

@article{Walker1931,
  author  = {Walker, Gilbert T.},
  title   = {On Periodicity in Series of Related Terms},
  journal = {Proceedings of the Royal Society A},
  volume  = {131},
  number  = {818},
  pages   = {518--532},
  year    = {1931},
  doi     = {10.1098/rspa.1931.0069}
}

@book{BoxJenkinsReinsel1994,
  author    = {Box, George E. P. and Jenkins, Gwilym M. and Reinsel, Gregory C.},
  title     = {Time Series Analysis: Forecasting and Control},
  edition   = {3rd},
  publisher = {Prentice Hall},
  address   = {Englewood Cliffs, NJ},
  year      = {1994}
}

@ARTICLE{Moreno2019,
       author = {{Moreno}, Jackeline and {Vogeley}, Michael S. and {Richards}, Gordon T. and {Yu}, Weixiang},
        title = "{Stochastic Modeling Handbook for Optical AGN Variability}",
      journal = {\pasp},
         year = 2019,
        month = jun,
       volume = {131},
       number = {1000},
        pages = {063001},
          doi = {10.1088/1538-3873/ab1597},
archivePrefix = {arXiv},
       eprint = {1811.00154},
 primaryClass = {astro-ph.IM},
       adsurl = {https://ui.adsabs.harvard.edu/abs/2019PASP..131f3001M}
}

@ARTICLE{oord2018,
       author = {{van den Oord}, Aaron and {Li}, Yazhe and {Vinyals}, Oriol},
        title = "{Representation Learning with Contrastive Predictive Coding}",
      journal = {arXiv e-prints},
         year = 2018,
          eid = {arXiv:1807.03748},
        pages = {arXiv:1807.03748},
       eprint = {1807.03748}
}

\clearpage
\appendix
\section{Yule-Walker Equations for estimating AR(2) coefficients}\label{appendix:YW_eq}
For the AR(2) process
\begin{equation}
    x_t= \phi_1 x_{t-1} + \phi_2 x_{t-2}+\epsilon_t,
\end{equation} the Yule-Walker equations can be derived by multiplying both sides by $x_{t-k}$ for some lag $k$ and comparing the expectation value. Then, 
\begin{equation}
    \langle x_t x_{t-k}\rangle= \phi_1 \langle x_{t-1}x_{t-k} \rangle + \phi_2 \langle x_{t-2} x_{t-k} \rangle+\langle\epsilon_t x_{t-k}\rangle
\end{equation} and we note that $\langle \epsilon_t\rangle=0.$ Choosing $k=1,2$ gives two equations which can be used to calculate $\phi_1, \phi_2.$

\section{Ablation study: the contrastive and consistency losses}\label{appendix:ablation}
All models used in the experiments of this paper are trained with the flow-matching loss only (Eq.~\ref{eq:fm_loss}). In addition, we studied two additional losses: a contrastive loss, which encourages the conditioning to better capture the parameters of the light curve, and a consistency loss, which encourages the generated light curves to be more faithful to the conditioning light curve. We define these losses below, and evaluate their effect on the conditional AR(1) generation task of Section~\ref{subsec:conditional}.

\subsection{Contrastive Loss: Learning Parametric Structure} \label{subsec: contrastive_loss}

To encourage the encoder to discover the latent parameters that govern the stochastic process, we employ a contrastive loss based on the principle that segments from the same light curve share the same underlying parameters, while segments from different light curves have different parameters.

At each training iteration, we randomly draw two non-overlapping segments of equal length (256 data points each) from each light curve with total length 2048 data points:
\begin{equation}
    \mathbf{x} \rightarrow \left\{ \mathbf{x}^{(1)}, \mathbf{x}^{(2)} \right\}.
\end{equation}

Segments from the same light curve form \textit{positive pairs}. They should produce similar representations since they share the same underlying stochastic process. Segments from different light curves form \textit{negative pairs}. They should generally produce dissimilar representations since they usually correspond to different stochastic processes, although two different light curves may have similar underlying parameters.

Let $\mathbf{x}^{(i)}$ and $\mathbf{x}^{(j)}$ be two segments from the same light curve (positive pair), and let $\mathbf{x}^{(k)}$ be a segment from a different light curve (negative pair). We compute their embeddings:
\begin{equation}
    \mathbf{z}_i = E_\phi(\mathbf{x}^{(i)}), \quad
    \mathbf{z}_j = E_\phi(\mathbf{x}^{(j)}), \quad
    \mathbf{z}_k = E_\phi(\mathbf{x}^{(k)}).
\end{equation}

We define the cosine similarity, which measures how closely the directions of two embeddings agree (1 for identical directions and $-1$ for opposite ones):
\begin{equation}
\mathrm{sim}(\mathbf{z}_a, \mathbf{z}_b)
=
\frac{\mathbf{z}_a^\top \mathbf{z}_b}{\|\mathbf{z}_a\| \|\mathbf{z}_b\|}.
\end{equation}
For a batch of $B$ light curves, let $\mathbf{z}_i$ and $\mathbf{z}_i^{+}$ be the embeddings of the two segments drawn from light curve $i$, and let $c_{ij}=\mathrm{sim}(\mathbf{z}_i,\mathbf{z}_j^{+})$. We use the InfoNCE contrastive loss \citep{oord2018}:
\begin{equation}\label{eq:contrastive_loss}
\begin{split}
\mathcal{L}_{\mathrm{cont}}=-\frac{1}{2B}\sum_{i}\Bigg[&
\log\frac{e^{c_{ii}/\tau_c}}{\sum_j e^{c_{ij}/\tau_c}}\\
&+\log\frac{e^{c_{ii}/\tau_c}}{\sum_j e^{c_{ji}/\tau_c}}\Bigg],
\end{split}
\end{equation}
where the temperature $\tau_c=0.1$ sets how sharply the loss distinguishes between similar and dissimilar pairs.

Intuitively, for each segment, the loss asks the encoder to pick out, among all the segments in the batch, the one that came from the same light curve (the positive pair), with the segments of the other light curves serving as alternatives (the negative pairs). Since only the \emph{relative} similarity matters, segments of different light curves are not required to be maximally dissimilar, which is important because two different light curves may have nearly identical underlying parameters. By optimizing this objective jointly with the generative model, gradients flow backward through the encoder to promote learning of the features that differentiate one stochastic process from another. This naturally incentivizes the encoder to learn statistical structure reflecting the hidden parameters that govern how variability differs across the population.

\subsection{Consistency Loss: Encouraging Adherence to Learned Structure}
\label{subsec: consistency_loss}

The generative model is trained conditionally using a representation $\mathbf{z} = E_\phi(\mathbf{x}^{(c)})$ obtained from a conditioning segment.
In standard conditional generative models, providing such conditioning is sufficient to guide generation. However, for stochastic noise processes, the posterior over valid light curves is inherently broad: many different realizations are consistent with the same underlying process. Conditioning alone does not guarantee that generated samples preserve the specific parametric properties encoded in $\mathbf{z}$.

To encourage the generative model to preserve and utilize the statistical structure learned by the encoder, we introduce a consistency loss that aligns generated samples with the conditioning representation in latent space.

A key property of flow matching is that the dynamics are defined along a linear interpolation between $\mathbf{x}_0$ and $\mathbf{x}_1$, allowing us to approximate the final sample without full ODE integration:
\begin{equation}
    \hat{\mathbf{x}}_1 \approx \mathbf{x}_t + (1 - t)\,\mathbf{v}_\theta(\mathbf{x}_t, t \mid \mathbf{z}).
\end{equation}

Given a conditioning segment $\mathbf{x}^{(c)}$ and the approximate generated sample $\hat{\mathbf{x}}_1$, we compute their embeddings:
\begin{equation}
    \mathbf{z}_c = E_\phi(\mathbf{x}^{(c)}), \quad
    \mathbf{z}_g = E_\phi(\hat{\mathbf{x}}_1),
\end{equation}
and define the consistency loss as:
\begin{equation}\label{eq:consistency_loss}
    \mathcal{L}_{\mathrm{cons}} = \tfrac12\mathbb{E}\left[1-\mathrm{sim}(\mathbf{z}_c, \mathbf{z}_g)\right].
\end{equation}

This objective encourages the generator to produce samples whose latent representation matches the conditioning signal. Intuitively, it encourages the model to generate realizations with statistical properties consistent with the conditioning signal, preserving the structure represented by the encoder.

\subsection{Effect of the additional losses}

We compare all four combinations of consistency and InfoNCE losses, always retaining flow matching, with three independent training runs per variant and the same 134.00M-parameter architecture. At each coefficient, 64 independent length-256 conditioning curves and 64 independent targets share the same true coefficient, with innovation SD 0.2 and burn-in 256. Each trained model generates 32 realizations per conditioning curve using 40 integration-grid points with Euler integration. The shared evaluation seed is 20261004 and sampling seed 4102.

\begin{figure*}\centering
\includegraphics[width=0.74\textwidth]{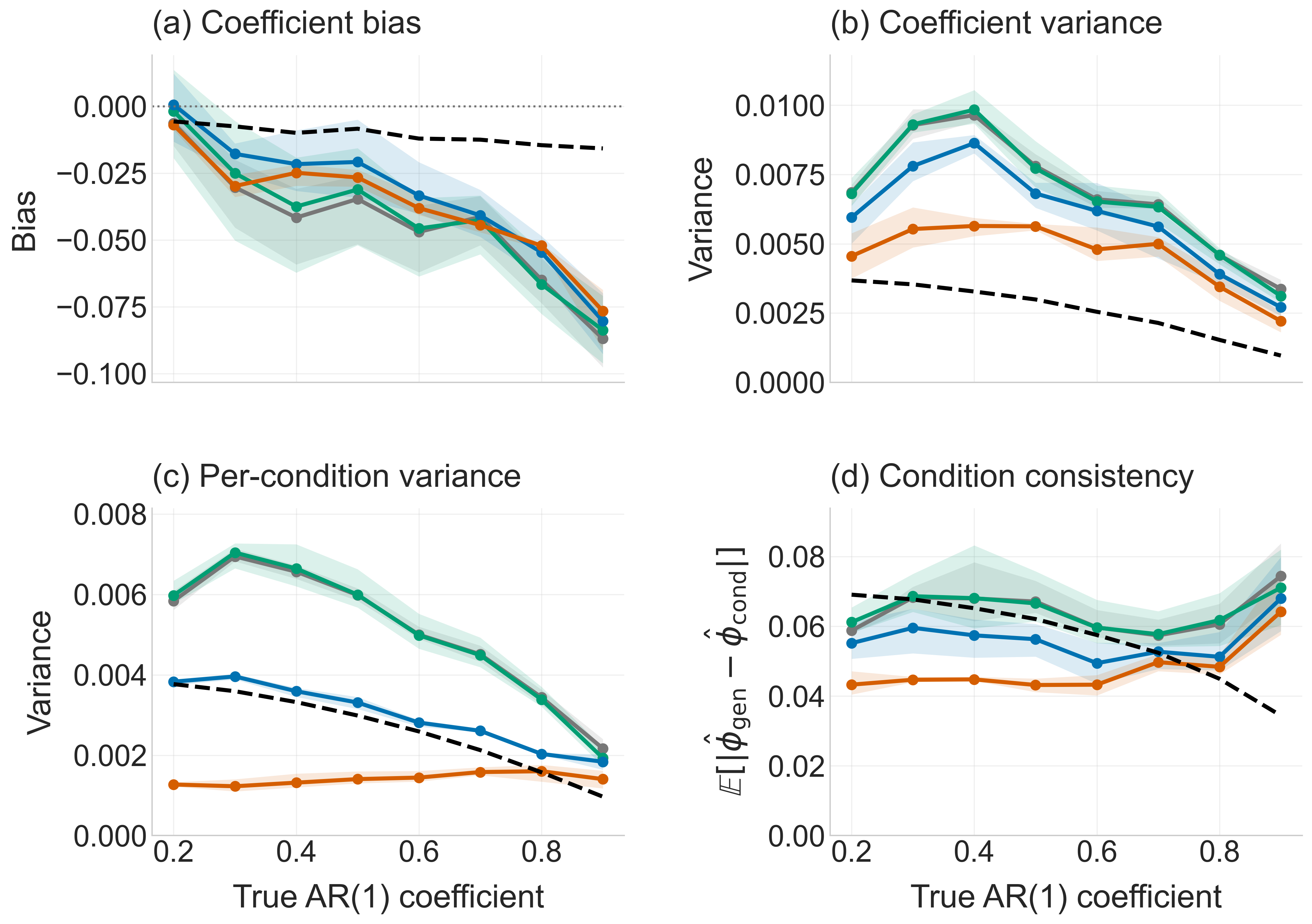}
\caption{Conditional AR(1) generation using \textcolor{ablationGrey}{flow matching alone (grey)}, \textcolor{ablationGreen}{with the consistency objective (green)}, \textcolor{ablationBlue}{with InfoNCE (blue)}, or \textcolor{ablationOrange}{with both auxiliary objectives (orange)}. At each true coefficient, 64 independent length-256 conditions each produce 32 draws. Panels show (a) coefficient bias from one realization per observation, (b) variance across observations using one realization per observation, (c) mean within-condition variance across 32 draws, and (d) mean absolute difference from the conditioning fit, using one realization per observation. All coefficients are estimated with centered Yule--Walker. The black dashed reference uses 4,096 independent groups of 32 known-coefficient AR(1) realizations per coefficient, with independent reference conditions; the \textcolor{ablationGrey}{grey dotted line} in (a) marks zero bias. Lines average three training seeds and shading spans their minimum--maximum. Innovation SD is 0.2. The title ``Condition consistency'' in panel (d) refers to fitted-coefficient agreement, not the embedding-based consistency objective. Exact definitions are in Appendix~\ref{appendix:ablation_metrics}; reduced spread alone does not establish predictive calibration.}
\label{fig:ablation}\end{figure*}

We estimate AR(1) coefficients from observed and generated curves using centered Yule--Walker fitting. Figure~\ref{fig:ablation} compares bias and variance across conditions, mean within-condition variance, and mean absolute deviation from the conditioning fit. The direct reference uses the true coefficient and 4,096 independent groups of 32 AR(1) realizations at each coefficient, together with 4,096 independently simulated reference conditioning curves. Its larger sample reduces Monte Carlo variation; it is not a posterior predictive inferred from an observation. Lines show the mean over three training runs; shading spans the lowest and highest values among those runs. Bias is the mean fitted coefficient minus its true value: negative values indicate systematic underestimation. Variance across observations measures the scatter of fitted coefficients from one generated realization per conditioning curve. Within-observation variance measures the spread across repeated realizations conditioned on the same curve. The mean absolute difference from the conditioning fit measures agreement with that noisy estimate, rather than accuracy relative to the true coefficient. Root-mean-square error (RMSE), used in the complementary evaluation below, summarizes the overall coefficient error, including both bias and scatter; smaller values indicate more accurate recovery. Exact definitions are given in Appendix~\ref{appendix:ablation_metrics}.

Figure~\ref{fig:ablation} shows that in all cases, the model captures the overall relationship between the conditioning signal and the underlying coefficient. Adding both losses (orange) strongly reduces the spread of the coefficients of the generated light curves: both the variance across conditioning light curves (panel b) and the variance among realizations generated from the same conditioning light curve (panel c) decrease, and the latter falls well below the variance of independent realizations of the true process (black dashed line). In contrast, the bias (panel a) is not reduced, and is similar for all four variants. In other words, the light curves generated with both losses are too similar to each other: they underestimate the range of light curves that the noise process can produce.

A separate evaluation of the predictive spread (Appendix~\ref{appendix:predictive}) confirms this: the 90\% predictive intervals of the model trained with both losses contain far fewer than 90\% of independent realizations, whereas the other variants come closer to the nominal coverage (Table~\ref{tab:coverage}).
\begin{table}\centering
\caption{Coverage of nominal 90\% predictive intervals for the AR(1) coefficient of a new realization: the fraction of independent target light curves whose fitted coefficient falls inside the interval derived from the generated light curves. A well-calibrated interval should contain about 90\%. Ranges are over three independently trained models.}\label{tab:coverage}
\begin{tabular}{lc}\hline Variant & Coverage \\ \hline
Both additional losses & 46.0--50.5\% \\
InfoNCE only & 66.5--72.0\% \\
Consistency only & 83.5--85.5\% \\
Flow matching only & 82.0--85.5\% \\
AR(1) posterior predictive & 88.0\% \\ \hline
\end{tabular}\end{table}

\paragraph*{Choice of training objective.}\label{subsec:loss_choice} Since a null hypothesis must represent the full range of noise realizations, all models used in the experiments of this paper are trained with the flow-matching loss only.

Table~\ref{tab:ablation_rmse} gives a separate coefficient-recovery evaluation using one generated realization per observation: 500 conditions at each of eight coefficients, one draw per condition, and 40 integration-grid points. It has a different evaluation design from the four-panel figure. Bias and RMSE are retained summaries from 4,000 coefficient estimates per training run (500 at each coefficient), and the table reports their means and sample SDs across three runs; the direct reference fits the corresponding conditioning curves. These values are retained from the recorded aggregate summary; its generating calculation is not included in the archived source. Root-mean-square error (RMSE) measures the overall coefficient error, combining systematic bias and random scatter: $\mathrm{RMSE}=[n^{-1}\sum_{i=1}^{n}(\hat\phi_i-\phi_i)^2]^{1/2}$, where $n$ is the number of evaluated curves, $\hat\phi_i$ is the fitted coefficient and $\phi_i$ its true value. Smaller RMSE and bias closer to zero indicate more accurate recovery. Squaring gives greater weight to large errors. A smaller error for one generated curve does not show whether the full spread of generated curves represents uncertainty correctly.
\begin{table}\centering
\caption{Error in the fitted AR(1) coefficient from one generated curve per observation in a separate evaluation. Bias is mean estimate minus truth; RMSE is the typical size of the error, including bias and scatter. Lower RMSE is better. Values average three training runs, with their sample standard deviation (SD); the direct reference fits the conditioning curves themselves rather than independent predictive curves.}\label{tab:ablation_rmse}
\begin{tabular}{lcc}\hline Variant & Bias & RMSE \\ \hline
Both auxiliary objectives & $-0.0356\pm0.0020$ & $0.0781\pm0.0020$ \\
InfoNCE only & $-0.0316\pm0.0081$ & $0.0858\pm0.0030$ \\
Consistency only & $-0.0403\pm0.0147$ & $0.0958\pm0.0077$ \\
Flow matching alone & $-0.0427\pm0.0104$ & $0.0962\pm0.0052$ \\
Direct reference & $-0.0105$ & $0.0521$ \\ \hline
\end{tabular}\end{table}

\section{Additional details of Experiments 1--3}\label{appendix:exp_details}
\subsection{Experiment 1}
The analytic target uses $\tau=-1/\log(0.95)\simeq19.496$ sample units and $\hat\sigma=0.08$. This spectral amplitude parameter differs from the innovation SD in the later AR experiments. We use a one-sided PSD convention: for an interior positive-frequency bin, the periodogram is $(2\Delta t/T)|\tilde x_k|^2$, with half that prefactor at DC and Nyquist. This continuous DRW spectrum is used as the target for finite Fourier synthesis, rather than identified with the exact spectrum of a sampled discrete AR(1) process. Within this construction, DC is fixed to zero and the Nyquist coefficient is real Gaussian with variance $T\,\mathrm{PSD}(f_{T/2})/\Delta t$. These special real-valued bins are excluded from the displayed marginal comparisons. We train a separate unconditional generator on each dataset, using complete 256-point curves and only the flow-matching objective. These models have no conditioning encoder or auxiliary losses; population statistics are evaluated without an observed conditioning curve. After training, we generate 3,000 new light curves per population for the PSD comparison 
and compute their average PSDs; separate marginal-distribution plots compare amplitudes $|\tilde{x}_k|$ across realizations. 
The amplitude-distribution plots have a recorded execution count of 10,000 generated curves per population; the retained-source provenance limitation is described at the end of Appendix~\ref{appendix:provenance}. These figures show training run 1; variation across training runs is not displayed.

\subsection{Experiment 2}\label{appendix:exp2_details}
We test population-level statistical reproduction without supplying AR(2) coefficients to the light-curve flow. Each dataset contains 100,000 parent curves of length 2,048 with Gaussian innovation SD 0.2 and no measurement noise. The simulator starts with an innovation-scaled Gaussian value and discards 100 initial samples before retaining each parent; the reference windows below therefore begin after this burn-in. Coefficients follow the mixture in Eqs.~\ref{eq:cen1}--\ref{eq:cen2}, with componentwise rejection outside $|\phi_1|+|\phi_2|\leq0.93$. Component selection precedes rejection. This diamond is a sufficient stability restriction, not the entire stationarity region. These models use the parent split and training construction in Appendix~\ref{subsec:construction}, and are trained with the flow-matching loss only. Population samples are drawn from the same trained model without supplying a conditioning observation.

We fit centered Yule--Walker AR(2) coefficients to the first 256 points of 4,000 curves excluded from training and to 4,000 curves generated without a supplied observation per training run, using 100 integration steps. This assesses the generated population rather than recovery for a particular conditioning observation. A disjoint set of 4,000 held-out parents supplies a finite data--data benchmark. Curve length and estimator are matched throughout. We quantify the difference with the Wasserstein distance $W_1$: the smallest average Euclidean distance needed to pair the two equally weighted samples in the $(\phi_1,\phi_2)$ plane. Smaller values mean closer coefficient distributions. We calculate the exact minimum-distance pairing on 2,000-point subsets, repeated ten times (ten random subset selections, using seeds 0--9). Reported $\pm$ values are population SD across subsets, not confidence intervals or training-seed uncertainty. The plotted kernel density estimates (KDEs) smooth the fitted coefficient pairs into density maps; $W_1$ is calculated from the pairs themselves, not from those smoothed maps. Comparing two independent held-out data samples shows the mismatch expected from finite sampling even without a generator.

\begin{table}\centering
\caption{Difference between fitted AR(2) coefficient distributions. $W_1$ is the smallest average distance needed to pair coefficient samples; lower is closer. The data--data row gives the finite-sampling mismatch without a generator. Values are mean $\pm$ SD over ten subset pairings, not confidence intervals.}\label{tab:wasserstein}
\begin{tabular}{lc}\hline Comparison & $W_1$ \\ \hline
Training run 1 vs reference data & $0.01679\pm0.00107$ \\
Training run 2 vs reference data & $0.01435\pm0.00095$ \\
Training run 3 vs reference data & $0.02349\pm0.00130$ \\
Independent reference samples & $0.01635\pm0.00117$ \\ \hline
\end{tabular}\end{table}

\subsection{Experiment 3}\label{appendix:conditional_details}
The evaluation of Fig.~\ref{fig:conditional} uses eight coefficients $\phi=0.2,0.3,\dots,0.9$, with 500 conditioning light curves at each coefficient. Each conditioning light curve is an AR(1) process with innovation standard deviation 0.2 and no observation noise; 256 initial samples are discarded (burn-in) and the following 256 samples are kept. All coefficients are estimated with the centered lag-1 Yule--Walker estimator $\hat\phi=\gamma(1)/\gamma(0)$, where $\gamma(1)$ is the lag-1 autocovariance of the mean-subtracted light curve and $\gamma(0)$ its variance. For finite light curves this estimator is biased low, by approximately $-(1+3\phi)/T$, i.e. $\approx -0.014$ at $\phi=0.9$ for $T=256$. In the reference panel, the coefficient fitted to each conditioning light curve is clipped to $[0,0.99]$ before simulating the new light curve with the same settings; the clipping is rarely active. In the model panel, the conditioning light curve is encoded and one new light curve is generated per conditioning light curve, using 200 integration steps. Using fewer integration steps visibly biases the generated coefficients (Table~\ref{tab:steps}).
\begin{table}\centering
\caption{Mean absolute error of the generated AR(1) coefficients in Fig.~\ref{fig:conditional} as a function of the number of integration steps.}\label{tab:steps}
\begin{tabular}{cc}\hline Integration steps & Mean absolute error \\ \hline
30 & 0.049 \\
200 & 0.027 \\
500 & 0.025 \\ \hline
\end{tabular}\end{table}

The displayed model was trained with the generative loss only, for 100,000 steps (training seed 11). It was selected among three independently trained models (seeds 11, 22, and 33) as the one whose mean estimates are closest to the identity line: seed 22 performs almost identically to seed 11, while seed 33 performs worse. The bands in Fig.~\ref{fig:conditional} are empirical quantiles over the 500 generated light curves at each coefficient, and combine the randomness of the conditioning light curve, the generative model and the final fit; they are not confidence intervals for the mean. The per-coefficient results are given in Table~\ref{tab:conditional}. Summary statistics averaged over the eight coefficients are given in Table~\ref{tab:conditional_summary}.
\begin{table}\centering
\caption{Summary of Fig.~\ref{fig:conditional}, averaged over the eight coefficients: mean absolute bias of the mean estimate, mean standard deviation (SD) and mean interquartile range (IQR) of the estimated coefficients.}\label{tab:conditional_summary}
\begin{tabular}{lccc}\hline & Mean $|$bias$|$ & Mean SD & Mean IQR \\ \hline
Reference & 0.020 & 0.071 & 0.096 \\
Generative model & 0.022 & 0.083 & 0.114 \\ \hline
\end{tabular}\end{table}
\begin{table}\centering
\caption{Estimated AR(1) coefficients after regeneration (Fig.~\ref{fig:conditional}), for 500 conditioning light curves per coefficient: mean, standard deviation (SD), and 5--95\% range of the estimates for the reference (fit and simulate) and the generative model.}\label{tab:conditional}
\footnotesize\setlength{\tabcolsep}{3pt}
\begin{tabular}{ccccccc}\hline
$\phi$ & \multicolumn{3}{c}{Reference} & \multicolumn{3}{c}{Generative model} \\
 & Mean & SD & 5--95\% & Mean & SD & 5--95\% \\ \hline
0.2 & 0.186 & 0.087 & 0.029--0.330 & 0.201 & 0.091 & 0.059--0.361 \\
0.3 & 0.288 & 0.085 & 0.145--0.419 & 0.293 & 0.102 & 0.118--0.444 \\
0.4 & 0.384 & 0.082 & 0.257--0.512 & 0.389 & 0.095 & 0.236--0.537 \\
0.5 & 0.482 & 0.077 & 0.362--0.604 & 0.479 & 0.098 & 0.310--0.638 \\
0.6 & 0.577 & 0.075 & 0.445--0.698 & 0.576 & 0.088 & 0.427--0.710 \\
0.7 & 0.678 & 0.063 & 0.570--0.783 & 0.674 & 0.079 & 0.541--0.791 \\
0.8 & 0.772 & 0.056 & 0.677--0.856 & 0.763 & 0.071 & 0.637--0.868 \\
0.9 & 0.872 & 0.043 & 0.790--0.932 & 0.850 & 0.043 & 0.771--0.911 \\ \hline
\end{tabular}\end{table}

\section{Calibration of the detection experiments}\label{appendix:calibration}
\subsection{Definitions}
The AR(2) evaluation uses 1,000 fresh independent 256-point observations, 500 with injected periodic signal and 500 null, evaluation seed 7 and the truncated coefficient mixture described in Section~\ref{subsec:exp2_latent_structure}. Signal-containing curves have random-phase sinusoidal periods of 30--80 sampling intervals. The centered sinusoid is normalized to unit RMS and scaled to a sampled ratio $r\sim U(0.8,1.2)$ times the centered unperturbed process RMS on the same finite window:
\begin{equation}\label{eq:snr}
\mathrm{SNR}_{\rm RMS}=\frac{\sqrt{T^{-1}\sum_t(s_t-\bar s)^2}}{\sqrt{T^{-1}\sum_t(n_t-\bar n)^2}}=r.
\end{equation}
Here $s_t$ is the added sinusoid and $n_t$ is the original stochastic curve; subtracting their means defines the variability RMS in the tested window. This normalization sets signal strength relative to the stochastic variability in each tested window.

Our detection statistic is the largest periodogram power anywhere in the searched spectrum. We subtract the mean, apply a Hann taper $w_t$ to reduce spectral leakage at the window edges, and take the squared Fourier amplitude in each nonzero-frequency bin, including Nyquist:
\begin{equation}
S(x)=\max_{k=1,\ldots,T/2}\left|\mathcal F[w_t(x_t-\bar x)]_k\right|^2.
\end{equation}
The search is not restricted to the injected-period interval. P-values are computed with Eq.~\ref{eq:mc_p}.

The true-positive rate, or detection power, is the fraction of signal-containing curves correctly detected. The false-positive rate (FPR) is the fraction of noise-only curves incorrectly flagged. The area under the ROC curve (AUC) summarizes ranking: 0.5 corresponds to random ranking and 1 to perfect separation. AUC does not establish whether significance estimates are calibrated. At a nominal threshold $\alpha$, a calibrated test should falsely flag approximately a fraction $\alpha$ of noise-only curves. Bootstrap confidence intervals describe finite-test-sample uncertainty by repeatedly resampling test curves with replacement; they do not measure variation from retraining the model. 

Because the tested observation also determines its simulated null, correctly calibrated p-values are not guaranteed. Under a calibrated continuous test, noise-only p-values are uniform; finite Monte Carlo samples allow only discrete values. Too many small p-values mean too many false detections, whereas p-values near one indicate an overly conservative test.

For AR(1), all four methods use $N=100$ null simulations per observation; the first four AR(2) methods also use $N=100$, while the AR(2) coefficient-distribution comparator uses $N=300$. For $N=100$, the minimum p-value is $1/101\simeq0.00990$, so rejection at 1\% requires zero exceedances. For $N=300$, the minimum is $1/301\simeq0.00332$, and rejection at 1\% permits at most two exceedances. This coarse resolution limits tail calibration. For AR(2), three light-curve models trained using flow matching alone (seeds 11, 22, and 33) are evaluated separately on the same observations. ROC curves and histogram densities are averaged across checkpoints; p-values and null draws are not pooled. AUC uses 1,000 observations (500 in each group), whereas FPR uses a separate 1,000 noise-only observations. Bootstrap intervals condition on the evaluated checkpoints and describe finite-test-sample uncertainty, while checkpoint variation describes training sensitivity. Detailed checkpoint settings and uncertainty calculations are given in Appendices~\ref{appendix:detection_checkpoints} and~\ref{appendix:detection_uncertainty}.

\subsection{AR(1) results}
Table~\ref{tab:ar1_calibration} gives the AUC and the false-positive rate (FPR) of each method. The FPR is measured on a separate set of 1,000 noise-only light curves. For the generative model, the nominal levels lie within the 95\% intervals of the measured FPRs. The p-value histograms in Fig.~\ref{fig:pvals_ar1} come from a separate evaluation. Full settings are given in Appendix~\ref{appendix:ar1_protocol}.

\begin{table*}\centering\footnotesize\setlength{\tabcolsep}{4pt}
\caption{Four-method AR(1) results. AUC uses the 1,000-total-curve ROC evaluation in Fig.~\ref{fig:roc_ar1}; FPR uses a separate set of 1,000 noise-only curves. Each p-value uses 100 simulated null curves. Parentheses give false-positive counts out of 1,000; FPR brackets give exact 95\% Clopper--Pearson intervals in percent; AUC brackets give 95\% stratified curve-bootstrap intervals. One model trained using flow matching alone is evaluated. AUC values correspond to Fig.~\ref{fig:roc_ar1}; the p-value histograms in Fig.~\ref{fig:pvals_ar1} come from a separate evaluation and are not used to calculate these FPRs.}
\label{tab:ar1_calibration}
\begin{tabular}{lccc}\hline
Method & AUC [95\% CI] & FPR at $\alpha=0.05$ & FPR at $\alpha=0.01$ \\ \hline
Fitted AR(1) & 0.9042 [0.8852, 0.9217] & 3.2\% (32) [2.20, 4.49] & 0.2\% (2) [0.02, 0.72] \\
Population AR(1), without fitting & 0.8291 [0.8019, 0.8546] & 5.3\% (53) [3.99, 6.88] & 1.5\% (15) [0.84, 2.46] \\
Oracle AR(1) & 0.9646 [0.9538, 0.9742] & 4.4\% (44) [3.21, 5.86] & 0.5\% (5) [0.16, 1.16] \\
Generative model (flow matching alone) & 0.9273 [0.9113, 0.9425] & 6.3\% (63) [4.87, 7.99] & 1.0\% (10) [0.48, 1.83] \\
\hline\end{tabular}\end{table*}

\subsection{AR(2) results}
\begin{table*}\centering\footnotesize
\caption{AR(2) detection ranking and false-positive rates. AUC uses 1,000 observations (500 in each group); FPR uses a separate set of 1,000 noise-only curves. The generative model is evaluated separately for each of three independently trained models (runs 1--3); the mean row averages the three evaluations.}
\label{tab:ar2_calibration}
\begin{tabular}{lccc}\hline
Method & AUC & FPR (\%), $\alpha=0.05$ & FPR (\%), $\alpha=0.01$ \\ \hline
Fitted AR(1), misspecified & 0.8505 & 41.60 & 23.70 \\
Fitted AR(2) & 0.9180 & 3.00 & 0.00 \\
Oracle AR(2) & 0.9855 & 5.50 & 0.70 \\
Generative model, run 1 & 0.9616 & 3.50 & 0.10 \\
Generative model, run 2 & 0.9662 & 3.40 & 0.10 \\
Generative model, run 3 & 0.9665 & 6.20 & 0.80 \\
Generative model, mean & 0.9648 & 4.37 & 0.33 \\
AR(2) coefficient flow & 0.9526 & 0.70 & 0.00 \\
\hline\end{tabular}\end{table*}

The mean FPRs of the generative model lie below the nominal levels, with some variation between training runs (Table~\ref{tab:ar2_calibration}). The fitted AR(2) null and the AR(2) coefficient flow are conservative at both levels, while the misspecified AR(1) null produces a large excess of false positives.

\subsection{The AR(2) coefficient flow and missing baselines}
The additional AR(2) comparator trains a small flow model on simulated light curves with known coefficient pairs to approximate $p(\phi_1,\phi_2\mid\mathbf x)$, the distribution of AR(2) coefficients compatible with a light curve. For each tested curve, it draws a new pair from this distribution and uses that pair to simulate an AR(2) curve for the Monte Carlo test. Unlike the single-fit method, these simulations therefore vary both the coefficients and the random noise realization. The figures label this method ``AR(2) coefficient flow''; it is the coefficient-distribution model described below. Its training uses known simulated coefficients; it is not the empirical-Bayes comparison that learns a coefficient prior from the paper's light-curve training set. The coefficient flow learns $p(\phi_1,\phi_2\mid\mathbf x)$ from labeled simulations and draws plausible coefficient pairs before simulating null curves. Its ranking performance is consistent with coefficient uncertainty contributing to the direct generator's performance, but does not establish that the light-curve encoder represents a coefficient posterior. No paired AUC-difference interval is supplied, so we do not claim a formal significance test of these ranking differences.

Population information is a plausible contributor to stronger ranking than independent plug-in fits, but it is not isolated here. In particular, a hierarchical AR(2) comparison would first learn how coefficient pairs vary across the training population, then combine that prior distribution with each observed curve to infer $p(\phi_1,\phi_2\mid x)$. Drawing coefficients from this distribution and simulating new curves propagates parameter uncertainty into the null. The available results do not include this empirical-Bayes comparison or Gaussian maximum-likelihood or frequency-domain Whittle fits. Therefore they cannot distinguish limitations of the AR(2) family from limitations of Yule--Walker estimation, or establish superiority to hierarchical AR(2) inference. Matching such a baseline would still support the interpretation of the flow as population-informed inference without a fixed autoregressive parameterization.

\section{Contamination experiments: details and additional results}\label{appendix:contamination}
We inject centered, RMS-normalized sinusoids into selected training parents in place, keeping the 90,000-parent training set fixed. Nominal fractions 0, 0.5, 1, 5, 10, and 20\% modify respectively 0, 450, 900, 4,500, 9,000, and 18,000 parents (contamination seed 2027). Held-out parents are not injected. Each modified parent receives a period drawn uniformly from 30--80 sampling intervals, a phase drawn uniformly from $[0,2\pi)$, and a sinusoid-to-parent RMS ratio drawn uniformly from 0.8--1.2. These draws use an independently seeded NumPy generator. The sine is centered over the full parent and normalized to unit RMS before scaling by the unperturbed parent's RMS. Sinusoids extend over the full 2,048-point parent; training uses nonoverlapping 256-point crops. Training SNR is defined over the parent and evaluation SNR over the 256-point observation, so their realized crop ratios can differ. These models use the flow-matching loss only, 134.00M parameters, 100,000 steps, and three independent training runs.

After training is complete, we simulate a new test set independently of the training data. None of these test light curves, or crops from them, is used during training. The same 400 independently generated 256-point test observations (200 periodic, 200 null) are then used to evaluate every contamination level, with evaluation seed 11, $\phi\sim U(0.1,0.9)$, innovation SD 0.2, no measurement noise, periods 30--80 samples, and RMS-SNR 0.8--1.2. Curve generation uses 100 integration steps with Euler integration. For each observation we combine 100 simulated curves from each of three independently trained models into one set of 300. We count how many of these curves have a maximum periodogram power above the observation and apply Eq.~\ref{eq:mc_p} with $N=300$. This defines the combined, or mixture, null. This is neither the mean of three p-values nor the mean of their AUCs. The fitted Yule--Walker AR(1) baseline uses 300 draws. Curve-bootstrap intervals and binomial FPR intervals follow Section~\ref{subsec:detection}; separate-seed AUC variation is reported in Table~\ref{tab:seed_auc}.

\begin{table*}\centering\footnotesize\setlength{\tabcolsep}{3pt}
\caption{Contamination results, including the 5\% level omitted from the figures. Each p-value compares an observation with 300 noise curves combined from three trained models. AUC measures separation; FPR measures false detections among 200 noise-only observations. A calibrated FPR should match the stated threshold. Intervals describe finite-test-sample uncertainty, not variation from retraining.}
\label{tab:current_calibration}
\begin{tabular}{lccc}\hline
Method & AUC [95\% CI] & FPR (\%), $\alpha=0.05$ [95\% CI] & FPR (\%), $\alpha=0.01$ [95\% CI] \\ \hline
AR(1) Baseline & 0.9699 [0.9568, 0.9819] & 4.00 [1.74, 7.73] & 0.00 [0.00, 1.83] \\
Nullasar (0\%) & 0.9904 [0.9815, 0.9965] & 10.50 [6.62, 15.60] & 1.00 [0.12, 3.57] \\
Nullasar (0.5\%) & 0.5735 [0.5173, 0.6309] & 9.50 [5.82, 14.44] & 1.50 [0.31, 4.32] \\
Nullasar (1\%) & 0.5557 [0.4952, 0.6114] & 9.50 [5.82, 14.44] & 1.50 [0.31, 4.32] \\
Nullasar (5\%) & 0.5387 [0.4807, 0.5967] & 7.50 [4.26, 12.07] & 1.00 [0.12, 3.57] \\
Nullasar (10\%) & 0.5218 [0.4638, 0.5778] & 9.00 [5.42, 13.85] & 1.50 [0.31, 4.32] \\
Nullasar (20\%) & 0.5430 [0.4861, 0.5990] & 8.50 [5.03, 13.26] & 1.00 [0.12, 3.57] \\
\hline\end{tabular}\end{table*}

Table~\ref{tab:current_calibration} gives the AUC and FPR for each contamination level, including the 5\% level that is omitted from the figures. The clean model ranks light curves better than the fitted AR(1) null, but has a higher FPR at the nominal 5\% level. Because the evaluation curves are newly simulated after training, the ranking gain cannot be attributed to reusing training light curves or their crops.

Already at 0.5\% contamination, the AUC drops close to chance, and it stays there, without a monotonic trend, at higher contamination levels. The results of each individual training run are given in Appendix~\ref{appendix:seed_results}.

\subsection{Contamination results by training run}\label{appendix:seed_results}
These checkpoint-specific results use 100 null draws per observation. Each AUC interval resamples the 400 shared test curves; each FPR interval uses the 200 null observations. They are distinct from the mixture-null results and do not estimate an interval over retraining.
\begin{table}\centering
\caption{Sensitivity to training randomness. For each contamination fraction, three independently trained models each use 100 simulations per observation. The table gives their mean AUC and sample standard deviation (SD). These are separate evaluations, unlike the figures that combine 300 simulations into one null before calculating p-values and AUC.}\label{tab:seed_auc}
\begin{tabular}{lcc}\hline Contamination & Mean AUC & SD across runs \\ \hline
0\% & 0.9802 & 0.0026 \\
0.5\% & 0.5771 & 0.0605 \\
1\% & 0.5659 & 0.0870 \\
5\% & 0.5419 & 0.0519 \\
10\% & 0.5309 & 0.0867 \\
20\% & 0.5472 & 0.0095 \\ \hline
\end{tabular}\end{table}

\begin{table*}\centering\footnotesize\setlength{\tabcolsep}{3pt}
\caption{Results for each individual training run, including 5\% contamination. Each p-value uses 100 simulations from that run. AUC measures signal/noise separation and FPR the noise-only false-detection rate. Brackets give 95\% test-curve resampling and exact binomial intervals, respectively; differences between rows reveal training variation.}
\label{tab:seed_full}
\begin{tabular}{lccc}\hline
Method & AUC [95\% CI] & FPR (\%), $\alpha=0.05$ [95\% CI] & FPR (\%), $\alpha=0.01$ [95\% CI] \\ \hline
Nullasar (0\%) / run 1 & 0.9813 [0.9681, 0.9925] & 9.00 [5.42, 13.85] & 2.00 [0.55, 5.04] \\
Nullasar (0\%) / run 2 & 0.9821 [0.9678, 0.9928] & 10.50 [6.62, 15.60] & 2.00 [0.55, 5.04] \\
Nullasar (0\%) / run 3 & 0.9773 [0.9616, 0.9903] & 14.00 [9.51, 19.59] & 3.00 [1.11, 6.42] \\
Nullasar (0.5\%) / run 1 & 0.6071 [0.5522, 0.6612] & 8.00 [4.64, 12.67] & 2.00 [0.55, 5.04] \\
Nullasar (0.5\%) / run 2 & 0.5074 [0.4488, 0.5641] & 13.00 [8.67, 18.47] & 2.50 [0.82, 5.74] \\
Nullasar (0.5\%) / run 3 & 0.6168 [0.5644, 0.6694] & 13.00 [8.67, 18.47] & 2.50 [0.82, 5.74] \\
Nullasar (1\%) / run 1 & 0.5240 [0.4657, 0.5803] & 9.50 [5.82, 14.44] & 2.00 [0.55, 5.04] \\
Nullasar (1\%) / run 2 & 0.5078 [0.4512, 0.5630] & 9.00 [5.42, 13.85] & 3.50 [1.42, 7.08] \\
Nullasar (1\%) / run 3 & 0.6660 [0.6088, 0.7162] & 13.00 [8.67, 18.47] & 1.50 [0.31, 4.32] \\
Nullasar (5\%) / run 1 & 0.5930 [0.5342, 0.6463] & 8.50 [5.03, 13.26] & 1.00 [0.12, 3.57] \\
Nullasar (5\%) / run 2 & 0.5433 [0.4876, 0.5977] & 9.50 [5.82, 14.44] & 2.00 [0.55, 5.04] \\
Nullasar (5\%) / run 3 & 0.4893 [0.4306, 0.5480] & 12.00 [7.84, 17.33] & 1.50 [0.31, 4.32] \\
Nullasar (10\%) / run 1 & 0.5702 [0.5114, 0.6251] & 11.00 [7.02, 16.18] & 2.50 [0.82, 5.74] \\
Nullasar (10\%) / run 2 & 0.5909 [0.5343, 0.6440] & 9.50 [5.82, 14.44] & 3.00 [1.11, 6.42] \\
Nullasar (10\%) / run 3 & 0.4315 [0.3748, 0.4867] & 9.00 [5.42, 13.85] & 1.00 [0.12, 3.57] \\
Nullasar (20\%) / run 1 & 0.5578 [0.4997, 0.6156] & 9.00 [5.42, 13.85] & 0.00 [0.00, 1.83] \\
Nullasar (20\%) / run 2 & 0.5447 [0.4877, 0.6027] & 9.00 [5.42, 13.85] & 1.50 [0.31, 4.32] \\
Nullasar (20\%) / run 3 & 0.5392 [0.4806, 0.5961] & 14.00 [9.51, 19.59] & 2.50 [0.82, 5.74] \\
\hline\end{tabular}\end{table*}

\section{Training and evaluation details}\label{appendix:protocols}
\subsection{Implementation details}\label{appendix:implementation}
Time is measured in sampling intervals and light-curve amplitudes are in simulation units; these populations are controlled variability benchmarks rather than a physical calibration to a particular quasar survey. To assess sensitivity to training randomness, we repeat the population-level AR(2), conditional-generation, and contamination experiments with three independently trained models. A random seed is a numerical setting that makes random choices reproducible; its value has no physical meaning. Different training seeds allow us to assess whether the results depend on a particular training realization. We use seeds 11, 22, and 33, hereafter runs 1, 2, and 3. Evaluation curves are shared across training runs where specified, so that differences reflect model variation rather than changes in the test sample. Seeds used to construct datasets or generate evaluation samples serve separate purposes. Some experiments use only one evaluated training run, as identified below; multiple-run checks therefore do not cover every result.

In the frequency-domain implementation, the flow coordinates are defined as follows. For the flow equations below, $\mathbf{x}_t$ denotes a real vector of model coordinates, with $t$ indexing generation rather than observing time. In the frequency-domain implementation, these coordinates pack the real parts of Fourier bins $0,\ldots,T/2$ followed by the imaginary parts of bins $1,\ldots,T/2-1$, giving $T$ real coordinates for even $T$. DC and Nyquist have zero imaginary part; the omitted negative-frequency coefficients follow by conjugate symmetry. We denote the Fourier transform followed by this packing by $\mathcal P$ and its inverse by $\mathcal P^{-1}$, which returns a real light curve. The forward Fourier transform is unnormalized and its inverse includes the factor $1/T$. Samples from the base distribution are denoted $\mathbf{x}_0\sim p_0=\mathcal N(\mathbf0,\mathbf I)$ in these working coordinates, while $\mathbf{x}_1\sim p_{\rm data}$ denotes a training curve represented in the same coordinates. Thus the Gaussian source is specified in the model domain, rather than assumed to be white noise in both domains.

Conditional models are trained with both an unconditional and a conditional flow-matching term:
\begin{equation}
\mathcal L_{\rm total}=\mathcal L_{\rm FM}^{\rm uncond}
+\mathcal L_{\rm FM}^{\rm cond}.
\end{equation}
Both terms have unit weight. Each flow-matching term averages squared velocity error over the batch and signal coordinates. Independent Gaussian starting states and generation times are drawn for the two terms. Omitting the conditioning summary at generation uses the unconditional branch trained by $\mathcal L_{\rm FM}^{\rm uncond}$. The Fourier-amplitude models have no conditioning encoder and minimize only this unconditional term.

For direct light-curve generation we use Euler integration with 100 integration steps in all experiments, except for the conditional generation experiment (Section~\ref{subsec:conditional}), which uses 200 steps. The coefficient-distribution comparator instead uses 40 midpoint updates (Appendix~\ref{appendix:detection_checkpoints}).
\subsection{Training construction and uncertainty summaries}\label{appendix:training}
\paragraph*{Training construction and model provenance.}\label{subsec:construction}
We refer to each full simulated light curve, before selecting training segments, as a \emph{parent curve}. For these runs, complete parent curves are permuted using split seed 2026 and divided into 90\% training and 10\% held-out parents before crops are selected. Each regular AR dataset has 100,000 parents of length 2,048: 90,000 training parents and 10,000 held-out parents. For each conditional training presentation, three distinct 256-point grid-aligned windows from the same parent supply the curve to be generated, the observed segment used to guide generation, and a second observed segment, which is used only by the contrastive loss of the ablation study in Appendix~\ref{appendix:ablation}. A regular parent has eight candidate windows. They do not overlap, but share latent parameters and may remain temporally correlated. In that ablation, other parents in the batch supply the negative pairs of the contrastive loss. Parents are revisited with resampled crops, so presentations and generated draws are not independent-parent counts. Unconditional amplitude runs use complete 256-point parents.

The regular conditional models contain 133,999,234 total parameters, including conditioning machinery. The U-Net has base width 128, multipliers $[1,2,4,8]$, three residual blocks per level, attention at resolution 64 with four heads, and conditioning/global dimensions 128. Training uses AdamW, learning rate $10^{-4}$, weight decay 0.05, batch size 500, and bfloat16. The standard training duration is 100,000 optimizer steps; the AR(2) detection models are evaluated at the shorter durations specified in Appendix~\ref{appendix:detection_checkpoints}. Seeds 11, 22, and 33 are used for population, ablation, and contamination evaluations. The Fourier-amplitude models have 104,680,449 parameters, no conditioning encoder. In conditional models, the encoder is trained jointly with the generator through the flow-matching loss.

Detection observations are freshly simulated and separate from training parents. In the contamination comparisons, the test set is generated only after training has finished; no test curve is reused from the training set or extracted from a training parent. Reusing this independent test set across trained models ensures a common evaluation sample, rather than overlap with training data. When testing for periodicity, the same observed curve is used both to construct its null (by conditioning the generator or fitting an AR model) and to measure the candidate signal. This reuse is separate from overlap with training data, and its effect on p-value calibration must be checked.

\paragraph*{Multiple training seeds and evaluation uncertainty.}\label{subsec:seeds}
The population-level AR(2), conditional-generation, and contamination experiments use independently trained models with random seeds 11, 22, and 33. Within each comparison, the architecture and train/test split are held fixed while the training seed changes random choices during training. Training settings are matched except where explicitly noted, including the different evaluated training durations in Appendix~\ref{appendix:detection_checkpoints}. A checkpoint is the set of learned network weights saved from one run. All runs are evaluated on the same curves, so differences between runs assess training variation; for the AR(2) detection models, differing training durations also contribute and are not isolated from seed effects. Figure~\ref{fig:ar2_recovery} displays training run 1 and Table~\ref{tab:wasserstein} reports all three. Figure~\ref{fig:ablation} plots equal-weight means across training runs, with shading spanning their full range. The contamination figures instead pool 100 null draws from each seed to define a 300-draw mixture null; Table~\ref{tab:seed_auc} and Appendix~\ref{appendix:seed_results} report checkpoint-specific variation separately. These summaries answer different questions and are not interchangeable.

The area under the receiver-operating-characteristic curve (AUC) summarizes detection ranking: 0.5 is chance and 1 is perfect separation. Its 95\% intervals are obtained by bootstrap resampling, repeatedly drawing test curves with replacement separately from the signal-present and noise-only groups. These intervals describe finite-test-sample uncertainty for the trained models and saved simulations. They do not measure uncertainty over retraining. Likewise, the Wasserstein subset SD measures sensitivity to finite transport samples, rather than training uncertainty. The Fourier-amplitude figures show training run 1 only, although three unconditional models were trained per population, and the AR(2) detection comparison averages separate evaluations of three light-curve models trained using flow matching alone and uses one coefficient-distribution training run. Multiple seeds therefore support the conditional-generation and contamination comparisons but do not supply multiseed uncertainty for every displayed experiment.

\subsection{Conditional-generation metrics}\label{appendix:ablation_metrics}
For reproducibility, at each fixed true $\phi$ and each training seed, write $c_p$ for the conditioning curve's centered Yule--Walker fit and $\hat\phi_{p,d}$ for the fitted coefficient of generated realization $d$ conditioned on observation $p$, with $P=64$ and $D=32$. Define $\overline{\hat\phi}_1=P^{-1}\sum_p \hat\phi_{p,1}$ and $\overline{\hat\phi}_p=D^{-1}\sum_d \hat\phi_{p,d}$. The four plotted statistics are
\begin{align*}
\text{(a)}\quad & P^{-1}\sum_p(\hat\phi_{p,1}-\phi),\\
\text{(b)}\quad & (P-1)^{-1}\sum_p(\hat\phi_{p,1}-\overline{\hat\phi}_1)^2,\\
\text{(c)}\quad & P^{-1}\sum_p\left[(D-1)^{-1}\sum_d(\hat\phi_{p,d}-\overline{\hat\phi}_p)^2\right],\\
\text{(d)}\quad & P^{-1}\sum_p|\hat\phi_{p,1}-c_p|.
\end{align*}
Thus the sample-variance denominators are 63 in (b) and 31 inside (c). We calculate each metric separately for each training run, then average with equal weight. For the shared black reference, the same definitions instead use $P=4096$ independent groups, $D=32$ direct known-$\phi$ draws per group, and independent conditioning-curve fits. Thus its between-group variance denominator is 4095, while the within-group denominator remains 31. The first draw in each group supplies (a), (b), and (d); all 32 draws supply (c). This reference uses simulation seed 5102 and does not involve a trained model or uncertainty in $\phi$. All fits use the same estimator. The grey dotted line in (a) marks zero bias.

\subsection{Evaluation of predictive uncertainty}\label{appendix:predictive}
A separate evaluation of predictive uncertainty uses 200 independent condition/target pairs with $\phi\sim U(0.1,0.9)$, 64 generated draws per condition, and known innovation SD 0.2. For each observed curve, the generated fits define a central 90\% interval; we then ask whether that interval contains the fit from a separately simulated curve with the same true coefficient. A well-calibrated predictive interval should succeed about 90\% of the time. This tests uncertainty for a new curve, not an interval for the underlying coefficient itself. The resulting coverage for each variant is given in Table~\ref{tab:coverage}. Thus the combined-loss model produces intervals that are too narrow or incorrectly centered for new realizations. Reassigning observations to unrelated conditions changes the predictions, confirming that the model uses its input, but does not resolve this uncertainty mismatch.

For the parametric predictive reference, the process mean is fixed at zero, the innovation SD is known to be 0.2, and $\phi$ has the same uniform prior on $[0.1,0.9]$ as the test population. We evaluate the stationary Gaussian AR(1) likelihood on 801 equally spaced coefficient values, including the stationary initial-value density with variance $0.2^2/(1-\phi^2)$ and the conditional transition densities. Normalized likelihood weights define the discretized posterior. We draw 64 coefficients per observation, simulate an independent length-256 curve for each with burn-in 256, and fit its coefficient with the same centered Yule--Walker estimator. The empirical 5th and 95th percentiles of these predictive fits define the central 90\% interval. The observation/target simulation seed is 2029, posterior coefficient sampling uses seed 7301, and predictive simulation uses seed 4102. This is a predictive distribution of finite-curve fitted coefficients, rather than an interval for the latent coefficient itself.

The encoder also assigns more similar summaries to pairs of segments from the same training curve than to pairs from a previously unseen curve (approximately 0.946--0.956 versus 0.449--0.487). This suggests that it may partly recognize training curves rather than only their stochastic properties. It does not establish the cause of the inadequate predictive intervals.

\subsection{AR(2) evaluation and uncertainty}\label{appendix:detection_uncertainty}
\paragraph*{Uncertainty.}
AUC uses a simulation call containing 1,000 ROC curves. FPR uses the noise-only second half of a separate 2,000-curve simulation call. Both calls reset the random-number generator to seed 7, so they should not be regarded as independent random streams across evaluations. The two summaries are evaluated separately and their samples are not combined into a larger effective test set. For each generative checkpoint we calculate p-values, ROC curves, and rejection rates separately on the same observations with matched Monte Carlo random seeds. The displayed ROC interpolates each seed's true-positive rate to a common false-positive-rate grid and averages those rates. Its legend reports the arithmetic mean of the individual AUCs, which may differ slightly from the area under the interpolated mean curve. Histograms average per-seed densities in common bins, and reported generative FPRs average the three rejection rates. We neither average p-values nor pool null draws into an ensemble.

The parametric nulls use the known innovation SD of 0.2 and zero observation noise; only the autoregressive coefficients are fitted. Their simulator discards 100 initial samples, whereas the coefficient-distribution comparator uses burn-in 256 after sampling each coefficient pair. Detection fits use centered sample covariances with lag-dependent denominators and a $10^{-8}$ diagonal regularizer in the Yule--Walker system. Fitted AR(2) coefficients are clipped individually to $[-0.999,0.999]$; no joint stationarity constraint is applied to these estimates.

Individual AUC intervals use stratified curve bootstraps and individual FPR intervals use exact binomial calculations. Intervals for mean AUC and FPR use 1,000 paired curve-bootstrap samples, resampling the same curve indices across seeds. They condition on the evaluated checkpoints and do not estimate uncertainty over all possible training runs. The 3,000 checkpoint--curve evaluations of noise-only observations are not 3,000 independent test curves. FPR is the fraction rejected at $p\leq\alpha$; nominal targets are 5\% and 1\%. Table~\ref{tab:ar2_calibration} reports the point estimates. With 1,000 independent null observations, zero rejections gives an upper endpoint of approximately 0.368\% for a two-sided 95\% Clopper--Pearson interval, rather than establishing zero population FPR.

\subsection{Detection checkpoints}\label{appendix:detection_checkpoints}
The direct light-curve generators use only the base flow-matching loss, following Section~\ref{subsec:loss_choice}. 
We evaluate three 133,999,234-parameter checkpoints with training seeds 11, 22, and 33, saved at 10,000, 10,000, and 15,000 optimizer steps, respectively. For these three runs, the most recently saved weights available to the evaluation script are used, rather than weights chosen by a validation score. These training durations differ across runs, so their variation cannot be attributed to random seeds alone. Each run uses 90,000 training and 10,000 held-out parents of length 2,048. Generation uses 100 integration steps. In the contamination evaluation, network evaluation uses bfloat16, flow states float32, and periodogram statistics float64. The coefficient-distribution model has 567,578 parameters and uses training seed 8401, AdamW with learning rate $10^{-3}$ and weight decay $10^{-4}$, and batches of 256 freshly simulated labeled curves. It was trained for 39,982 optimization steps (10,235,392 examples) under a 900-second training budget. A fixed independent validation set of 4,096 curves determines the retained weights by minimum validation flow-matching loss; the recorded best epoch is 292. Its encoder has channel widths 36, 72, and 144 and a 288-dimensional context. The output transform $\phi_1=0.93\tanh u_1$, $\phi_2=(0.93-|\phi_1|)\tanh u_2$ enforces $|\phi_1|+|\phi_2|<0.93$. Generation uses 40 midpoint updates. The complete 100,000-step baseline runs used for population, ablation, and contamination comparisons instead use their final saved weights, without validation-based model selection.

\subsection{AR(1) evaluation protocol}\label{appendix:ar1_protocol}
\paragraph*{AR(1) evaluation and false-positive rates.}
The AR(1) evaluation uses the same four comparison methods and the model trained using flow matching alone (training run 1). This model has 133,999,234 parameters and 100,000 training steps, with the parent split described in Appendix~\ref{subsec:construction}. Only one training seed is evaluated here. The ROC evaluation uses 1,000 independent length-256 observations in total, including periodic and noise-only curves, evaluation seed 17, $\phi\sim U(0.1,0.9)$, noise SD 0.2, and no additional observation noise. Injected periods are 30--100 samples and centered sinusoid-to-realized-noise RMS ratios are 0.5--0.8. Generation uses 100 integration steps. All four methods use the maximum non-DC Hann-windowed periodogram statistic and 100 null draws per observation, with raw $p=(b+1)/101$ and no post-hoc calibration. The resulting AUCs are given in Table~\ref{tab:ar1_calibration}.

FPR is measured separately on 1,000 noise-only observations, the noise-only half of a 2,000-curve simulation call with evaluation seed 17, using the same checkpoint, settings, and 100 null draws per observation. Table~\ref{tab:ar1_calibration} reports the rejection counts and exact 95\% binomial intervals.  Uncertainty here describes finite noise-only test counts, not variation across training seeds. The ROC sample of 1,000 curves in total, the separate FPR sample of 1,000 noise-only curves, and the 100 Monte Carlo draws per observation are distinct sample sizes.

\subsection{Source and retained-figure provenance}\label{appendix:provenance}
The Fourier-marginal figure execution notes record 10,000 generated curves per population, whereas the archived rendering notebook defaults to 50,000 and permits a sample-count override. Since the generated arrays and executed override are not retained in the source snapshot, 10,000 is a recorded execution count rather than a count independently verified from saved arrays. Likewise, Table~\ref{tab:ablation_rmse} retains the exported single-realization errors: its original aggregation code is absent, so the distinction between a pooled RMSE and an average of per-coefficient RMSEs cannot be independently reconstructed. We do not interpret these entries as predictive-distribution calibration. Missing raw p-value arrays also prevent reconstruction of the mean-model confidence intervals for Table~\ref{tab:ar2_calibration}; only retained point estimates are reported there.

\label{lastpage}
\end{document}